\documentstyle[a4p,12pt,epsfig,array,cite,pennames,rotating]{article}
\newcommand{\PPEnum} {CERN-PPE/97-101}

\newcommand{\Date}      {29th July 1997}
\newcommand{\inmath}[1] {\ifmmode#1\else$#1$\fi}
\newcommand{\definmath}[2] {\def#1{\ifmmode#2\else$#2$\fi}}
\newcommand{\alphas} {\alpha_{\mathrm{s}}}
\newcommand{\alphaem} {\mbox{$\alpha_{\mathrm{em}}$}}
\newcommand{\alphaemsq} {\mbox{$\alpha^{2}_{\mathrm{em}}$}}

\newcommand{\PZ}   {\mbox{$\mathrm{Z}$}}  % Z without a zero preferred
\definmath{\PWpm} {\mathrm{W}^{\pm}}      % W+-
\definmath{\Plp} {\ell^{+}}        % l+
\definmath{\Plm} {\ell^{-}}        % l-
\definmath{\Plpm}   {\ell^{\pm}}         % l+-
\definmath{\Pgtp} {\tau^{+}}        % tau+
\definmath{\Pgtm} {\tau^{-}}        % tau-
\definmath{\Pgtpm}   {\tau^{\pm}}         % tau+-
\definmath{\Pgn}  {\nu}          % neutrino
\definmath{\Pagn} {\overline{\nu}}     % anti-neutrino
\definmath{\Pf}      {\mathrm{f}}
\definmath{\Paf}  {\overline{\mathrm{f}}}
\definmath{\Pq}      {\mathrm{q}}
\definmath{\Paq}  {\overline{\mathrm{q}}}
\definmath{\Pu}      {\mathrm{u}}
\definmath{\Pau}  {\overline{\mathrm{u}}}
\definmath{\Pd}      {\mathrm{d}}
\definmath{\Pad}  {\overline{\mathrm{d}}}
\definmath{\Ps}      {\mathrm{s}}
\definmath{\Pas}  {\overline{\mathrm{s}}}
\definmath{\Pc}      {\mathrm{c}}
\definmath{\Pac}  {\overline{\mathrm{c}}}
\definmath{\Pb}      {\mathrm{b}}
\definmath{\Pab}  {\overline{\mathrm{b}}}
\definmath{\Pt}      {\mathrm{t}}
\definmath{\Pat}  {\overline{\mathrm{t}}}
\definmath{\Pap}  {\overline{\mathrm{p}}}
\definmath{\Pan}  {\overline{\mathrm{n}}}
\definmath{\PaD}  {\overline{\mathrm{D}}}
\definmath{\PaDz} {\overline{\mathrm{D}}^{0}}
\definmath{\PaB}  {\overline{\mathrm{B}}}
\definmath{\PaBz} {\overline{\mathrm{B}}^{0}}
\definmath{\PsDpm}   {\mathrm{D}^{\pm}_{\mathrm{s}}}  % Ds+-
\definmath{\PcgLpm}  {\Lambda^{\pm}_{\mathrm{c}}}  % Lambda_c+-
\definmath{\PD} {\mathrm{D}}     % D
\definmath{\PDst} {\mathrm{D}^{*}}     % D*
\definmath{\PgLz} {\Lambda^{0}}        % Lambda0
\newcommand{\PKL} {\mbox{$\mathrm{K}^{0}_{\mathrm{L}}$}}

\newcommand {\gluino}        {\tilde{\mathrm g}}
\newcommand {\squark}        {\tilde{\mathrm q}}
\newcommand{\snu}{\tilde{\nu}}
\newcommand{\chp}{\tilde{\chi}^+}
\newcommand{\chm}{\tilde{\chi}^-}
\newcommand{\chpm}{\tilde{\chi}^\pm}
\newcommand{\nt}{\tilde{\chi}^0}
\newcommand{\massof}[1] {m_{\smash{#1}\mathstrut}}
\newcommand{\mtop}   {\massof{\mathrm{top}}}
\newcommand{\mHiggs} {\massof{\mathrm{Higgs}}}

\newcommand{\mPW} {\massof{\mathrm{W}}}
\newcommand{\mPZ} {\massof{\mathrm{Z}}}
\newcommand{\mX}  {\massof{\mathrm{X}}}
\newcommand{\Rb}  {\mbox{$R_{\rm b}$}}
\newcommand{\Rc}  {R_{\rm c}}

\newcommand{\AFB}    {A_{\mathrm{FB}}}
\newcommand{\AFBSM}  {A_{\mathrm{FB}}^{\mathrm{SM}}}
\newcommand{\jtoth}  {\mathrm{j^{tot}_{had}}}
\newcommand{\jfbl}   {\mathrm{j^{fb}_{\ell}}}

\newcommand{\thacol}{\theta_{\mathrm{acol}}}

\newcommand{\Lsig}      {L/\sigma_{L}}
\newcommand{\epem}   {\Pep\Pem}
\newcommand{\gamgam} {\Pgg\Pgg}
\newcommand{\mumu}   {\Pgmp\Pgmm}
\newcommand{\tautau} {\Pgtp\Pgtm}
\newcommand{\ff}     {\Pf\Paf}
\newcommand{\fpfp}   {\mathrm{f'}\overline{\mathrm{f'}}}

\newcommand{\qqbar}  {\Pq\Paq}
\newcommand{\uubar}  {\Pu\Pau}
\newcommand{\ddbar}  {\Pd\Pad}

\newcommand{\ccbar}  {\Pc\Pac}
\newcommand{\bbbar}  {\Pb\Pab}

\newcommand{\eetogg}    {\epem\to\gamgam}
\newcommand{\eetoee}    {\epem\to\epem}

\newcommand{\eetomumu}     {\epem\to\mumu}
\newcommand{\eetotautau}   {\epem\to\tautau}

\newcommand{\eetoqq}    {\epem\to\qqbar}

\newcommand{\dsdcc}        {{\rm d}\sigma/{\rm d}\cos\theta}
\newcommand{\dsdabscc}     {{\rm d}\sigma/{\rm d}|\cos\theta|}
\newcommand{\roots} {\sqrt{s}}

\newcommand{\Evis}   {\mbox{$E_{\mathrm{vis}}$}}
\newcommand{\Rvis}   {\mbox{$R_{\mathrm{vis}}$}}
\newcommand{\Rbal}   {\mbox{$R_{\mathrm{bal}}$}}
\newcommand{\Eclus}  {E_{\mathrm{clus}}}

\newcommand {\ct}      {\mbox{$\cos \theta$}}
\newcommand {\absct}   {\mbox{$|\cos \theta |$}}
\newcommand {\ctem}    {\mbox{$\cos \theta_{{\rm e}^-}$}}
\newcommand {\absctem} {\mbox{$|\cos \theta_{{\rm e}^-} |$}}

\newcommand {\absctep} {\mbox{$|\cos \theta_{{\rm e}^+} |$}}

\newcommand{\epsb}   {(\epsilon_{\rm \Pb}-\overline{\epsilon}_{\rm \Pb})}
\newcommand{\epsc}   {(\epsilon_{\rm \Pc}-\overline{\epsilon}_{\rm \Pc})}
\newcommand{\epsuds} {(\epsilon_{\rm \Pu\Pd\Ps}-
                      \overline{\epsilon}_{\rm \Pu\Pd\Ps})}
\definmath{\GeV}  {\mathrm{GeV}}
\definmath{\GeVc} {\mathrm{GeV}\!/c}
\definmath{\GeVcc}   {\mathrm{GeV}\!/c^2}
\definmath{\MeV}  {\mathrm{MeV}}
\definmath{\MeVc} {\mathrm{MeV}\!/c}
\definmath{\MeVcc}   {\mathrm{MeV}\!/c^2}
\definmath{\MVm}  {\mathrm{MV}\!/\mathrm{m}}
\definmath{\keV}  {\mathrm{keV}}
\definmath{\keVcm}   {\mathrm{keV}\!/\mathrm{cm}}
\definmath{\kV}      {\mathrm{kV}}
\definmath{\km}      {\mathrm{km}}
\definmath{\meter}   {\mathrm{m}}
\definmath{\cm}      {\mathrm{cm}}
\definmath{\mm}      {\mathrm{mm}}
\definmath{\micron}  {\mu\mathrm{m}}
\definmath{\nm}      {\mathrm{nm}}
\definmath{\kg}      {\mathrm{kg}}
\definmath{\gram} {\mathrm{g}}
\definmath{\second}  {\mathrm{s}}
\definmath{\microsec}   {\mu\mathrm{s}}
\definmath{\degree}  {^\circ}
\definmath{\degC} {^\circ\mathrm{C}}
\definmath{\ohm}  {\Omega}
\definmath{\Mohm} {\mathrm{M}\Omega}
\definmath{\rad}  {\mathrm{rad}}
\definmath{\mrad} {\mathrm{mrad}}
\definmath{\nb}      {\mathrm{nb}}
\newcommand{\eqref}[1]  {(\ref{#1})}
\newcommand{\PhysLett}  {Phys.~Lett.}
\newcommand{\PRL} {Phys.~Rev.\ Lett.}
\newcommand{\PhysRep}   {Phys.~Rep.}
\newcommand{\PhysRev}   {Phys.~Rev.}
\newcommand{\NPhys}  {Nucl.~Phys.}
\newcommand{\NIM} {Nucl.~Instrum.\ Methods}
\newcommand{\ZPhys}  {Z.~Phys.}
\newcommand{\IEEENS} {IEEE Trans.\ Nucl.~Sci.}
\newcommand{\CPC} {Comput. Phys. Commun.}
\newcommand{\LEPone}    {LEP\,1}

\newcommand{\ALEPHColl}   {ALEPH Collab.}

\newcommand{\LthreeColl}  {L3 Collab.}
\newcommand{\OPALColl}    {OPAL Collab.}
\newcolumntype{L} {>{$}l<{$}}
\newcolumntype{C} {>{$}c<{$}}
\newcolumntype{R} {>{$}r<{$}}

\newcommand{\ee}       {\mathrm{e}^+\mathrm{e}^-}   \newcommand{\lept}
{\ell^+\ell^-}   

\newcommand{\epsz}  {\varepsilon_0}   \newcommand{\lamm}   {\Lambda_-}
\newcommand{\lamp} {\Lambda_+} 

\begin{document}
%%%%%%%%%%%%%%%%%%%%%%%%%%%%%%%%%%%%%%%%%%%%%%%%%%%%%%%%%%%%%%%%%%%%%%%%
%
%  Title Page
%
\begin{titlepage}
%     Header
%
\begin{center}
    \Large EUROPEAN LABORATORY FOR PARTICLE PHYSICS
\end{center}
\bigskip 
\begin{flushright}
    \large \PPEnum\\ \Date \\ 
\end{flushright}
%
%     Main title
%
\begin{center}
    \huge\bf\boldmath Tests of  the Standard Model and Constraints  on
    New  Physics  from  Measurements of Fermion-pair  Production  at
    130--172~\GeV\ at LEP 
\end{center}
\vspace{0.5cm} 
%
%     Author names
%
\begin{center}
    \LARGE   The  OPAL   Collaboration   \\  
\vspace{0.5cm} 
%
%     Abstract
%
\begin{abstract}%=======================================================
Production of events with hadronic and leptonic final states has been
measured in $\epem$ collisions at centre-of-mass energies of 130--172~GeV,
using the OPAL detector at LEP. Cross-sections and leptonic forward-backward  
asymmetries are presented,  both including and excluding the dominant 
production  of radiative \PZ$\gamma$ events, and compared to Standard Model
expectations. The ratio \Rb\ of the cross-section for \bbbar\
production to the hadronic cross-section has been measured. In a
model-independent fit to the \PZ\ lineshape, the data have been used to
obtain an improved precision on the measurement of $\gamma$-\PZ\
interference. The energy dependence of $\alpha_{\mathrm{em}}$ has been 
investigated. The measurements have also been used to obtain limits on 
extensions of the Standard Model described by
effective four-fermion contact interactions, to search for $t$-channel
contributions from new   massive particles  and  to  place limits   on
chargino pair production with subsequent decay  of the chargino into a
light gluino and a quark pair. 
\end{abstract}%=========================================================
\end{center}
\vspace{0.5cm}
%\begin{center}
%{\bf \large DRAFT VERSION, DO NOT QUOTE}
%\end{center}
%{\bf DISCLAIMER: This note describes preliminary OPAL results and is 
%                 intended primarily for members of the collaboration.}
%\begin{center}
% {\em Submitted to Physics Letters B}
%\end{center}
%\vspace{0.5cm}  
%    \large {\bf
%    Authors:}\\ \normalsize LEP2 Standard  Model Group \\ \large  {\bf
%    Editorial Board:} \\  \normalsize   Aldo Michelini, Tara   Shears,
%    David Miller, Peter Schleper\\ 
%\vspace{0.5cm}
%{\bf Comments to {\tt Patricia.Ward@cern.ch and Michael.Kobel@cern.ch} \\
%     before 17:00 Friday 25th July }
%\end{center}
\begin{center}
{\large Submitted to \ZPhys\ C}
\end{center}
%
%  End of title page
%
\end{titlepage}
\begin{center}{\Large        The OPAL Collaboration
}\end{center}\bigskip
\begin{center}{
%begin authorlist
K.\thinspace Ackerstaff$^{  8}$,
G.\thinspace Alexander$^{ 23}$,
J.\thinspace Allison$^{ 16}$,
N.\thinspace Altekamp$^{  5}$,
K.J.\thinspace Anderson$^{  9}$,
S.\thinspace Anderson$^{ 12}$,
S.\thinspace Arcelli$^{  2}$,
S.\thinspace Asai$^{ 24}$,
D.\thinspace Axen$^{ 29}$,
G.\thinspace Azuelos$^{ 18,  a}$
% ...
%end authorlist
}\end{center}\bigskip
\bigskip
%begin institutes
% ...
$^{  2}$Dipartimento di Fisica dell' Universit\`a di Bologna and INFN,
I-40126 Bologna, Italy
\newline
% ...
$^{  5}$Cavendish Laboratory, Cambridge CB3 0HE, UK
\newline
% ...
$^{  8}$CERN, European Organisation for Particle Physics,
CH-1211 Geneva 23, Switzerland
\newline
$^{  9}$Enrico Fermi Institute and Department of Physics,
University of Chicago, Chicago IL 60637, USA
\newline
% ...
$^{ 12}$Indiana University, Department of Physics,
Swain Hall West 117, Bloomington IN 47405, USA
\newline
% ...
$^{ 16}$Department of Physics, Schuster Laboratory, The University,
Manchester M13 9PL, UK
\newline
% ...
$^{ 18}$Laboratoire de Physique Nucl\'eaire, Universit\'e de Montr\'eal,
Montr\'eal, Quebec H3C 3J7, Canada
\newline
% ...
$^{ 23}$Department of Physics and Astronomy, Tel Aviv University,
Tel Aviv 69978, Israel
\newline
$^{ 24}$International Centre for Elementary Particle Physics and
Department of Physics, University of Tokyo, Tokyo 113, and
Kobe University, Kobe 657, Japan
\newline
% ...
$^{ 29}$University of British Columbia, Department of Physics,
Vancouver BC V6T 1Z1, Canada
\newline
% ...
$^{  a}$ and at TRIUMF, Vancouver, Canada V6T 2A3
\newline
% ...
\newline

\clearpage
%=======================================================================
%       Main Text
%=======================================================================
 
%-----------------------------------------------------------------------
\section{Introduction}           \label{sec:intro}
%-----------------------------------------------------------------------
Fermion-pair production in \Pep\Pem\ collisions is
one of the basic processes of the Standard Model, and deviations
of measured cross-sections from the  predicted values could be a first
indication of  new physics beyond  the Standard Model. Measurements up
to 161~GeV  centre-of-mass energy~\cite{bib:OPAL-SM130,bib:OPAL-SM161}
have  shown  no     significant  deviations   from  Standard     Model
expectations.  In this paper we  present  new measurements of hadronic
and leptonic final states in \Pep\Pem\  collisions at a centre-of-mass
energy   $\sqrt{s}$  of     172~GeV, and improved results for the same
final states at 130, 136, and 161~GeV,  using the    OPAL   detector  at
LEP. Cross-sections have been measured for hadronic, \bbbar, \Pep\Pem,
\Pgmp\Pgmm,  and     \Pgtp\Pgtm\  final states,   together    with the
forward-backward asymmetries for the leptonic final states. We present
values   both  including  and  excluding the  production  of radiative
\PZ$\gamma$ events. 
In  general,  we  define     a
`non-radiative' sample as   events   with   $s'/s >  0.8$,     whereas
`inclusive' measurements are corrected to $s'/s > 0.01$, where
$\sqrt{s'}$  is defined as the centre-of-mass  energy of the \Pep\Pem\
system after initial-state radiation. 

In these analyses, we have introduced a well-defined treatment of the
interference between initial- and final-state photon radiation, and an
improved method of taking account of the contributions from four-fermion
production. While both of these effects are small (${\cal O}(1\%)$) 
compared with the statistical precision of the current data, they will
become significant with the increased luminosity expected at LEP in the 
future, especially  when combining results with other  
experiments~\cite{bib:ADL-SM}. We have reanalysed our data at
130--136~GeV~\cite{bib:OPAL-SM130} and 161~GeV~\cite{bib:OPAL-SM161}
using the same treatment of interference and four-fermion effects, 
in order to provide a uniform  sample of measurements for comparison with
Standard Model predictions.

The revised results at 130--136~GeV  also benefit from several improvements
to the analysis.   In   particular,  we  benefit   from  an   improved
understanding of the  background  in  the inclusive  hadronic  samples
arising   from  two-photon events.  The  separation of `non-radiative'
hadronic events has been  improved. The main  changes to the lepton
analyses include increased  efficiency for the  selection of tau  pair
events, and the  use of a Monte Carlo  generator with multiple  photon
emission for  simulating    the  $\eetoee(n\gamma)$ process  instead of one
containing    only  single photon  production.  There   have also been
improvements to the  detector calibration,  which particularly benefit
the measurement of  $\Rb$, the ratio  of the cross-section for \bbbar\
production to the hadronic  cross-section. Most of these  improvements
are      already  included       in   the   published    results    at
161~GeV~\cite{bib:OPAL-SM161}. 

As has been shown previously~\cite{bib:OPAL-SM130,bib:OPAL-SM161}, the
comparable size of the photon exchange  and \PZ\ exchange amplitudes at
these centre-of-mass energies allows  constraints to be placed on  the
size of the interference terms between them.  In this paper we improve
our previous constraints  by including  the  data  at 172~GeV.  In  an
alternative   treatment,  we assume the  Standard    Model form of the
amplitudes and use the data  to  investigate the energy dependence  of
the electromagnetic coupling   constant, \alphaem.  We have  also used
the data to search for evidence for physics beyond the Standard
Model. Firstly we do this within a general framework in which possible
contributions  from extensions of  the Standard Model are described by
an  effective four-fermion   contact interaction.  This  analysis   is
essentially the        same      as    those     performed      in
references~\cite{bib:OPAL-CI,bib:OPAL-SM161}, but the inclusion of data at
172~GeV centre-of-mass   energy gives significant improvements  to the
limits presented  there.  In  the previous   analysis  of the hadronic
cross-section we assumed   the contact interaction was  flavour-blind;
here we extend the  study to include  the case where  the new  physics couples
exclusively to one up-type quark or one down-type quark.  In a second,
more specific analysis,  we set limits on  the coupling  strength of a
new heavy particle   which might be  exchanged  in  $t$-channel production
of hadronic final states.
Such a particle could  be a squark,  the  supersymmetric partner of a
quark, in theories where $R$-parity  is violated, or a leptoquark, which
is  predicted  in many theories  which   connect the quark  and lepton
sector of the Standard Model. In this  analysis we assume that the new
physics involves only one isomultiplet of heavy particles coupling with
defined helicity. 
These studies are of topical interest in view of the indication of an 
anomaly at large momentum transfers in e$^+$p collisions reported by the 
HERA experiments~\cite{bib:HERA}.
Contact interactions or production of a heavy particle have both been 
suggested as possible explanations~\cite{bib:lqsig,bib:lqalt,bib:dreiner}.   
Finally, we place limits on gaugino pair production with
subsequent decay of the chargino or neutralino into a light gluino and a 
quark pair in supersymmetric extensions to the Standard Model.

The  paper  is organized as follows. In section~\ref{sec:theory} we
describe  Monte Carlo simulations, the treatment of interference
effects between initial- and final-state radiation and of the
contributions from four-fermion final states.  In section~\ref{sec:data} 
detailed descriptions of the luminosity measurement and the analysis of   
hadronic events,  of each lepton channel and of the measurement of \Rb\
are given.  In section~\ref{sec:sm} we
compare   measured cross-section and   asymmetry values  with Standard
Model predictions, and use them to place constraints $\gamma$-\PZ\
interference and $\alphaem$.
Finally, in section~\ref{sec:new_phys}   we  use our measurements   to
place limits on extensions of the Standard Model. 

%-----------------------------------------------------------------------
\section{Theoretical Considerations and Simulation}   \label{sec:theory}
%-----------------------------------------------------------------------

%-----------------------------------------------------------------------
\subsection{Monte Carlo Simulations}                  \label{sec:MC}
%-----------------------------------------------------------------------
The estimation of efficiencies and background processes makes extensive 
use of Monte Carlo simulations of many different final states. For studies
of $\epem\to\mathrm{hadrons}$  we used the PYTHIA5.7~\cite{bib:pythia}
program with input parameters that have  been optimized by a study of
global  event  shape variables and  particle production  rates in \PZ\
decay data~\cite{bib:OPAL-qg}.     For   $\eetoee$   we    used    the
BHWIDE~\cite{bib:bhwide} Monte Carlo  program, and for $\eetomumu$ and
$\eetotautau$  the KORALZ4.0 program~\cite{bib:koralz}.   Four-fermion
events   were      modelled   with    the      grc4f~\cite{bib:grc4f},
FERMISV~\cite{bib:fermisv}      and     EXCALIBUR~\cite{bib:excalibur}
generators, with PYTHIA used to  check the separate contributions from
WW and We$\nu$  diagrams.  Two-photon background processes  with
hadronic  final  states      were  simulated    using   PYTHIA     and
PHOJET~\cite{bib:phojet}  at      low  $Q^2$. At     high   $Q^2$  the
TWOGEN~\cite{bib:twogen}      program    with   the         `perimiss'
option~\cite{bib:OPAL-f2gam} was found to give the best description of
data;  PYTHIA  and   HERWIG~\cite{bib:herwig}   were also   used   for
comparison.   The  Vermaseren generator~\cite{bib:vermaseren} was used
to simulate purely leptonic final states in two-photon processes.  The
$\eetogg$ background in  the \Pep\Pem\ final  state was modelled  with
the   RADCOR~\cite{bib:radcor} program,  while  the  contribution from
$\epem\gamma$  where the photon  and one  of  the charged particles are
inside     the    detector   acceptance       was     modelled    with
TEEGG~\cite{bib:teegg}.  All samples  were processed  through the OPAL
detector simulation program~\cite{bib:gopal}  and reconstructed as for
real data. 
For the measurement of the luminosity, the cross-section for 
small-angle Bhabha scattering was calculated using the Monte Carlo program 
BHLUMI~\cite{bib:bhlumi}, using generated events processed through a 
simulation program for the forward calorimetry.

%-----------------------------------------------------------------------
\subsection{Initial-final State Photon Interference}  \label{sec:ifsr}
%-----------------------------------------------------------------------
A feature of \Pep\Pem\ collision data at  energies well above the \PZ\
mass  is a tendency  for radiative return  to the \PZ.  If one or more
initial-state radiation photons are emitted which reduce the effective
centre-of-mass    energy  of  the    subsequent \Pep\Pem\   collision
$\sqrt{s'}$ to the region of the \PZ\ resonance, the cross-section is
greatly enhanced.  In order to test the  Standard Model at the highest
possible   energies, we separate  clearly radiative  events from those
with $\sqrt{s'} \sim \sqrt{s}$   using  methods similar to  those   in
previous  analyses~\cite{bib:OPAL-SM161}.     In    this   separation,
$\sqrt{s'}$  is defined as the centre-of-mass  energy of the \Pep\Pem\
system after initial-state radiation. The  existence of  interference between
initial- and final-state radiation means that there is an ambiguity in
this definition. The Monte Carlo generators used to determine
experimental efficiencies and acceptances do not include interference
between initial- and final-state radiation, but these programs are
used to correct the data, which do include interference. Therefore
further corrections have to be applied to the data before measurements can
be compared with theoretical predictions.

For  Standard Model  predictions (for  all channels
except \Pep\Pem,     which    is  described   below)     we   use  the
ZFITTER~\cite{bib:zfitter} program, which has an   option either to
enable or to disable interference between  initial-   and final-state
radiation. We choose to use the option  with interference disabled for
our comparisons,  and correct our measurements to  account for this as
explained  below. This choice has the advantage of making the definition
of $s'$ unambiguous, and is more suitable for interpreting the measurements
in terms of theoretical parameters. For example, the S-matrix ansatz
used to fit the data, described in section~\ref{sec:blob}, is unsuitable
when the non-resonant part of the interference between initial- and
final-state radiation contributes~\cite{bib:stuart}.

To determine corrections to the measured cross-sections, we define a 
differential `interference cross-section'
(d$^2\sigma_{\mathrm{IFSR}}$ / d$m_{\ff}$ d\ct) 
as  the difference between the differential
cross-section including initial-final  state interference,
\mbox{(d$^2\sigma_{\mathrm{int}}$ / d$m_{\ff}$ d\ct),} 
and that excluding interference, 
(d$^2\sigma_{\mathrm{noint}}$ / d$m_{\ff}$ d\ct), 
as calculated by ZFITTER using the appropriate
flag settings\footnote{Cross-sections including interference are
obained  by  setting {\tt  INTF}=1,  those  excluding  interference by
setting {\tt INTF}=0.  For hadrons, we also set {\tt INCL}=0 to enable
the {\tt INTF}  flag.}.  The differential interference 
cross-section  may be either positive or
negative, depending on the values of the cosine of the angle
$\theta$ between the fermion and the electron beam direction, and the
invariant mass of  the fermion pair $m_{\ff}$.  
We  then estimate the fraction of
this cross-section accepted by our selection cuts by assuming that, 
as a  function of $\ct$ and $m_{\ff}$, its selection efficiency
\begin{equation}
\epsilon_{\mathrm{IFSR}}(\ct,m_{\ff}) = 
                               \epsilon_{\mathrm{noint}}(\ct,m_{\ff}), 
\label{eq:ifsreff}
\end{equation} 
where $\epsilon_{\mathrm{noint}}$ has been  
determined  from Monte Carlo events  which do not include interference.
The corrected cross-section $\sigma_{\mathrm{corr}}$ is
obtained from the measured cross-section after background subtraction
and efficiency correction $\sigma_{\mathrm{meas}}$ as: 
\begin{equation}
%     \sigma_{\mathrm{corr}}(\ct) = \sigma_{\mathrm{meas}}(\ct) -
%     \frac{\sum_{m_{\ff}}\sigma_{\mathrm{int}}(\ct,m_{\ff}) *
%     \epsilon_{\mathrm{int}}(\ct,m_{\ff})}
%    {\epsilon_{\mathrm{noint}}(\ct)}.
     \frac{{\rm d}\sigma_{\mathrm{corr}}}{{\rm d}\ct} 
   = \frac{{\rm d}\sigma_{\mathrm{meas}}}{{\rm d}\ct} 
   - \frac{{\rm d}\sigma_{\mathrm{noint}}}{{\rm d}\ct} 
     \frac{\int \epsilon_{\mathrm{IFSR}}(\ct,m_{\ff})
       \frac{{\rm d}^2\sigma_{\mathrm{IFSR}}}{{\rm d}m_{\ff}{\rm d}\cos\theta} 
       {\rm d}m_{\ff}}     
       {\int \epsilon_{\mathrm{noint}}(\ct,m_{\ff})
       \frac{{\rm d}^2\sigma_{\mathrm{noint}}}{{\rm d}m_{\ff}{\rm d}\cos\theta} 
       {\rm d}m_{\ff}}.     
\end{equation}
In practice the integrals were evaluated in appropriate
bins of \ct\ and $m_{\ff}$. 
As the accepted cross-section is estimated as a function of $\ct$, the
method    is  easily   applied   to   total  cross-sections,   angular
distributions or asymmetry measurements. 

The systematic error on this procedure was assessed by repeating the
estimate modifying the assumption of eq.~\ref{eq:ifsreff} to
\begin{equation}
 \epsilon_{\mathrm{IFSR}}(\ct,m_{\ff}) =
  \frac{\epsilon_{\mathrm{noint}}(\ct,m_{\ff}) + 
        \epsilon_{\mathrm{noint}}(\ct,\sqrt{s})}{2}, 
\end{equation}
i.e.\ for each bin of $\ct$ and
$m_{\ff}$ the average of the efficiency in that bin
and the efficiency in the bin including
$m_{\ff}=\sqrt{s}$ for the same
$\ct$ range was used. This was motivated as follows. The efficiencies
$\epsilon_{\mathrm{noint}}$ used
for the interference correction are an average over events with
initial-state radiation and events with final-state radiation, and are a 
good approximation to the true $\epsilon_{\mathrm{IFSR}}$ 
if these efficiencies are similar. For large $m_{\ff}$
this is the case, but for small $m_{\ff}$ the 
efficiency for the relatively rare events with final-state radiation
may be significantly higher than that for events with initial-state
radiation. To account for this, the error  on  $\epsilon_{\mathrm{IFSR}}$
 is taken as half the difference between the average efficiency  
$\epsilon_{\mathrm{noint}}(\ct,m_{\ff})$, 
and  the  largest possible efficiency,    
$\epsilon_{\mathrm{noint}}(\ct,\sqrt{s})$ at a given value of \ct. 

For  the
hadrons there is an additional uncertainty due to QCD effects. We
have  taken this additional uncertainty to  be 100\% of the correction
without  an   $s'$   cut. The   basic  assumption here    is that  the
near-cancellation between virtual  (box) and real interference effects
without  cuts (Kinoshita-Lee-Nauenberg theorem~\cite{bib:kln_theorem})
is not completely destroyed by large QCD corrections to both. This has
been proven for pure final-state  radiation~\cite{bib:yr}, but not yet
for initial-final state interference.  In the absence of a theoretical
calculation we allow the asymptotic value to change by 100\%. 

The corrections to inclusive ($s'/s > 0.01$) cross-sections  are
small, reflecting the Kinoshita-Lee-Nauenberg cancellation.
They typically amount to
$(-0.36\pm0.04)$\%  for muon pairs,  $(-0.5\pm0.1)$\%
for tau pairs    and (+0.1$\pm$0.1)\% for hadrons, where the
statistical errors are small compared to the 
systematic errors, derived as described above. 
For  non-radiative
events    ($s'/s   >  0.8$)   the corrections     are  rather larger, 
%  typically
%$(-1.8\pm0.5)$\%  for muon pairs, $(-1.4\pm0.4)$\%  for  tau pairs and
%(+1.1$\pm$0.5)\% for hadrons. 
and are given in detail in table~\ref{tab:ifsr}.
The differences between the muon and tau
corrections reflect the different acceptance cuts in $\ct$ used in the
event selection;  the  hadron corrections are of opposite sign from
those of the  leptons because of  the quark charges.  The  corrections
change the lepton asymmetry  values by typically  $-0.006\pm0.001$ for
$s'/s > 0.01$ and $-0.015\pm0.005$ for $s'/s > 0.8$. 
All corrections depend only very weakly on $\sqrt{s}$.

\begin{table}[bt]%-------------------------------------------------------
\centering
\begin{tabular}{|l|l|l|l|l|}
\hline%-----------------------------------------------------------------
\hline%-----------------------------------------------------------------
\multicolumn{5}{|c|}{\bf Interference Corrections \boldmath  ($s'/s>0.8$)} \\
\hline%-----------------------------------------------------------------
           &130.25~GeV &136.22~GeV &161.34~GeV &172.12~GeV \\
\hline%-----------------------------------------------------------------
$\Delta\sigma_{\rm had}/\sigma_{\rm SM}$ (\%)  
& $+1.0\pm0.3\pm0.4$ 
         & $+1.2\pm0.4\pm0.5$  & $+1.3\pm0.4\pm0.6$ & $+1.2\pm0.3\pm0.5$ \\
$\Delta\sigma_{\mu\mu}/\sigma_{\rm SM}$ (\%)   
& $-1.6\pm0.5$  & $-1.8\pm0.6$ & $-1.7\pm0.5$       & $-1.8\pm0.5$       \\ 
$\Delta\sigma_{\tau\tau}/\sigma_{\rm SM}$ (\%) 
& $-1.3\pm0.4$  & $-1.4\pm0.4$ & $-1.4\pm0.4$       & $-1.4\pm0.4$       \\
\hline%-----------------------------------------------------------------
\hline%-----------------------------------------------------------------
\end{tabular}
\caption[]{Corrections $\Delta\sigma$, which have to be applied to 
the measured non-radiative cross-sections in order to remove the 
contribution from initial-final state interference. They are
expressed as a percentage fraction of the expected Standard Model cross-section.
The first error reflects the uncertainty from  modelling the selection
efficiency for the interference cross-section, the second one
our estimate of possible additional QCD corrections for the hadrons.
}
\label{tab:ifsr}
\end{table}%------------------------------------------------------------

We  have  checked the  results of the  above  correction  procedure by
comparing   them      to   an   independent     estimate    using  the
KORALZ~\cite{bib:koralz}  Monte  Carlo   generator.  We  generated two
samples of muon pair events at 171 GeV, and subjected them to the full
detector    simulation,       reconstruction  and   event    selection
procedures. Both samples    were  generated with   only single  photon
emission.   In  the first sample   there  was no  interference between
initial-  and  final-state  radiation, while   in   the second  sample
interference  was   enabled.  The   differences  between  the observed
cross-sections and asymmetries agreed  with the estimates from ZFITTER
described above within one standard deviation,  for both the inclusive
($s'/s > 0.01$) and non-radiative ($s'/s > 0.8$) cases. 

The above correction procedure has been applied to all cross-sections,
asymmetry measurements and angular distributions, except for those for
the \epem\ final state. In this case,  we do not use a  cut on $s'$ so
there  is no  ambiguity  in its  definition.  Both  the Standard Model
calculations and the Monte  Carlo program used to calculate efficiency
and acceptance corrections include  interference between initial-  and
final-state   radiation. Results for    \epem\ are therefore presented
including such effects. 

%-----------------------------------------------------------------------
\subsection{Four-fermion Effects}           \label{sec:4f}
%-----------------------------------------------------------------------

Contributions from four-fermion production $\ff\fpfp$ to the process
$\epem\rightarrow\ff$ pose non-trivial problems both experimentally
and theoretically. While four-fermion final states arising from  the
`two-photon' (multiperipheral) diagrams, for example, can be considered  
background  to
two-fermion  production, those arising from   the emission of low mass
$\fpfp$ pairs in $s$-channel diagrams  may  in some circumstances   be
considered signal, in the same way as  is emission of photons. A clean
separation is not  possible because of  interference between diagrams
contributing to the same final state. 

The correct  experimental treatment  of the  four-fermion contribution
depends on  whether or not the theoretical  calculation with which the
experimental   measurement  is to  be  compared   includes emission of
fermion  pairs.  For   example, ALIBABA~\cite{bib:alibaba}   does  not
include such emission. By default  ZFITTER includes initial-state pair
emission via virtual photons,  although this can be disabled. However,
pair emission  via virtual \PZ\ bosons  is  not included. 
By  comparing the predictions  of
ZFITTER with and without pair emission\footnote{We have modified the 
ZFITTER code so that the $s'$ cut acts on fermion
pairs as well as photons, since by 
default hard pair emission leading to 
$s' < 0.5s$ is not included.}, we 
estimate that the effect of
including it increases  the cross-sections for $s'/s  > 0.01$ by about
1\% and decreases   those  for $s'/s  > 0.8$   by about 0.1\%   at the
energies considered here. Similar  values are obtained for hadrons and
lepton pairs.  Final-state pair radiation is not explicitly treated in
ZFITTER.  The dominant    part of  its  (very small)   effect  on  the
cross-section is  covered in  the  inclusive treatment of final  state
radiation\footnote{Inclusive treatment of final-state radiation is obtained 
by setting the flag {\tt  FINR}=0.}.
For corrections  to the selection efficiency, however, both initial- and  
final-state pair radiation have to be considered, as described below. 

None of the theoretical calculations to which we compare  our data  has an
option to separate real from virtual fermion pair effects\footnote{ZFITTER
with the flag  {\tt FOT2}=2, like  ALIBABA, includes neither real pair
emission nor  vertex corrections involving  virtual pairs. The default
setting {\tt  FOT2}=3 includes  both,  summing up beforehand the  soft
part  of real  pair  emission  and the   vertex corrections.}.  
Therefore two-   and four-fermion events have to  be
treated together in the  data  analysis. Considering all  four-fermion
events  as background, for example, would  not account for the virtual
vertex corrections, which can be even  larger than the effects of real
pair emission.  Therefore some  four-fermion events always have to  be
excluded from background estimates. 

In   general, we compare  our   measurements with ZFITTER  predictions
including pair  emission. This means   that pair emission  via virtual
photons  from both the  initial and  final  state must be included  in
efficiency calculations, and be excluded from background estimates. In
order to perform the separation,  we ignore interference between  $s$-
and $t$-channel diagrams contributing to  the same four-fermion  final
state, and generate separate   Monte Carlo samples for   the different
diagrams for each final state.  For a two-fermion  final state \ff\ we
then  include as   signal   those four-fermion   events   arising from
$s$-channel  processes for which $m_{\ff}  > m_{\fpfp}$, $m_{\fpfp} <$
70~GeV   and $m_{\ff}^{2}/s > 0.01$  ($m_{\ff}^{2}/s    > 0.8$ in  the
non-radiative case).  This kinematic classification closely models the
desired classification of $\ff\fpfp$  in terms of intermediate bosons,
in that pairs  arising from virtual photons  are generally included as
signal whereas those arising from virtual Z bosons are not. All events
arising  from $s$-channel processes failing   the above cuts, together
with those arising from the  $t$-channel process (Zee) and  two-photon
processes are regarded as background. Four-fermion processes involving
WW or single W production  are also background in  all cases.  The
overall efficiency, $\epsilon$, is calculated as 
\begin{equation}
  \epsilon = (1 -
    \frac{\sigma_{\ff\fpfp}}{\sigma_{\mathrm{tot}}})\epsilon_{\ff}
  + \frac{\sigma_{\ff\fpfp}}{\sigma_{\mathrm{tot}}}\epsilon_{\ff\fpfp}
\end{equation}
where $\epsilon_{\ff}$, $\epsilon_{\ff\fpfp}$ are the efficiencies 
derived from the two-fermion and four-fermion signal Monte Carlo events
respectively, $\sigma_{\ff\fpfp}$ is the generated four-fermion 
cross-section, and $\sigma_{\mathrm{tot}}$ is the total cross-section
from ZFITTER including pair emission. Using this definition of efficiency,
effects of cuts on soft pair emission in the four-fermion generator
are correctly summed with vertex corrections involving virtual pairs.
The inclusion of the four-fermion part of the signal produces
negligible changes to the efficiencies for hadronic events and for
lepton pairs with $s'/s > 0.8$. The efficiencies for lepton pairs
with $s'/s > 0.01$ are decreased by about 0.5\%.

The discussion in the above paragraph applies to hadronic, muon pair
and tau pair final states. In the case of electron pairs, the situation
is slightly different. In principle the $t$-channel process with a second
fermion pair arising from the conversion of a virtual photon emitted from 
an initial- or final-state electron should be included as signal. 
As this process is not included in any program we use for
comparison we simply ignore such events: they are not included as 
background as this would underestimate the cross-section.

%-----------------------------------------------------------------------
\section{Data Analysis}           \label{sec:data}
%-----------------------------------------------------------------------
 
The OPAL detector\footnote{OPAL uses a right-handed coordinate system in
which the $z$ axis is along the electron beam direction and the $x$
axis is horizontal. The polar angle, $\theta$, is measured with respect
to the $z$ axis and the azimuthal angle, $\phi$, with respect to the
$x$ axis.}, trigger and data acquisition system are fully described 
elsewhere~\cite{bib:OPAL-detector,bib:OPAL-SI,bib:OPAL-SW,bib:OPAL-TR,
bib:OPAL-DAQ}.
Data from three separate data-taking periods are used in this analysis:
\begin{itemize}
\item
 Integrated luminosities of 2.7~pb$^{-1}$ and 2.6~pb$^{-1}$ recorded at
 \epem\ centre-of-mass energies of 130.25 and 136.22~GeV, respectively, in 
 1995 (LEP1.5). The energy measurements have a common systematic uncertainty 
 of 0.05~GeV~\cite{bib:ELEP}. 
\item
 An integrated luminosity of 10.1~pb$^{-1}$ recorded at an \epem\
 centre-of-mass energy of 161.34$\pm$0.05~GeV~\cite{bib:ELEP} during 
 1996.
\item
 An integrated luminosity of approximately 9.3~pb$^{-1}$ at an \Pep\Pem\ 
 centre-of-mass energy of 172.3~GeV and 1.0~pb$^{-1}$ at an energy of 
 170.3~GeV, recorded during 1996. The data from these two energies have been 
 analysed together; the luminosity-weighted mean centre-of-mass energy has 
 been determined to be 172.12$\pm$0.06~GeV~\cite{bib:ELEP}.
\end{itemize}
 
%-----------------------------------------------------------------------
\subsection{Measurement of the Luminosity}           \label{sec:lumi}
%-----------------------------------------------------------------------
 
\begin{figure}
  \epsfxsize=\textwidth
  %\epsfbox[0 0 567 567]{lumi_plots.eps}
  \epsfbox[0 0 567 567]{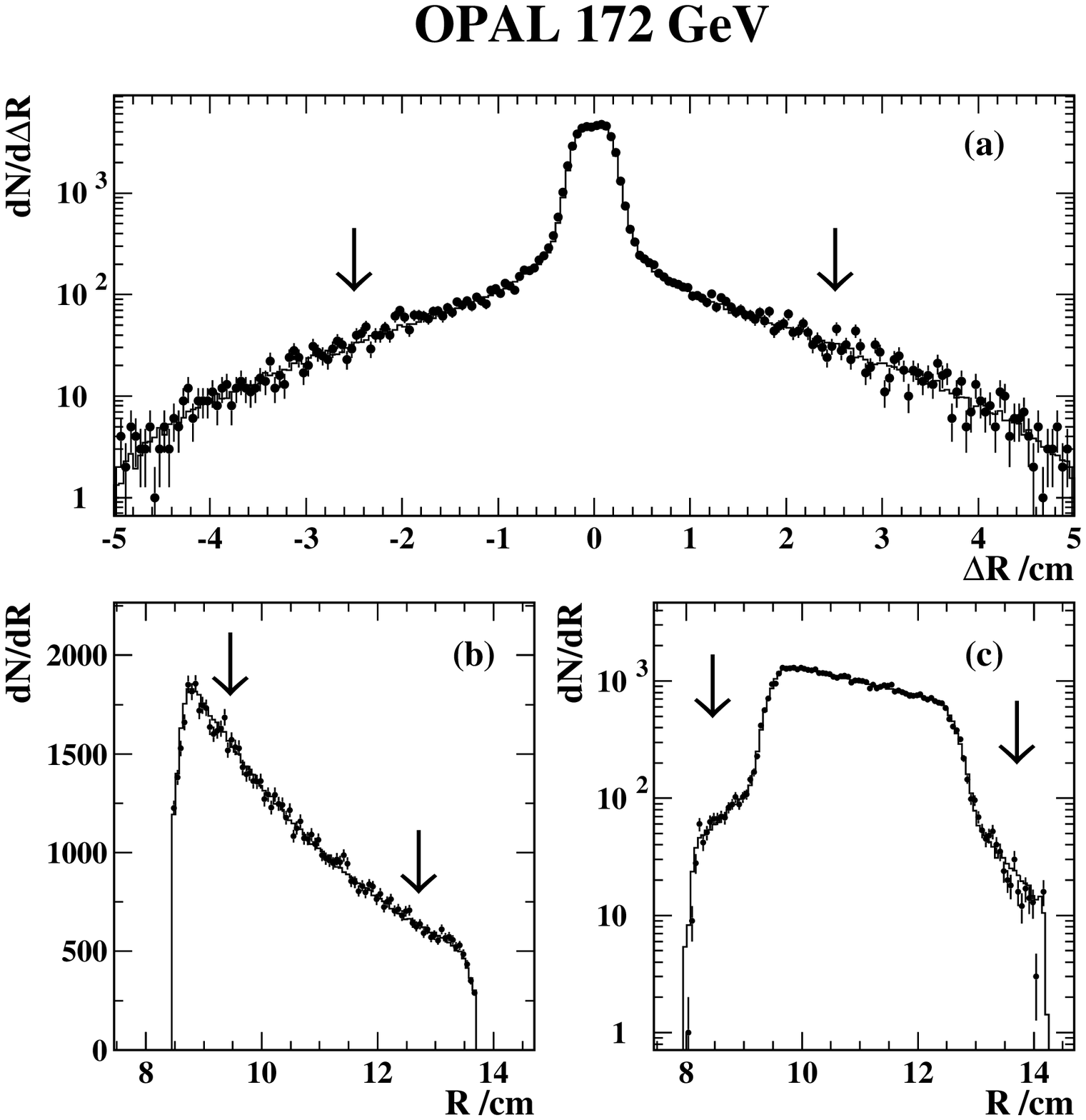}
  \caption
{
   (a) The distribution of the difference in radial coordinate between
   the two clusters in Bhabha scattering events used for the silicon-tungsten
   luminosity measurement at 172~GeV. Distributions of the radial coordinates 
   of clusters for (b) the `narrow' side and (c) the `wide' side calorimeter.
   Distributions are shown after all cuts except the acollinearity cut in (a) 
   and the inner and outer radial acceptance 
   cuts, on that side, in (b) and (c). Points show the data, while the 
   histograms show the Monte Carlo expectation. The arrows show the 
   positions of the cuts which define the acceptance.
}
\label{fig:swlumi}
\end{figure}

The integrated luminosity was measured using small-angle Bhabha scattering 
events, $\epem\to\epem$, recorded in the forward calorimetry. The
primary detector is a silicon-tungsten luminometer~\cite{bib:OPAL-SW} which
consists of two finely segmented silicon-tungsten calorimeters placed around
the beam pipe, symmetrically on the left and right sides on the OPAL
detector, 2.4~m from the interaction point. Each calorimeter covers 
angles from the beam between 25 and 59~mrad. Bhabha scattering events were 
selected by requiring a high energy cluster in each end of the detector, using
asymmetric acceptance cuts. The energy in each calorimeter
had to be at least half the beam energy, and the average energy 
had to be at least three quarters of the beam energy.
The two highest energy clusters were required to be back-to-back in $\phi$,
$||\phi_{R} - \phi_{L}| - \pi| <$ 200~mrad, where $\phi_{R}$ and $\phi_{L}$
are the azimuthal angles of the cluster in the right- and left-hand 
calorimeter respectively. They were also required to be collinear, by
placing a cut on the difference between the radial positions, 
$\Delta R\equiv |R_{R} - R_{L}| <$ 25~mm, where $R_{R}$ and $R_{L}$ are the 
radial coordinates of the clusters on a plane approximately 7 radiation 
lengths into the calorimeter. This cut, corresponding to an acollinearity 
angle of about 10.4~mrad, effectively defines the acceptance for 
single-photon 
radiative events, thus reducing the sensitivity of the measurement to the 
detailed energy response of the calorimeter. The distribution of 
$\Delta R$ for the 172~GeV data is shown in figure~\ref{fig:swlumi}(a).

For the 130--136~GeV data, the inner and outer radial acceptance cuts
delimited a region between 31 and 52~mrad on one side of the calorimeter,
while for the opposite calorimeter a wider zone between 27 and 56~mrad 
was used. Two luminosity measurements were formed with the narrower
acceptance on one side or the other side. The final measurement was
the average of the two and has no first order dependence on beam offsets
or tilts. Before data-taking started at $\sqrt{s}$=161~GeV, 
tungsten shields designed
to protect the tracking detectors from synchrotron radiation were 
installed around the beam pipe. The shields, 5~mm thick and 334~mm long,
present roughly 50~radiation lengths to particles originating from
the interaction region, almost completely absorbing electromagnetically
showering particles between 26 and 33~mrad from the beam axis. The
fiducial regions for accepting Bhabha events for the 161 and 172~GeV data
were therefore reduced, to between 38 and 52~mrad on one side and between 
34 and 56~mrad on the opposite side. The distributions of the radial
coordinates of the clusters for the 172~GeV data are shown in
figure~\ref{fig:swlumi}(b,c).

%The cross-section for Bhabha scattering accepted by these cuts was
%calculated using the Monte Carlo program BHLUMI~\cite{bib:bhlumi}.
The error on the luminosity measurement is dominated by data statistics.
For the 130 and 136~GeV data, the acceptance of the luminometer was
reduced at the trigger level by a prescaling factor of 16 in order to
increase the experimental live time as far as possible, giving a
statistical error of 0.9\% on the combined 130 and 136~GeV data. For
the two higher energies this prescaling factor was reduced to 2 or 4,
and the statistical error amounts to 0.42\% (0.43\%)
at 161 (172)~GeV. The largest
systematic uncertainty arises from theoretical knowledge of the
cross-section (0.25\%), with detector effects amounting to a further
0.20\% (0.23\%) at 161 (172)~GeV.

A second luminosity measurement was provided by the forward detector, a
lead-scintillator sampling calorimeter covering angles from the beam
between 40 and 150~mrad. The selection of Bhabha events within the calorimeter
acceptance is unchanged from reference~\cite{bib:OPAL-LS91}, but the acceptance
was reduced to the region between 65 and 105~mrad from the beam because
of the addition of the silicon-tungsten luminometer on the inside front
edge of the device. The overall acceptance of the calorimeter was measured
by normalizing to the precisely known cross-section for hadronic events
at the \PZ\ peak, and applying small corrections derived from Monte
Carlo simulations to reflect changes in acceptance with centre-of-mass
energy. To allow for changes in acceptance between years,
this normalization was performed separately for 1995 and 1996 using 
data recorded at the \PZ\ in each year. Knowledge of the hadronic
acceptance for the \PZ\ data is the main source of systematic error
in the forward detector luminosity measurement, which amounts to 
0.8\% (1.0\%) for the data taken in 1995 (1996). 

The luminosity measured by the forward detector agreed with that
measured by the silicon-tungsten luminometer to within one standard
deviation of the combined error for all data samples. For the 130 and
136~GeV data, where the precision of the two measurements was similar,
the average luminosity was used; the overall error on this average
measurement is 0.7\%. At 161 and 172~GeV the silicon-tungsten luminosity
was preferred as the more precise; the overall error on this
measurement amounts to 0.53\% (0.55\%) at 161 (172)~GeV. The errors
on luminosity are included in the systematic errors on all cross-section
measurements presented in this paper. Correlations between cross-section
measurements arising from errors in the luminosity have been taken into account 
in the interpretation of the results.

%-----------------------------------------------------------------------
\subsection{Hadronic Events}           \label{sec:mh}
%-----------------------------------------------------------------------
\subsubsection{\boldmath Inclusive Events ($s'/s >$ 0.01)}
The criteria used to select an inclusive sample of hadronic 
events with $s'/s > 0.01$ were based on energy clusters in the 
electromagnetic calorimeter and the charged track multiplicity. 
Clusters in 
the barrel region were required to have an energy of at least 100~MeV, and 
clusters in the endcap detectors were required to contain at least two 
adjacent lead glass blocks and have an energy of at least 200~MeV. Tracks 
were required to have at least 20 measured space points. The point of
closest approach to the nominal beam axis was found, and required to
lie less than 2~cm in the $r$--$\phi$ plane and less than 40~cm along the beam 
axis from the nominal interaction point.
Tracks were also required to have a minimum momentum component transverse 
to the beam direction of 50 MeV.
 
\begin{figure}
  \epsfxsize=\textwidth
  %\epsfbox[0 0 567 567]{disk$cbhpae:[work.drw.duck]rvis_rbal2.eps}
  \epsfbox[0 0 567 567]{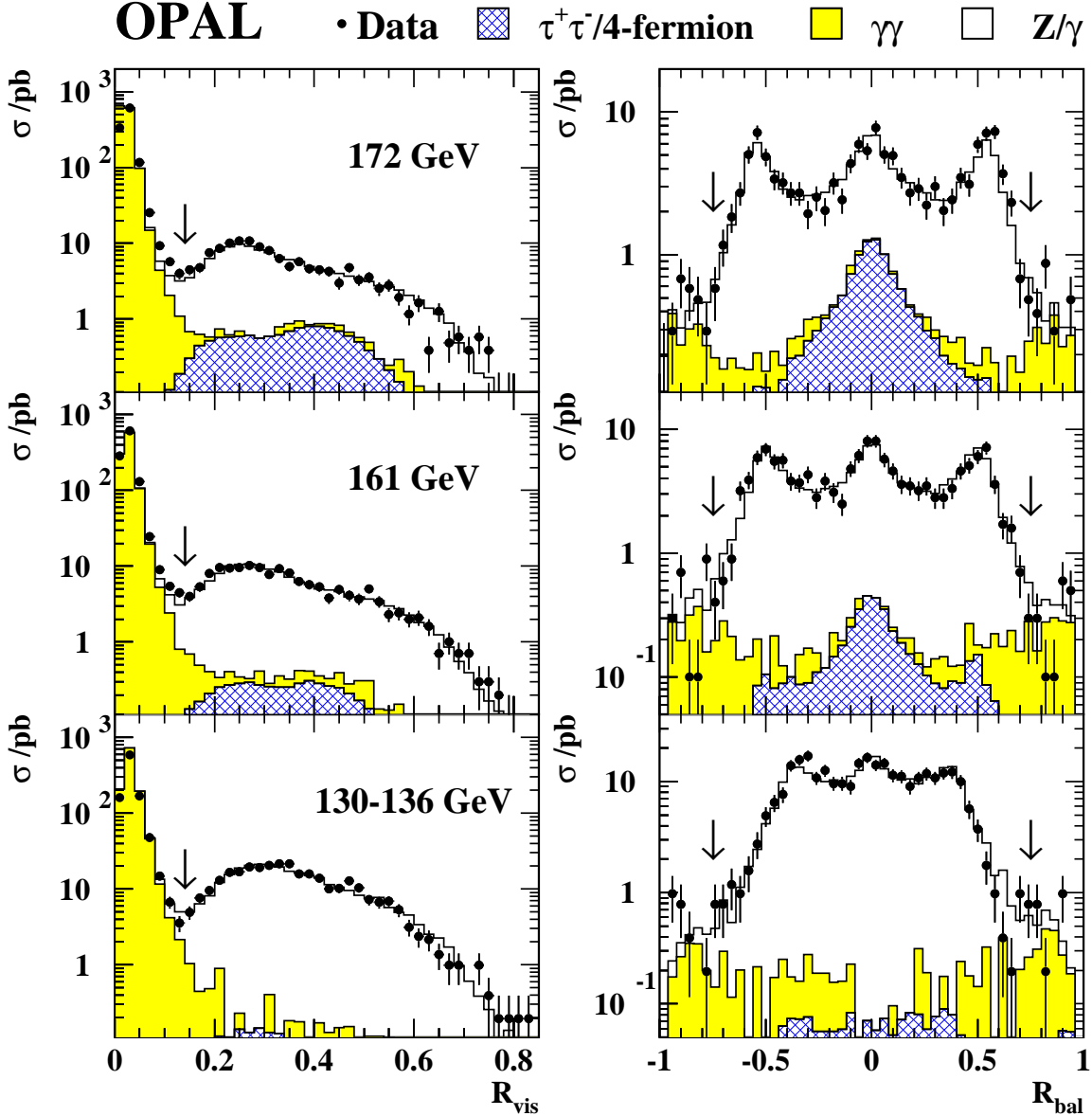}
  \caption
{
  The distributions of \Rvis, the ratio of the total energy deposited
  in the electromagnetic calorimeter to the centre-of-mass energy,
  and \Rbal, the energy imbalance along the beam direction,
  for hadronic events at each energy. Distributions are shown after
  all event selection cuts except the cut on that quantity. The positions
  of the cuts are indicated by the arrows. Points show the data and
  histograms the Monte Carlo expectations, normalized to the integrated
  luminosity of the data, with light shading indicating
  background from two-photon events and cross-hatching indicating
  other backgrounds, mainly four-fermion (including WW) and \tautau\ events. 
}
\label{fig:mh_rvis_rbal2}
\end{figure}

The following requirements were used to select hadronic candidates.
\begin{itemize}
 \item{}
  To reject leptonic final states, events were required to have high 
  multiplicity: at least 7~electromagnetic clusters and at least 5~tracks.
 \item{} Background from
  two-photon events was reduced by requiring a total energy deposited in 
  the electromagnetic calorimeter of at least 14\% of the centre-of-mass 
  energy: $ \Rvis \equiv {\Sigma \Eclus }/\sqrt{s} > 0.14, $
  where $\Eclus$ is the energy of each cluster.
 \item{} 
  Any remaining background from beam-gas and beam-wall interactions was 
  removed, and two-photon events further reduced,
  by requiring an energy balance along the beam direction which satisfied
  \mbox{$ \Rbal \equiv \mid \Sigma (\Eclus \cdot \cos \theta) \mid /
  \Sigma \Eclus < 0.75$}, where $\theta$ is the polar angle of the cluster.
\end{itemize}
These criteria are identical to those used previously at 
161~GeV~\cite{bib:OPAL-SM161}, but the cut on $\Rbal$ is somewhat 
looser than that used previously at 130--136~GeV, resulting in a slightly
higher efficiency for radiative return events. Distributions of $\Rvis$
and $\Rbal$ for each centre-of-mass
energy are shown in figure~\ref{fig:mh_rvis_rbal2}.
The efficiency of the selection cuts was determined from Monte Carlo 
simulations, and the value for each centre-of-mass energy is given in 
table~\ref{tab:mh}. From comparisons of the data distributions of $\Rbal$
and $\Rvis$ with Monte Carlo, at these energies and at energies
around the Z peak (LEP1), 
we estimate the systematic error on the selection efficiency to be 1\%.

Above the W-pair threshold, the largest single contribution to the
background arises from WW events. No cuts have been
applied to reject W-pair events; the expected contribution from these
to the visible cross-section
has been subtracted, and amounts to (2.4$\pm$0.2)\% at 161~GeV and
(9.6$\pm$0.2)\% at 172~GeV, where the error arises mainly from the
uncertainty in the W mass~\cite{bib:PDG96}.
Backgrounds to the inclusive hadron samples at all energies arise 
from other four-fermion events which are not considered part of the signal, 
in particular two-photon events and the channels Zee and We$\nu$,
and tau pairs. These amount to 1.9\% at 130~GeV, rising to 4.1\%
at 172~GeV. The main uncertainty on this background arises from the
two-photon events; we assign a 50\% error to this contribution, which
covers the predictions from all the generators discussed in 
section~\ref{sec:MC}. 

The numbers of selected events and the resulting cross-sections are
shown in table~\ref{tab:mh}.

\begin{table}[htb]%-------------------------------------------------------
\centering
\begin{tabular}{|l|c|c|c|c|}
\hline%-----------------------------------------------------------------
\hline%-----------------------------------------------------------------
\multicolumn{5}{|c|}{\bf \boldmath Hadrons ($s'/s>0.01$)} \\
\hline%-----------------------------------------------------------------
&130.25~GeV &136.22~GeV &161.34~GeV &172.12~GeV \\
\hline%-----------------------------------------------------------------
Events &832  &673  &1472 &1368 \\
Efficiency (\%)  &95.8$\pm$1.0 &95.2$\pm$1.0 &92.3$\pm$0.9 &91.2$\pm$0.9  \\
Background (pb)  &5.9$\pm$2.1  &5.9$\pm$2.1  &8.5$\pm$1.7  &15.6$\pm$1.5   \\
$\sigma_{\mathrm{meas}}$ (pb) &317$\pm$11$\pm$5     &264$\pm$10$\pm$4   
                              &150$\pm$4$\pm$2      &127$\pm$4$\pm$2   \\
$\sigma_{\mathrm{corr}}$ (pb) &317$\pm$11$\pm$5     &264$\pm$10$\pm$4 
                              &150$\pm$4$\pm$2      &127$\pm$4$\pm$2 \\
$\sigma_{\mathrm{SM}}$ (pb) &330  &273  &150  &125  \\
\hline%-----------------------------------------------------------------
\hline%-----------------------------------------------------------------
\multicolumn{5}{|c|}{\bf \boldmath Hadrons ($s'/s>0.8$)} \\
\hline%-----------------------------------------------------------------
&130.25~GeV &136.22~GeV &161.34~GeV &172.12~GeV \\
\hline%-----------------------------------------------------------------
Events &174  &166  &370  &339  \\
Efficiency (\%)  &91.0$\pm$0.7 &91.0$\pm$0.7 &91.8$\pm$0.5  &91.8$\pm$0.5  \\
Feedthrough (\%) &8.9$\pm$1.9  &7.9$\pm$1.9  &5.2$\pm$1.9   &4.8$\pm$1.9  \\
Background (pb)  &1.3$\pm$0.1  &1.3$\pm$0.1  &2.73$\pm$0.22 &6.82$\pm$0.26 \\
$\sigma_{\mathrm{meas}}$ (pb) &63.5$\pm$4.9$\pm$1.5 &63.1$\pm$5.0$\pm$1.5 
                              &35.1$\pm$2.0$\pm$0.8 &26.7$\pm$1.8$\pm$0.6 \\
$\sigma_{\mathrm{corr}}$ (pb) &64.3$\pm$4.9$\pm$1.5 &63.8$\pm$5.0$\pm$1.5 
                              &35.5$\pm$2.0$\pm$0.8 &27.0$\pm$1.8$\pm$0.6 \\
$\sigma_{\mathrm{SM}}$ (pb) &77.6 &62.9 &33.7 &27.6 \\
\hline%-----------------------------------------------------------------
\hline%-----------------------------------------------------------------
\end{tabular}
\caption[]{Numbers of selected events, efficiencies, backgrounds, 
feedthrough of events from lower $s'$ to the $s'/s > 0.8$ samples and
measured cross-sections for hadronic events. The errors on efficiencies,
feedthrough and background include Monte Carlo statistics and systematic 
effects, with the latter dominant. The cross-sections labelled
$\sigma_{\mathrm{meas}}$ are the measured values without the correction
for interference between initial- and final-state radiation, those
labelled $\sigma_{\mathrm{corr}}$ are with this correction. For the
inclusive measurements, the results are the same to the quoted precision. The 
first error on each measured cross-section is statistical, the second 
systematic.
The Standard Model predictions, $\sigma_{\mathrm{SM}}$, are from the
ZFITTER~\cite{bib:zfitter} program.
}
\label{tab:mh}
\end{table}%------------------------------------------------------------

\subsubsection{\boldmath Non-radiative events ($s'/s > 0.8$)}
\begin{figure}
  \epsfxsize=\textwidth
  %\epsfbox[0 0 567 567]{disk$cbhpae:[drw1.lep171]2fpaper.eps}
  \epsfbox[0 0 567 567]{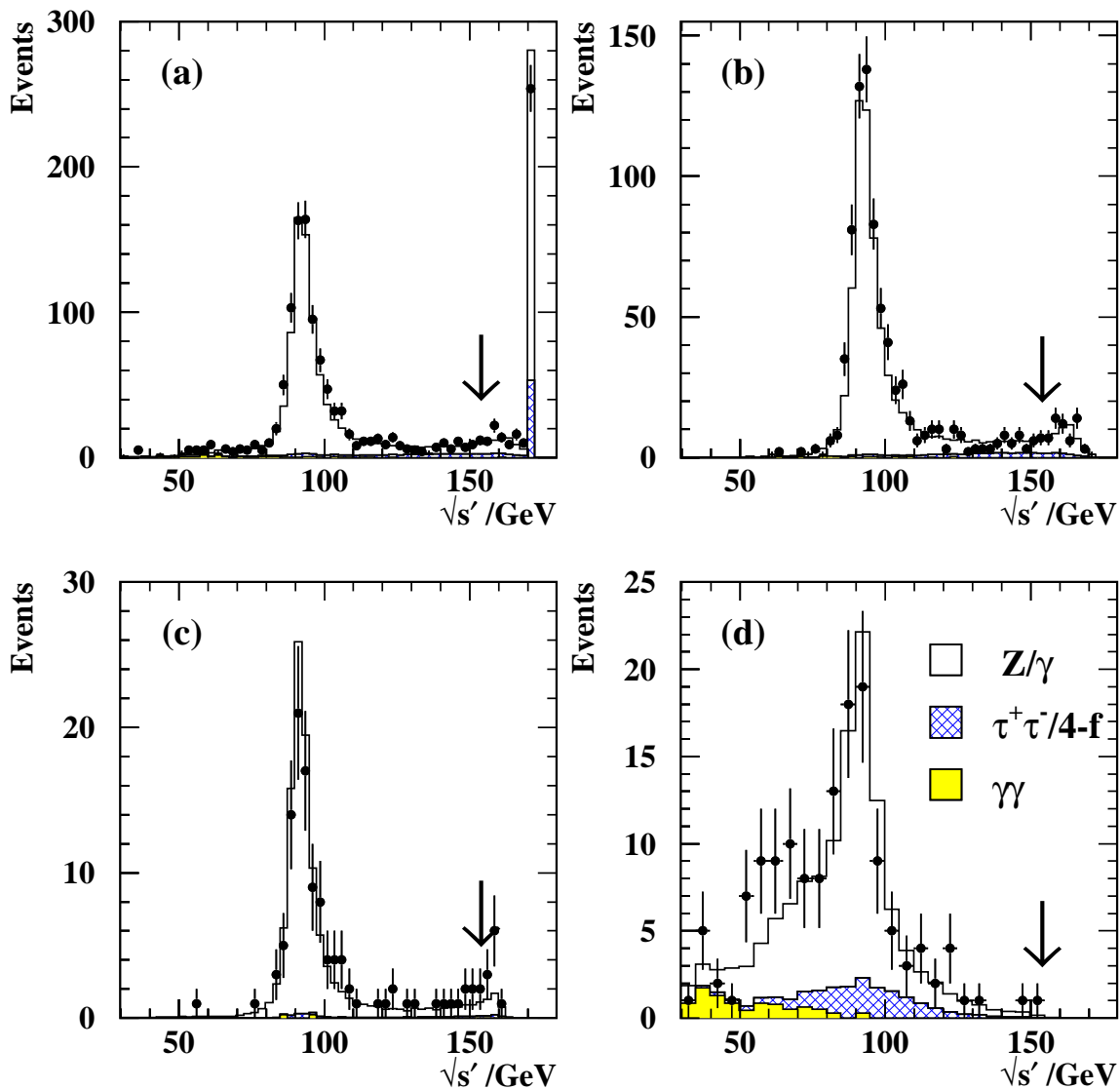}
  \caption
{
   Distributions of effective centre-of-mass energy $\protect\sqrt{s'}$
   reconstructed for hadronic events at 172~GeV. The distributions
   are shown (a) for all events, (b) for events with one photon along
   the beam axis, (c) for events with one photon in the electromagnetic
   calorimeter and (d) for events with more than one photon.
   In each case the points show the data and the histogram the Monte
   Carlo prediction normalized to the luminosity of the data. The
   arrows indicate the position of the cut used to select `non-radiative'
   events.
}
\label{fig:mh_sp}
\end{figure}

The effective centre-of-mass energy $\sqrt{s'}$ of the \epem\ collision
for hadronic events selected as above was estimated as follows.
The method is the same as that used in reference~\cite{bib:OPAL-SM161}.
Isolated photons in the electromagnetic calorimeter
were identified, and the remaining tracks, electromagnetic and hadron
calorimeter clusters formed into jets using the Durham ($k_{T}$) 
scheme~\cite{bib:durham} with a jet resolution parameter $y_{\rm cut}=0.02$.
If more than four jets were found the number was forced to be four. 
The jet energies and angles were corrected for double counting using the 
algorithm described in reference~\cite{bib:GCE}. The jets and observed photons 
were then subjected to a series of kinematic fits with the constraints of 
energy and momentum conservation, in which zero, one, or two additional 
photons emitted close to the beam direction were allowed. The fit with 
the lowest number of extra photons which gave an acceptable $\chi^2$ was 
chosen. The value of $\sqrt{s'}$ was then computed from the fitted 
four-momenta of the jets, i.e.\ excluding photons identified in the 
detector or those close
to the beam direction resulting from the fit. If none of the kinematic fits 
gave an acceptable $\chi^{2}$, $\sqrt{s'}$ was estimated directly from the 
angles of the jets as in reference~\cite{bib:OPAL-SM130}. 
Figure~\ref{fig:mh_sp} shows $\sqrt{s'}$
distributions at 172~GeV for events with different numbers of photons.
Note that this algorithm results in $s'$ equal to $s$ for
events which give a good kinematic fit with no photon either in the detector 
or along the beam direction.

Non-radiative events were selected by demanding $s'/s > 0.8$. The numbers
of events selected at each energy are shown in table~\ref{tab:mh}, 
together with the corresponding efficiencies and the fractions of the
$s'/s > 0.8$ sample arising from feedthrough of events with lower $s'/s$,
determined from Monte Carlo simulations.

The estimation of background in the non-radiative samples is less 
problematic than in the inclusive case, because the contribution from
two-photon events is tiny. The largest contribution arises from W-pair
events (above the W-pair threshold), and as in the inclusive case the
expected contribution has been subtracted. This amounts to (6.0$\pm$0.5)\%
at 161~GeV and (19.8$\pm$0.2)\% at 172~GeV, where again the dominant error
reflects the uncertainty in the W mass. Additional small backgrounds
arise from four-fermion production and tau pair events. The
total background at each energy is shown in table~\ref{tab:mh},
together with the final non-radiative hadronic cross-sections.

The main systematic error arises from the modelling of the separation 
of non-radiative from clearly radiative events, and was 
estimated by comparing eight different methods of separation. For
example, the algorithm was changed to allow for only a single radiated
photon, the photon identification algorithm was modified, the hadron
calorimeter was removed from the analysis or the jet resolution
parameter was altered. In each case, the modified algorithm was
applied to data and Monte Carlo, and the corrected cross-section
computed. The changes observed were in all cases compatible with
statistical fluctuations, but to be conservative the largest change
(averaged over all beam energies) was taken to define the systematic
error, amounting to 2.0\%. This error is expected to decrease in
future with improved data statistics. The error 
arising from the subtraction of W-pair background was investigated by
performing an alternative analysis in which events identified as W-pairs
according to the criteria in reference~\cite{bib:OPAL-WW161} were rejected. 
The resulting cross-sections after correcting to no interference, 
35.4$\pm$1.9$\pm$0.8 (26.5$\pm$1.7$\pm$0.6)~pb at 161 (172)~GeV are in
excellent agreement with those obtained by subtracting the expected
W-pair contribution.

To measure the angular distribution of the primary quark in the
hadronic events, we have used as an estimator the thrust axis for each event
determined from the observed tracks and clusters. The angular distribution of
the thrust axis was then corrected to the primary quark level using
bin-by-bin corrections determined from Monte Carlo events. No attempt
was made to identify the charge in these events, and thus we measured
the folded angular distribution. The measured values for the
$s'/s > 0.8$ sample are given in table~\ref{tab:mh_angdis}.
%and are shown in figure~\ref{fig:mh_angdis}.

%Overall, satisfactory agreement is observed between data and predictions
%for the hadronic final state.
%The largest deviation is observed for the non-radiative cross-section at
%$\sqrt{s}=130.25$~GeV, being 2.6 standard deviations below the 
%expectation, which is also apparent in its angular distribution in
%figure~\ref{fig:mh_angdis}.

\begin{table}[htbp]%-------------------------------------------------------
\begin{center}
\begin{tabular} {|c|c|c|c|c|}
\hline
\hline
\multicolumn{5}{|c|}{\bf \boldmath Hadrons ($s'/s>0.8$)} \\
\hline
$\absct$   &\multicolumn{4}{c|}{$\dsdabscc$ (pb)} \\
\hline%-----------------------------------------------------------------
           &130.25~GeV &136.22~GeV &161.34~GeV &172.12~GeV \\
\hline%-----------------------------------------------------------------
$ [0.0,0.1]$ & 44$\pm$13 & 35$\pm$11 &28.0$\pm$5.5  &22.3$\pm$5.2 \\
$ [0.1,0.2]$ & 47$\pm$12 & 57$\pm$14 &26.9$\pm$5.4  &21.7$\pm$5.3 \\
$ [0.2,0.3]$ & 58$\pm$15 & 44$\pm$13 &36.8$\pm$6.3  &18.0$\pm$4.9 \\
$ [0.3,0.4]$ & 51$\pm$13 & 58$\pm$15 &28.4$\pm$5.5  &27.5$\pm$5.9 \\
$ [0.4,0.5]$ & 52$\pm$14 & 38$\pm$12 &24.3$\pm$5.1  &26.5$\pm$5.8 \\
$ [0.5,0.6]$ & 53$\pm$14 & 53$\pm$14 &29.9$\pm$5.6  &13.8$\pm$4.4 \\
$ [0.6,0.7]$ & 70$\pm$16 &114$\pm$20 &42.5$\pm$6.6  &36.1$\pm$6.4 \\
$ [0.7,0.8]$ & 66$\pm$15 & 57$\pm$14 &45.5$\pm$6.8  &32.5$\pm$5.9 \\
$ [0.8,0.9]$ & 84$\pm$17 & 69$\pm$16 &38.5$\pm$6.2  &34.6$\pm$6.1 \\
$ [0.9,1.0]$ &141$\pm$31 &122$\pm$29 &57.7$\pm$10.3 &35.6$\pm$8.2 \\
\hline%-----------------------------------------------------------------
\hline%-----------------------------------------------------------------
\end{tabular}     
\caption{
Differential cross-sections for $\qqbar$ production.  The values are
corrected to no interference between initial- and final-state radiation 
as described in the text. Errors include statistical and systematic effects 
combined, with the former dominant.}
\label{tab:mh_angdis}
\end{center}  
\end{table}%------------------------------------------------------------

%-----------------------------------------------------------------------
\subsection{\boldmath Electron Pairs}           \label{sec:ee}
%-----------------------------------------------------------------------
The production of electron pairs is dominated by $t$-channel photon
exchange, for which a definition of $s'$ as for the other channels is
less natural. In addition, the increased probability for final-state 
radiation relative to initial-state radiation renders the separation between 
initial- and final-state photons more difficult.
Events with little radiation were therefore selected by a cut 
on $\thacol$, the acollinearity angle between electron and positron. A cut 
of $\thacol < 10^{\circ}$ roughly corresponds to a cut on the effective 
centre-of-mass energy of $s'/s > 0.8$, for the $s$-channel contribution.
We measure cross-sections for three different acceptance regions, defined
in terms of the angle of the electron, $\theta_{\mathrm{e^-}}$, or positron,
$\theta_{\mathrm{e^+}}$, with respect to the incoming electron direction,
and the acollinearity angle: 
\begin{itemize}
\item {\bf A:} $\absctem < 0.9$, $\absctep < 0.9$, $\thacol < 170^{\circ}$;
      this is a loose `inclusive' measurement;
\item {\bf B:} $\absctem < 0.7$, $\thacol < 10^{\circ}$; this acceptance
      region is expected to be enriched in the $s$-channel contribution, 
      and is used for asymmetry measurements;
      in addition, we measure the angular distribution for $\absctem < 0.9$ 
      and $\thacol < 10\degree$;
\item {\bf C:} $\absctem < 0.96$, $\absctep < 0.96$, $\thacol < 10^{\circ}$; 
      this `large acceptance' selection acts as a check on the luminosity 
      measurements.
\end{itemize} 

The selection of electron pair events is identical to previous 
analyses~\cite{bib:OPAL-SM161}.
Events selected as electron pairs are required to have at least two and not 
more than eight clusters in the electromagnetic calorimeter, and not more than
eight tracks in the central tracking chambers. At least two clusters
must have an energy exceeding 20\% of the beam energy, and the total
energy deposited in the electromagnetic calorimeter must be at least
50\% of the centre-of-mass energy. For selections A and B, at least two of 
the three highest energy clusters must each have an associated central 
detector track. If a cluster has more than one associated track, the highest 
momentum one is chosen. If all three clusters have an associated track, the 
two highest energy clusters are chosen to be the electron and positron. For 
the large acceptance selection, C, no requirement is placed on the association
of tracks to clusters, but the requirement on the total electromagnetic 
energy is increased to 70\% of the centre-of-mass energy.

\begin{figure}
  \epsfxsize=\textwidth
  %\epsfbox[0 0 567 567]{ee_esum_new.eps}
  \epsfbox[0 0 567 567]{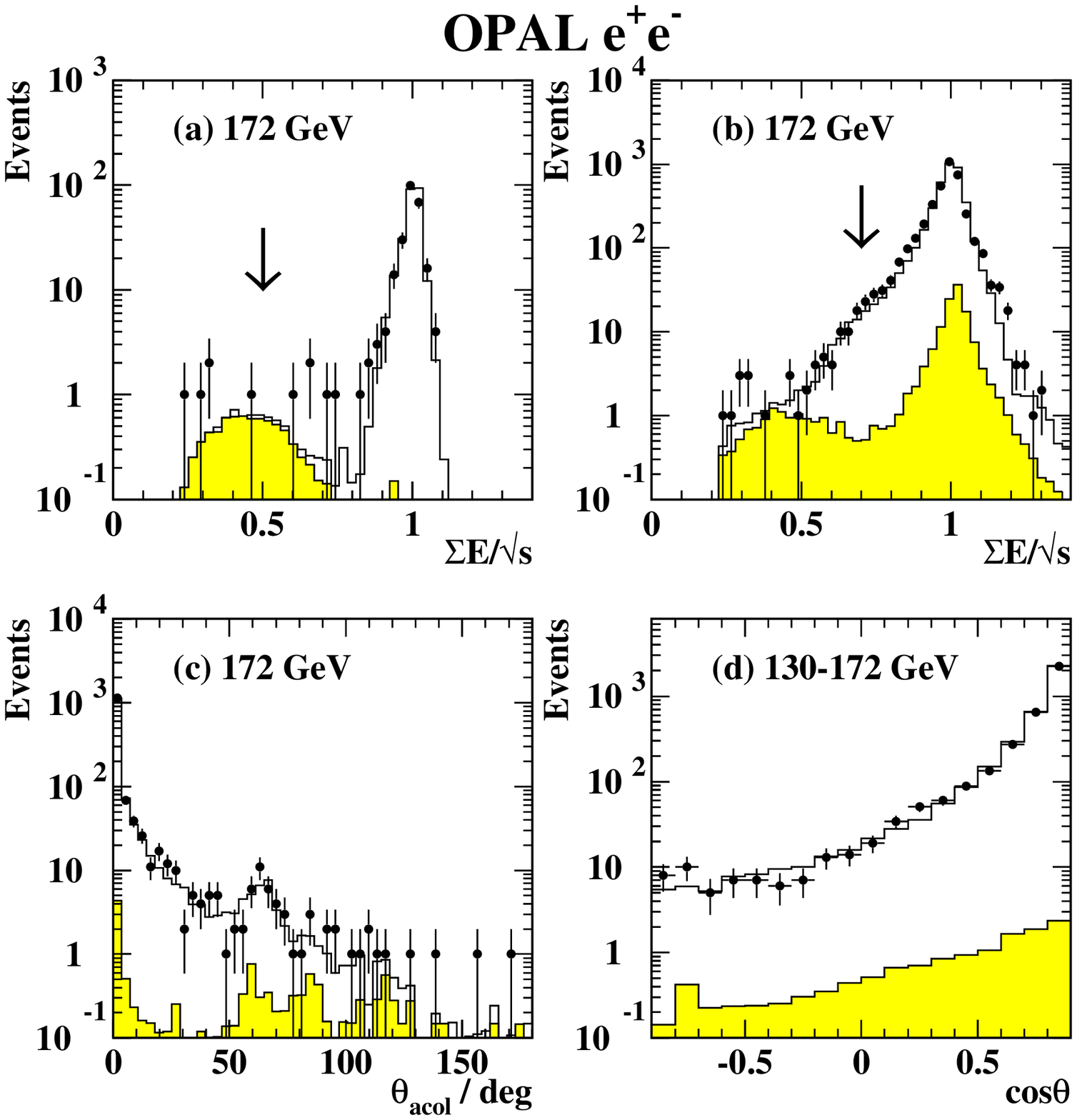}
  \caption
{
  (a) The distribution of the ratio of total electromagnetic calorimeter
  energy to the centre-of-mass energy for $\eetoee$ events in 
  acceptance region B, $\absctem < 0.7$ and $\thacol < 10\degree$, at
  172~GeV.
  (b) The same distribution for the large acceptance selection, C,
  $\absct < 0.96$ and $\thacol < 10\degree$, at 172~GeV. 
  Distributions are shown after all cuts except the one on total 
  electromagnetic calorimeter energy; the arrows indicate the positions 
  of the cuts on this quantity.
  (c) The acollinearity angle distribution for events satisfying the
  inclusive selection, in the acceptance region A, $\absctep < 0.9$,
  $\absctem < 0.9$, at 172~GeV. The cut on acollinearity angle has not 
  been applied.
  (d) Observed distribution of $\ct$ of the outgoing electron in \epem\
  events with $\thacol < 10\degree$. All centre-of-mass energies have been 
  summed for this plot.
  In each case, the points show the data and the histograms the Monte
  Carlo expectations, normalized to the integrated luminosity of the
  data, with the background contributions shaded.
}
\label{fig:ee_esum}
\end{figure}

These cuts have a very high efficiency for \epem\ events while providing
excellent rejection of backgrounds, which either have high multiplicity or
lower energy deposited in the electromagnetic calorimeter. The efficiency of 
the selection cuts, and small acceptance corrections, have been determined 
using Monte Carlo events generated with the BHWIDE~\cite{bib:bhwide} program. 
These are found to be independent of energy over the range considered here. 
Remaining backgrounds arise from \tautau\ events and, in the case of the 
loose acollinearity cut, also from electron pairs in two-photon events and 
from radiative Bhabha scattering events in which one electron is outside the 
detector acceptance but the photon is within the acceptance.
In the case of the large acceptance selection, C, which does not require
tracks, the main background arises from \gamgam\ final states. The efficiencies
and backgrounds at the three energies are summarized in table~\ref{tab:ee}.
In figure~\ref{fig:ee_esum}(a,b) we show distributions of total 
electromagnetic calorimeter
energy, after all other cuts, for acceptance regions B and C at 172~GeV, 
showing reasonable agreement between data and Monte Carlo. The degraded
energy resolution in acceptance region C arises from the increased amount
of material in front of the electromagnetic calorimeter at large
$\absct$, where the events are concentrated.
 The acollinearity
angle distribution for the inclusive selection, A, is shown in 
figure~\ref{fig:ee_esum}(c), and we see good agreement between data and
Monte Carlo expectation, including the peak corresponding to 
radiative $s$-channel return to the \PZ.

The numbers of selected events and resulting cross-sections are shown in
table~\ref{tab:ee}. 
The following sources of systematic error in the cross-section
measurements have been considered.
\begin{itemize}
\item
 Deficiencies in the simulation of the selection cuts. As shown in
 figure~\ref{fig:ee_esum}(b), the total calorimeter energy distribution
 is slightly broader in data than Monte Carlo for the large acceptance
 selection, C. The effect of this on the efficiency of this selection
 has been estimated by varying the cut in the range 40\% to 75\% of
 the centre-of-mass energy. In the other two selections,
 a more important effect is the efficiency for finding two 
 tracks, which has been investigated using events in which only one
 cluster has an associated track.
\item
 Knowledge of the acceptance correction and how well the edge of the 
 acceptance is modelled. Because of the steeply falling distribution,
 any bias in the measurement of $\theta$ has a significant effect on
 the cross-sections, particularly for the large acceptance selection.
 This has been investigated by comparing measurements of $\theta$ made
 using central detector tracks, calorimeter clusters and the outer muon 
 chambers. 
 In addition, in each case the full size of the acceptance correction
 derived from Monte Carlo has been included as a systematic error.
\item
 Uncertainties in the background contributions. For selections A and B
 these have been assessed by considering the numbers of events failing 
 the total energy cut. Data and Monte Carlo
 are consistent within the statistical precision of 30\%. The background
 in the large acceptance selection is almost all from $\gamma\gamma$
 final states, which is much less uncertain.
\end{itemize}
The total systematic error in selections A and B amounts 
to 1.4\% and 0.8\% respectively, of which the largest contribution arises 
from uncertainty in the track matching efficiency (0.8\% and 0.5\% 
respectively). In the large acceptance selection, the largest component in 
the total systematic error of 1.1\% arises from uncertainty in modelling the 
edge of the acceptance (0.9\%).

\begin{table}[htbp]%-------------------------------------------------------
\centering
\begin{tabular}{|l|c|c|c|c|}
\hline%-----------------------------------------------------------------
\hline%-----------------------------------------------------------------
\multicolumn{5}{|c|}{\bf \boldmath \epem (A: $\absct<0.9$, 
$\thacol<170\degree$)} \\
\hline%-----------------------------------------------------------------
           &130.25~GeV &136.22~GeV &161.34~GeV &172.12~GeV \\
\hline%-----------------------------------------------------------------
Events &591  &514  &1587 &1397 \\
Efficiency (\%)  &\multicolumn{4}{c|}{98.2$\pm$1.3}  \\
Background (pb)  &3.7$\pm$1.1  &3.4$\pm$1.0  &2.3$\pm$0.7  &1.9$\pm$0.6   \\
$\sigma_{\mathrm{meas}}$ (pb) &220$\pm$9$\pm$3   &197$\pm$9$\pm$3   
                              &158$\pm$4$\pm$2   &135$\pm$4$\pm$2   \\
$\sigma_{\mathrm{SM}}$ (pb) &237  &217  &154  &135  \\
\hline%-----------------------------------------------------------------
\hline%-----------------------------------------------------------------
\multicolumn{5}{|c|}{\bf \boldmath \epem (B: $\absctem<0.7$, 
$\thacol<10\degree$)} \\
\hline%-----------------------------------------------------------------
           &130.25~GeV &136.22~GeV &161.34~GeV &172.12~GeV \\
\hline%-----------------------------------------------------------------
Events &112  &98   &285  &246 \\
Efficiency (\%)  &\multicolumn{4}{c|}{99.2$\pm$0.7}  \\
Background (pb)  &0.6$\pm$0.2  &0.5$\pm$0.2  &0.4$\pm$0.1  &0.3$\pm$0.1   \\
$\sigma_{\mathrm{meas}}$ (pb) &41.3$\pm$4.0$\pm$0.5 &37.3$\pm$3.8$\pm$0.4 
                              &28.1$\pm$1.7$\pm$0.3 &23.5$\pm$1.5$\pm$0.2 \\
$\sigma_{\mathrm{SM}}$ (pb) &43.1 &39.5 &28.1 &24.7 \\
\hline%-----------------------------------------------------------------
\hline%-----------------------------------------------------------------
\multicolumn{5}{|c|}{\bf \boldmath \Pep\Pem (C: $\absct<0.96$, 
$\thacol<10\degree$)} \\
\hline%-----------------------------------------------------------------
           &130.25~GeV &136.22~GeV &161.34~GeV &172.12~GeV \\
\hline%-----------------------------------------------------------------
Events &1686 &1542 &4446 &3870 \\
Efficiency (\%)  &\multicolumn{4}{c|}{98.5$\pm$1.1}  \\
Background (pb)  &21.1$\pm$2.1  &19.3$\pm$1.9  &13.9$\pm$1.4  &12.2$\pm$1.2 \\
$\sigma_{\mathrm{meas}}$ (pb) &615$\pm$16$\pm$8   &580$\pm$15$\pm$8   
                              &434$\pm$7$\pm$5   &365$\pm$6$\pm$5   \\
$\sigma_{\mathrm{SM}}$ (pb) &645  &592  &425  &375  \\
\hline%-----------------------------------------------------------------
\hline%-----------------------------------------------------------------
\end{tabular}
\caption[]{Numbers of selected events, efficiencies, backgrounds and 
measured cross-sections for \epem\ events. The efficiencies are effective 
values combining the efficiency of selection cuts for events within the 
acceptance region and the effect of acceptance corrections. The 
errors on the efficiencies and backgrounds include Monte Carlo statistics
and all systematic effects, the latter being dominant. 
The first error on each measured cross-section is statistical, the
second systematic. The Standard Model predictions, $\sigma_{\mathrm{SM}}$, 
are from the ALIBABA~\cite{bib:alibaba} program. Unlike all other channels, 
values for $\epem$ include the effect of interference between initial- and
final-state radiation.
}
\label{tab:ee}
\end{table}%------------------------------------------------------------

The measurement of the angular distribution and asymmetry uses the same
event selection as above, with the further requirement that the two tracks 
have opposite charge. This extra requirement reduces the efficiency by
about 2.5\% in the region $\absct < 0.9$. In addition, to reduce the
effect of charge misassignment, events with $\ctem < -0.8$ must satisfy
two extra criteria: both electron and positron tracks must have momenta
of at least 25\% of the beam momentum, and there must be only one good
track associated with each cluster. These extra criteria reduce the
overall correction factor to the angular distribution for $\ctem < -0.8$
from about 25\% to 5\%.

The observed angular distribution of the electron, for events with
$\thacol < 10\degree$, is shown in figure~\ref{fig:ee_esum}(d). 
As the variation 
of the angular distribution with energy is small over the range considered 
here, we have summed data from all energies for this comparison with Monte 
Carlo expectation. The corrected distributions in $\cos\theta$
at each energy are given in
table~\ref{tab:angdis}. Systematic errors, arising mainly from 
uncertainty in the efficiency for finding two tracks with opposite
charge, amount to 1.2\% and are included in the errors in 
table~\ref{tab:angdis}.
The forward-backward asymmetries for the $\thacol < 10\degree$ sample at 
each energy within the 
angular range $\absctem < 0.7$ were evaluated by counting  
the numbers of events in the forward and backward $\ctem$ hemispheres.
The measured values are shown in table~\ref{tab:ee_afb}. Again, the errors
are predominantly statistical, with small systematic effects arising
from charge misassignment, acceptance definition and background included
in the values given.

\begin{figure}
  \epsfxsize=\textwidth
  %\epsfbox[0 0 567 567]{disk$cbhpae:[work.drw.talbot]spr172.eps}
  \epsfbox[0 0 567 567]{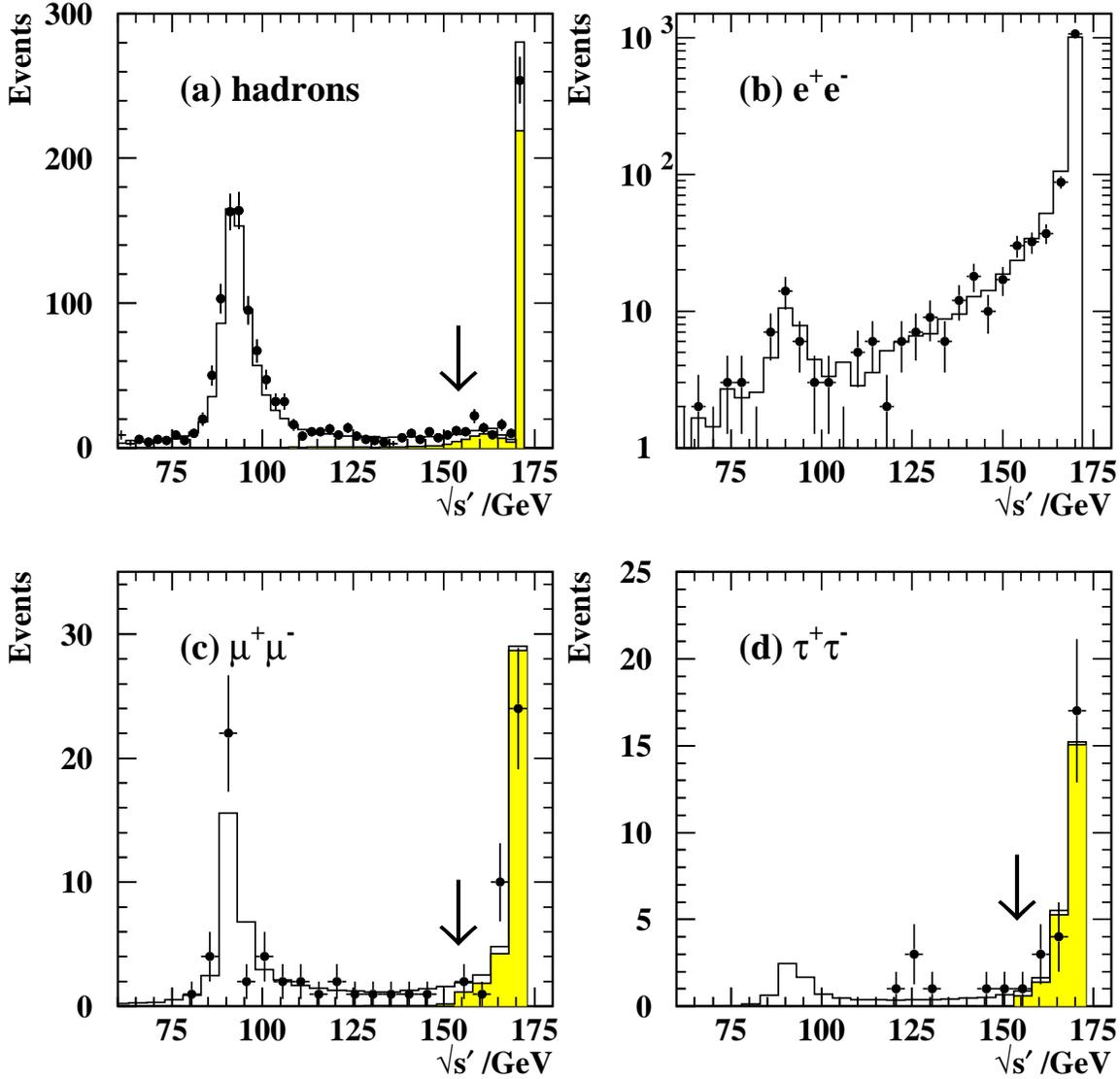}
  \caption
{
 The distributions of $\protect\sqrt{s'}$ 
 %measured from the lepton angles 
 for (a) hadronic events, (b) electron pair events with $\absctep < 0.9$, 
 $\absctem < 0.9$ and $\thacol < 170\degree$,
 (c) muon pair and (d) tau pair events at 172~GeV. In each case, the points 
 show the data and the histogram the Monte Carlo prediction, normalized
 to the integrated luminosity of the data, with the 
 contribution from events with true $s'/s > 0.8$ shaded in (a), (c) and (d).
 The arrows in (a), (c) and (d) show the position of the cut used to
 select `non-radiative' events.
}
\label{fig:emutau_sp}
\end{figure}

In figure~\ref{fig:emutau_sp}(b) we show the distribution of $\sqrt{s'}$ for
the inclusive electron pair events at 172~GeV.
The value of $s'$ for each event was estimated from the polar angles,
$\theta_{1}$ and $\theta_{2}$, of the two electrons, assuming massless 
three-body kinematics to calculate the energy of a possible undetected 
initial-state photon along the beam direction as 
$\sqrt{s}\cdot |\sin(\theta_{1} + \theta_{2})| / 
         (|\sin(\theta_{1} + \theta_{2})| + \sin\theta_{1} +\sin\theta_{2}) $.
A similar technique was used to calculate $s'$ for muon pairs and tau
pairs.
In contrast to the other final states, the radiative return peak forms
only a very small contribution to this channel.
%Good agreement is observed between data and predictions for all
%measurements performed for the electron pair final state.

%\clearpage
\begin{table}[p]%-------------------------------------------------------
\begin{center}
\small
\begin{tabular} {|c|r@{$\pm$}l|r@{$\pm$}l|r@{$\pm$}l|r@{$\pm$}l|}
\hline%-----------------------------------------------------------------
\hline%-----------------------------------------------------------------
\multicolumn{9}{|c|}{\boldmath $\epem$} \\
\hline
$[\ct_{\rm min},\ct_{\rm max}]$  &\multicolumn{8}{c|}{$\dsdcc$ (pb)} \\
\hline%-----------------------------------------------------------------
                          &\multicolumn{2}{c|}{130.25~GeV}
                          &\multicolumn{2}{c|}{136.22~GeV}
                          &\multicolumn{2}{c|}{161.34~GeV}
                          &\multicolumn{2}{c|}{172.12~GeV} \\
\hline%-----------------------------------------------------------------
$[-0.9,-0.7]$ &  6&$^{6}_{3}$  &  6&$^{6}_{3}$  &4.6&$^{2.2}_{1.6}$ 
&1.5&$^{1.5}_{0.8}$ \\
$[-0.7,-0.5]$ &  4&$^{5}_{2}$  &  4&$^{5}_{3}$  &2.4&$^{1.7}_{1.1}$ 
&1.4&$^{1.4}_{0.8}$ \\
$[-0.5,-0.3]$ &  7&$^{6}_{4}$  & 10&$^{7}_{4}$  &1.4&$^{1.5}_{0.8}$ 
&0.4&$^{1.1}_{0.4}$ \\
$[-0.3,-0.1]$ &  6&$^{6}_{3}$  &  8&$^{6}_{4}$  &2.4&$^{1.7}_{1.1}$ 
&3.8&$^{1.9}_{1.4}$ \\
$[-0.1,\;\;\;0.1]$ & 11&$^{7}_{5}$  &  8&$^{6}_{4}$  &6.0&$^{2.3}_{1.8}$ 
&5.3&$^{2.2}_{1.6}$ \\
$[\;\;\;0.1,\;\;\;0.3]$ & 19&$^{8}_{6}$  & 15&$^{8}_{5}$  & 16&3   
                             & 18&3   \\
$[\;\;\;0.3,\;\;\;0.5]$ & 49&10          & 23&$^{9}_{7}$  & 32&4   
                             & 23&3   \\
$[\;\;\;0.5,\;\;\;0.7]$ &112&15          &121&15    & 79&6   & 65&6 \\
$[\;\;\;0.7,\;\;\;0.9]$ &795&40          &701&38    &588&19  &506&17  \\
\hline%-----------------------------------------------------------------
\hline%-----------------------------------------------------------------
\multicolumn{9}{|c|}{\boldmath $\mumu$} \\
\hline
$[\ct_{\rm min},\ct_{\rm max}]$  &\multicolumn{8}{c|}{$\dsdcc$ (pb)} \\
\hline%-----------------------------------------------------------------
                          &\multicolumn{2}{c|}{130.25~GeV}
                          &\multicolumn{2}{c|}{136.22~GeV}
                          &\multicolumn{2}{c|}{161.34~GeV}
                          &\multicolumn{2}{c|}{172.12~GeV} \\
\hline%-----------------------------------------------------------------
$[-1.0,-0.8]$ &$-1$&$^{3}_{0}$ &$0$&$^{3}_{0}$   &$1.2$&$^{1.8}_{0.9}$ 
&$-0.1$&$^{0.8}_{0.0}$ \\
$[-0.8,-0.6]$ &$5$&$^{6}_{3}$  &$0$&$^{2}_{0}$   &$0.4$&$^{1.2}_{0.5}$ 
&$1.0$&$^{1.4}_{0.7}$ \\
$[-0.6,-0.4]$ &$0$&$^{4}_{2}$  &$0$&$^{2}_{0}$   &$2.5$&$^{1.8}_{1.2}$ 
&$0.4$&$^{1.2}_{0.4}$ \\
$[-0.4,-0.2]$ &$7$&$^{6}_{4}$  &$6$&$^{6}_{3}$   &$0.1$&$^{1.2}_{0.4}$ 
&$0.3$&$^{1.2}_{0.4}$ \\
$[-0.2,\;\;\;0.0]$      &$1$&$^{4}_{2}$  &$3$&$^{5}_{3}$   
                            &$1.9$&$^{1.6}_{1.0}$ &$2.4$&$^{1.7}_{1.1}$ \\
$[\;\;\;0.0,\;\;\;0.2]$ &$2$&$^{5}_{2}$       &$5$&$^{6}_{3}$ 
                            &$0.8$&$^{1.4}_{0.7}$ &$0.3$&$^{1.2}_{0.4}$ \\
$[\;\;\;0.2,\;\;\;0.4]$ &$9$&$^{6}_{4}$       &$4$&$^{5}_{3}$   
                            &$2.8$&$^{1.8}_{1.2}$ &$2.8$&$^{1.8}_{1.2}$ \\
$[\;\;\;0.4,\;\;\;0.6]$ &$3$&$^{5}_{3}$       &$14$&$^{8}_{5}$  
                            &$2.1$&$^{1.8}_{1.2}$ &$3.6$&$^{2.1}_{1.5}$ \\
$[\;\;\;0.6,\;\;\;0.8]$ &$5$&$^{6}_{3}$       &$10$&$^{7}_{5}$  
                            &$5.5$&$^{2.4}_{1.8}$ &$1.9$&$^{1.8}_{1.2}$ \\
$[\;\;\;0.8,\;\;\;1.0]$ &$14$&$^{9}_{6}$      &$20$&$^{11}_{8}$ 
                            &$4.9$&$^{2.9}_{2.1}$ &$5.1$&$^{2.9}_{2.0}$ \\
\hline%-----------------------------------------------------------------
\hline%-----------------------------------------------------------------
\multicolumn{9}{|c|}{\boldmath $\tautau$} \\
\hline
$[\ct_{\rm min},\ct_{\rm max}]$  &\multicolumn{8}{c|}{$\dsdcc$ (pb)} \\
\hline%-----------------------------------------------------------------
                          &\multicolumn{2}{c|}{130.25~GeV}
                          &\multicolumn{2}{c|}{136.22~GeV}
                          &\multicolumn{2}{c|}{161.34~GeV}
                          &\multicolumn{2}{c|}{172.12~GeV} \\
\hline%-----------------------------------------------------------------
$[-1.0,-0.8]$ &$-1$&$^{19}_{0}$ &$0$&$^{21}_{0}$  &$-0.1$&$^{3.5}_{0.0}$
&$-0.4$&$^{3.8}_{0.0}$ \\
$[-0.8,-0.6]$ &$0$&$^{3}_{0}$   &$0$&$^{3}_{0}$   &$1.4$&$^{1.9}_{1.0}$ 
&$0.5$&$^{1.7}_{0.6}$ \\
$[-0.6,-0.4]$ &$0$&$^{3}_{0}$   &$0$&$^{3}_{0}$   &$0.7$&$^{1.6}_{0.6}$ 
&$0.7$&$^{1.6}_{0.6}$ \\
$[-0.4,-0.2]$ &$0$&$^{3}_{0}$   &$0$&$^{3}_{0}$   &$1.6$&$^{2.1}_{1.2}$ 
&$0.3$&$^{1.6}_{0.6}$ \\
$[-0.2,\;\;\;0.0]$ &$2$&$^{6}_{2}$       &$0$&$^{3}_{0}$   
                       &$3.4$&$^{2.3}_{1.5}$ &$2.3$&$^{2.1}_{1.3}$ \\
$[\;\;\;0.0,\;\;\;0.2]$ &$5$&$^{7}_{3}$       &$6$&$^{7}_{4}$   
                            &$1.9$&$^{2.0}_{1.2}$ &$2.2$&$^{2.1}_{1.3}$ \\
$[\;\;\;0.2,\;\;\;0.4]$ &$1$&$^{6}_{2}$       &$3$&$^{6}_{2}$   
                            &$6.1$&$^{2.9}_{2.1}$ &$2.6$&$^{2.2}_{1.3}$ \\
$[\;\;\;0.4,\;\;\;0.6]$ &$10$&$^{8}_{5}$      &$11$&$^{9}_{5}$  
                            &$1.5$&$^{2.0}_{1.2}$ &$1.9$&$^{2.0}_{1.1}$ \\
$[\;\;\;0.6,\;\;\;0.8]$ &$9$&$^{9}_{5}$       &$8$&$^{8}_{5}$   
                            &$5.4$&$^{2.9}_{2.1}$ &$1.8$&$^{2.3}_{1.4}$ \\
$[\;\;\;0.8,\;\;\;1.0]$ &$-2$&$^{14}_{0}$     &$5$&$^{31}_{11}$ 
                            &$5.8$&$^{8.8}_{4.3}$ &$2.4$&$^{8.1}_{2.9}$ \\
\hline
\hline
\end{tabular}     
\caption{
Differential cross-sections for lepton pair production. The values for
$\epem$ are for $\thacol < 10\degree$; those for $\mumu$ and $\tautau$
are for $s'/s > 0.8$ and are corrected to no interference between 
initial- and final-state radiation as described in the text. Errors include 
statistical and systematic effects combined, with the former dominant.}
\label{tab:angdis}
\end{center}  
\end{table}%------------------------------------------------------------
\clearpage

\begin{table}[htbp]%-------------------------------------------------------
\centering
\begin{tabular}{|l|c|c|c|c|}
\hline%-----------------------------------------------------------------
\hline%-----------------------------------------------------------------
\multicolumn{5}{|c|}{\bf \boldmath \epem ($\absctem<0.7$, 
$\thacol<10\degree$)} \\
\hline%-----------------------------------------------------------------
           &130.25~GeV &136.22~GeV &161.34~GeV &172.12~GeV \\
\hline%-----------------------------------------------------------------
$N_{\mathrm{F}}$       & 98  &84  &257  &222 \\
$N_{\mathrm{B}}$       & 12  &13  &17   &17  \\
$\AFB^{\mathrm{meas}}$ &0.79$\pm$0.06 &0.73$\pm$0.07 &0.88$\pm$0.03
                       &0.86$\pm$0.04 \\
$\AFBSM$               &0.80 &0.80 &0.81 &0.81 \\
\hline%-----------------------------------------------------------------
\hline%-----------------------------------------------------------------
\end{tabular}
\caption[]{The numbers of forward ($N_{\mathrm{F}}$) and backward 
($N_{\mathrm{B}}$) events and measured asymmetry values for electron
pairs. The measured asymmetry values include corrections for background and 
efficiency. The errors shown are the combined statistical and systematic 
errors; in each case the statistical error is dominant. The Standard Model 
predictions, $\AFBSM$, are from the {\mbox ALIBABA~\cite{bib:alibaba}} program. 
Unlike all other channels, values for $\epem$ include the effect of 
interference between initial- and final-state radiation.
}
\label{tab:ee_afb}
\end{table}%------------------------------------------------------------

%-----------------------------------------------------------------------
\subsection{\boldmath Muon Pairs}           \label{sec:mumu}
%-----------------------------------------------------------------------
The selection of muon pair events is essentially identical to previous
analyses~\cite{bib:OPAL-SM161}, except that the cut on visible energy has
been made dependent on the centre-of-mass energy, to reduce loss of
radiative return events at higher energies. Muon pair events were required
to have at least two tracks with momentum greater than 6~GeV and
$\absct < 0.95$, separated in azimuthal angle by more than 320~mrad,
and identified as muons. These tracks must have at least 20 hits
in the central tracking chambers and the point of closest approach to the 
nominal beam axis must lie less than 1~cm in the $r$--$\phi$ plane 
and less than 50~cm along the beam axis from the nominal interaction
point. 
To be identified as a muon, a track
had to satisfy any of the following conditions:
\begin{itemize}
\item
At least 2 muon chamber hits associated with the track
within \hbox{$\Delta\phi=(100+100/p)$~mrad,} with the
momentum~$p$ in~GeV;
\item
At least 4 hadron calorimeter
strips associated with the track
within \hbox{$\Delta\phi=(20+100/p)$~mrad,} with $p$ in~GeV.
The average number of strips in layers
containing hits had to be less than 2
to discriminate against hadrons.
For $|\cos\theta|<0.65$,
where tracks traverse all 9 layers of strips in
the barrel calorimeter,
a hit in one of the last
3~layers of strips was required;
\item
Momentum
$p>15$~GeV and the electromagnetic energy
associated to the track within $\Delta\phi<70$~mrad 
less than 3~GeV.
\end{itemize}
If more than one pair of tracks satisfied the above conditions, the pair
with the largest total momentum was chosen. Background from high multiplicity
events was rejected by requiring that there be no more than one other
track in the event with a transverse momentum greater than 0.7~GeV.

\begin{figure}
  \epsfxsize=\textwidth
  %\epsfbox[0 0 567 567]{disk$cbhpae:[work.drw.talbot]praaa_mu.eps}
  \epsfbox[0 0 567 567]{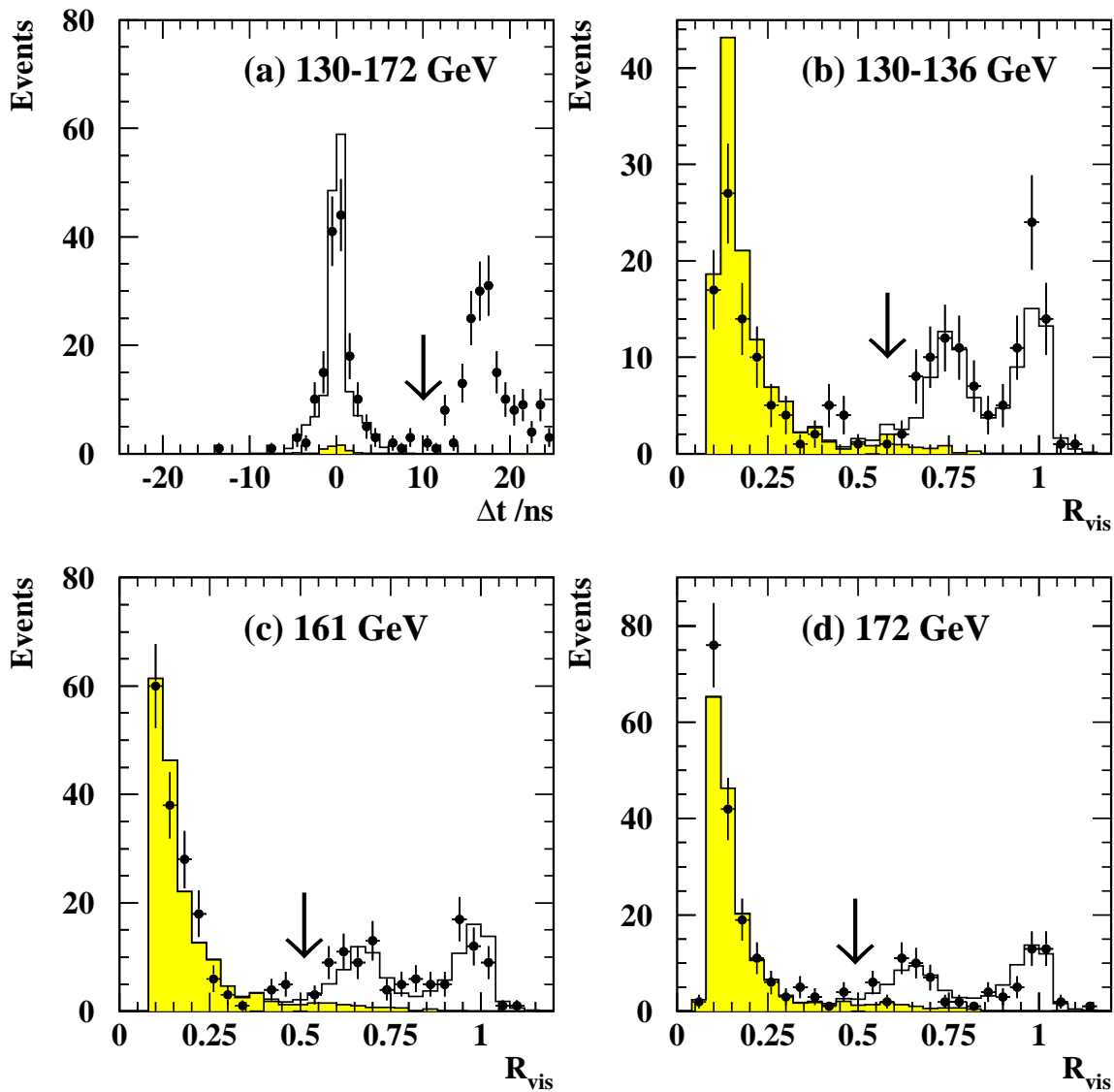}
  \caption
{
 (a) The time difference between hits in back-to-back TOF counters for 
 muon pair candidates; data from all energies have been included in this
 figure. (b) The distribution of the ratio of the visible energy to the 
 centre-of-mass energy for muon pair events at 130--136~GeV. (c) and (d) 
 show similar distributions at 161 and 172~GeV. 
 In each case, the points show the data and the histogram the Monte Carlo 
 expectation, normalized to the integrated luminosity of the data, with 
 the background contribution shaded. The arrows show the positions of the 
 cuts, which for the $\Rvis$ distributions are positioned 0.15 below 
 the expected value for radiative return events where the photon is
 undetected.
}
\label{fig:mu_xtot}
\end{figure}

Background from cosmic ray events was removed using the time-of-flight
(TOF) counters and vertex cuts.
In the barrel region, at least one TOF measurement was required
within 10~ns of that expected for a particle coming from the
interaction point. In addition, back-to-back pairs of TOF counters
were used to reject cosmic rays which had traversed the detector.
Figure~\ref{fig:mu_xtot}(a) shows the distribution of time difference, $\Delta t$,
between pairs of back-to-back TOF counters for muon pair events, before 
applying this cut, clearly showing one peak at the origin from muon
pairs and a second peak at about 15~ns from cosmic rays.
In the forward region, for which TOF information was not available,
the matching of the central detector tracks to the interaction
vertex was used in order to remove cosmic ray background.
The cosmic ray contamination after all cuts is low. There are no events
remaining close to the cosmic ray rejection cut boundaries
%after other cuts have been applied, 
in the 130 and 136~GeV samples, and one event 
remaining in each of the 161 and 172 GeV samples.

Background from two-photon events was rejected by placing a cut on the
total visible energy, $\Evis$, defined as the scalar sum of the momenta of the
two muons plus the energy of the highest energy cluster in the 
electromagnetic calorimeter:
\[
   \Rvis \equiv \Evis / \sqrt{s} > 0.5 (\mPZ^{2} / s) + 0.35.
\] 
The value of this cut is 0.15 below the expected value of \Rvis\ for
muon pairs in radiative return events where the photon escapes detection,
visible as secondary peaks in figure~\ref{fig:mu_xtot}(b-d).

The value of $s'$ for each event was estimated from the polar angles
of the two muons, as described in section~\ref{sec:ee} for electrons.
The observed distribution of $\sqrt{s'}$ at 172~GeV is shown in 
figure~\ref{fig:emutau_sp}(c). 
A non-radiative sample of events was selected by requiring $s'/s > 0.8$.
The selection efficiencies, and feedthrough of events from lower $s'/s$
into the non-radiative sample, were determined from Monte Carlo simulations, 
and are shown in table~\ref{tab:mumu}.

The residual background in the inclusive sample, of around 4\% at 130~GeV
increasing to 11\% at 172~GeV, arises mainly from \epem\mumu\ final states, 
while that in the non-radiative sample is predominantly \tautau\ events
and amounts to about 5\% in total. Total backgrounds are shown in
table~\ref{tab:mumu}, together with the numbers of selected events
and resulting cross-section measurements.

Systematic errors on the cross-section measurements, which arise from 
uncertainties in efficiency and backgrounds, are small compared to the 
statistical errors. In all cases, the dominant systematic error arises from 
the uncertainty in the background contamination.

\begin{table}[htbp]%-------------------------------------------------------
\centering
\begin{tabular}{|l|c|c|c|c|}
\hline%-----------------------------------------------------------------
\hline%-----------------------------------------------------------------
\multicolumn{5}{|c|}{\bf \boldmath \mumu\ ($s'/s>0.01$)} \\
\hline%-----------------------------------------------------------------
           &130.25~GeV &136.22~GeV &161.34~GeV &172.12~GeV \\
\hline%-----------------------------------------------------------------
Events &55   &56   &110  &82   \\
Efficiency (\%)  &82.6$\pm$0.8 &82.2$\pm$0.8 &79.9$\pm$0.7 &78.8$\pm$0.7  \\
Background (pb)  &0.8$\pm$0.3  &0.7$\pm$0.3  &0.8$\pm$0.2  &0.9$\pm$0.2   \\
$\sigma_{\mathrm{meas}}$ (pb) &23.7$\pm$3.2$\pm$0.5 &25.5$\pm$3.4$\pm$0.5 
                              &12.8$\pm$1.2$\pm$0.3 & 9.2$\pm$1.0$\pm$0.3 \\
$\sigma_{\mathrm{corr}}$ (pb) &23.6$\pm$3.2$\pm$0.5 &25.5$\pm$3.4$\pm$0.5 
                              &12.8$\pm$1.2$\pm$0.3 & 9.2$\pm$1.0$\pm$0.3 \\
$\sigma_{\mathrm{SM}}$ (pb) &22.0 &18.8 &11.3 & 9.6 \\
\hline%-----------------------------------------------------------------
\hline%-----------------------------------------------------------------
\multicolumn{5}{|c|}{\bf \boldmath \mumu\ ($s'/s>0.8$)} \\
\hline%-----------------------------------------------------------------
           &130.25~GeV &136.22~GeV &161.34~GeV &172.12~GeV \\
\hline%-----------------------------------------------------------------
Events &26   &30  &45   &37   \\
Efficiency (\%)  &90.1$\pm$0.7 &89.7$\pm$0.7 &89.6$\pm$0.6 &89.8$\pm$0.6  \\
Feedthrough (\%) &10.7$\pm$0.4 & 8.9$\pm$0.3 & 6.5$\pm$0.2 & 6.1$\pm$0.1  \\
Background (pb)  &0.3$\pm$0.2  &0.3$\pm$0.2  &0.15$\pm$0.06 &0.20$\pm$0.07  \\
$\sigma_{\mathrm{meas}}$ (pb) & 9.2$\pm$1.8$\pm$0.2 &11.5$\pm$2.1$\pm$0.2 
                              & 4.6$\pm$0.7$\pm$0.1 & 3.6$\pm$0.6$\pm$0.1 \\
$\sigma_{\mathrm{corr}}$ (pb) & 9.0$\pm$1.8$\pm$0.2 &11.4$\pm$2.1$\pm$0.2 
                              & 4.5$\pm$0.7$\pm$0.1 & 3.6$\pm$0.6$\pm$0.1 \\
$\sigma_{\mathrm{SM}}$ (pb) & 8.0 & 7.0 & 4.4 & 3.8 \\
\hline%-----------------------------------------------------------------
\hline%-----------------------------------------------------------------
\end{tabular}
\caption[]{Numbers of selected events, efficiencies, backgrounds, 
feedthrough of events from lower $s'$ to the $s'/s > 0.8$ samples and
measured cross-sections for muon pair events. The errors on efficiencies
and background include Monte Carlo statistics and systematic effects. 
The cross-sections labelled $\sigma_{\mathrm{meas}}$ are the measured values 
without the correction for interference between initial- and final-state 
radiation, those labelled $\sigma_{\mathrm{corr}}$ are with this correction. 
In some cases, the results are the same to the quoted precision. The first 
error on each measured cross-section is statistical, the second systematic.
The Standard Model predictions, $\sigma_{\mathrm{SM}}$, are from the
ZFITTER~\cite{bib:zfitter} program.
}
\label{tab:mumu}
\end{table}%------------------------------------------------------------

\begin{figure}
  \epsfxsize=\textwidth
  %\epsfbox[0 0 567 567]{disk$cbhpae:[work.drw.talbot]qcos2.eps}
  \epsfbox[0 0 567 567]{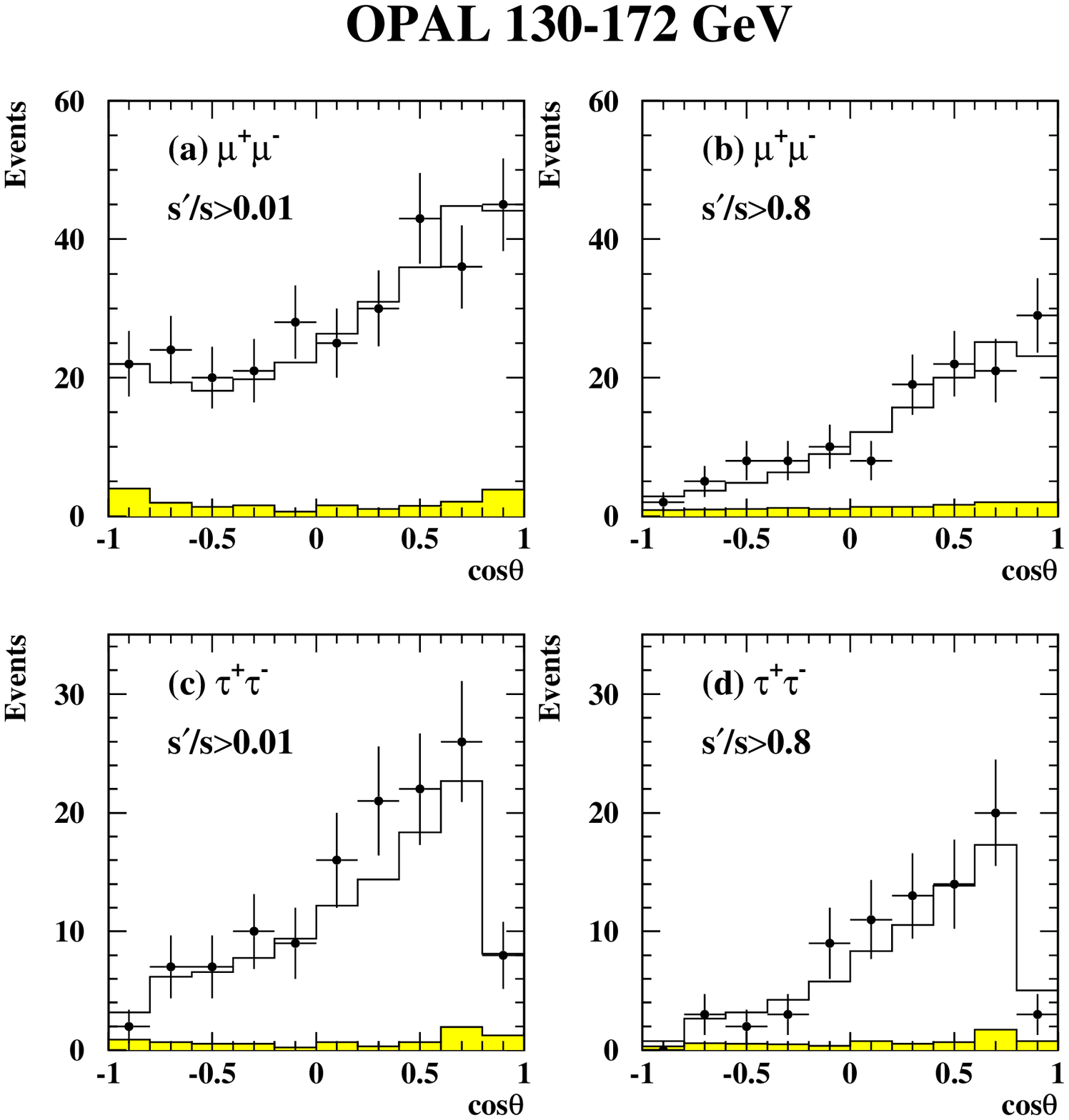}
  \caption
{
 Observed distributions of $\ct$ of the outgoing negative lepton in muon
 and tau pair events. (a) and (b) show muon pair events with
 $s'/s > 0.01$ and $s'/s > 0.8$ respectively, (c) and (d) tau
 pair events with  $s'/s > 0.01$ and $s'/s > 0.8$. The points
 show data for all energies combined, while the histograms show the 
 Monte Carlo expectations, formed by summing the predictions at
 each energy normalized to the measured integrated luminosity values.
 The background contributions (including feedthrough from lower $s'$
 in the $s'/s > 0.8$ cases) are shaded.
}
\label{fig:mutau_angdis}
\end{figure}

The observed angular distribution of the $\Pgmm$ is shown in 
figure~\ref{fig:mutau_angdis}(a) for the $s'/s > 0.01$ sample and 
figure~\ref{fig:mutau_angdis}(b) for the $s'/s > 0.8$ sample, for all 
centre-of-mass energies combined. The angular 
distributions at each energy were corrected for efficiency and background, 
including feedthrough of muon pair events from lower $s'/s$ into the 
non-radiative samples, using Monte Carlo events. The corrected angular
distributions are shown in table~\ref{tab:angdis}. The final values have
been obtained by averaging the distribution measured using the negative
muon with that using the positive muon; although this averaging does
not reduce the statistical errors on the measurements, it is expected
to reduce most systematic effects. The forward-backward asymmetries at
each energy were obtained by counting the numbers of events in the 
forward and backward hemispheres, after correcting for background and
efficiency. The data at 130 and 136~GeV have been combined for the
asymmetry measurements.
Systematic errors were assessed by comparing results obtained 
using different combinations of tracking and muon chambers to measure the muon 
angles. The total systematic error, including the contribution from
the correction for interference between initial- and final-state
radiation, is below 0.01 in all cases, much smaller than the 
statistical errors. The measured 
asymmetry values are compared with the Standard Model predictions in 
table~\ref{tab:mutau_afb}.

\begin{table}[htbp]%-------------------------------------------------------
\centering
\small
\begin{tabular}{|l|c|c|c|}
\hline%-----------------------------------------------------------------
\hline%-----------------------------------------------------------------
\multicolumn{4}{|c|}{\bf \boldmath \mumu\ ($s'/s>0.01$)} \\
\hline%-----------------------------------------------------------------
       &133.17~GeV &161.34~GeV &172.12~GeV \\
\hline%-----------------------------------------------------------------
$N_{\mathrm{F}}$       &71  &63   &47    \\
$N_{\mathrm{B}}$       &38  &43   &32.5  \\
$\AFB^{\mathrm{meas}}$ &0.31$\pm$0.09 &0.16$\pm$0.10 &0.18$\pm$0.11 \\
$\AFB^{\mathrm{corr}}$ &0.31$\pm$0.09 &0.16$\pm$0.10 &0.17$\pm$0.11 \\
$\AFBSM$               &0.29          &0.28          &0.28 \\
\hline%-----------------------------------------------------------------
\hline%-----------------------------------------------------------------
\multicolumn{4}{|c|}{\bf \boldmath \mumu\ ($s'/s>0.8$)} \\
\hline%-----------------------------------------------------------------
       &133.17~GeV &161.34~GeV &172.12~GeV \\
\hline%-----------------------------------------------------------------
$N_{\mathrm{F}}$       &42  &31     &27  \\
$N_{\mathrm{B}}$       &12  &12.5   & 9  \\
$\AFB^{\mathrm{meas}}$ &0.64$\pm$0.11 &0.47$\pm$0.14 &0.57$\pm$0.15 \\
$\AFB^{\mathrm{corr}}$ &0.63$\pm$0.11 &0.45$\pm$0.14 &0.55$\pm$0.15 \\
$\AFBSM$               &0.69          &0.60          &0.59 \\
\hline%-----------------------------------------------------------------
\hline%-----------------------------------------------------------------
\multicolumn{4}{|c|}{\bf \boldmath \tautau\ ($s'/s>0.01$)} \\
\hline%-----------------------------------------------------------------
       &133.17~GeV &161.34~GeV &172.12~GeV \\
\hline%-----------------------------------------------------------------
$N_{\mathrm{F}}$       &37  &35.5   &17  \\
$N_{\mathrm{B}}$       &12  &17.5   & 9  \\
$\AFB^{\mathrm{meas}}$ &0.43$\pm$0.13 &0.31$\pm$0.13 &0.21$\pm$0.19 \\
$\AFB^{\mathrm{corr}}$ &0.43$\pm$0.13 &0.30$\pm$0.13 &0.21$\pm$0.19 \\
$\AFBSM$               &0.29          &0.28          &0.28 \\
\hline%-----------------------------------------------------------------
\hline%-----------------------------------------------------------------
\multicolumn{4}{|c|}{\bf \boldmath \tautau\ ($s'/s>0.8$)} \\
\hline%-----------------------------------------------------------------
       &133.17~GeV &161.34~GeV &172.12~GeV \\
\hline%-----------------------------------------------------------------
$N_{\mathrm{F}}$       &21  &24.5   &15  \\
$N_{\mathrm{B}}$       &1   &10.5   & 6  \\
$\AFB^{\mathrm{meas}}$ &--   &0.51$\pm$0.15 &0.55$\pm$0.20 \\
$\AFB^{\mathrm{corr}}$ &--   &0.51$\pm$0.15 &0.55$\pm$0.20 \\
$\AFBSM$               &0.69 &0.60          &0.59 \\
\hline%-----------------------------------------------------------------
\hline%-----------------------------------------------------------------
\multicolumn{4}{|c|}{\bf \boldmath Combined \mumu\ and \tautau\ ($s'/s>0.01$)}
 \\
\hline%-----------------------------------------------------------------
       &133.17~GeV &161.34~GeV &172.12~GeV \\
\hline%-----------------------------------------------------------------
$\AFB^{\mathrm{meas}}$ &0.35$\pm$0.08 &0.21$\pm$0.08 &0.18$\pm$0.10 \\
$\AFB^{\mathrm{corr}}$ &0.35$\pm$0.08 &0.21$\pm$0.08 &0.18$\pm$0.10 \\
$\AFBSM$               &0.29          &0.28          &0.28 \\
\hline%-----------------------------------------------------------------
\hline%-----------------------------------------------------------------
\multicolumn{4}{|c|}{\bf \boldmath Combined \mumu\ and \tautau\ ($s'/s>0.8$)}
 \\
\hline%-----------------------------------------------------------------
       &133.17~GeV &161.34~GeV &172.12~GeV \\
\hline%-----------------------------------------------------------------
$\AFB^{\mathrm{meas}}$ &0.71$\pm$0.08 &0.49$\pm$0.10 &0.57$\pm$0.12 \\
$\AFB^{\mathrm{corr}}$ &0.70$\pm$0.08 &0.48$\pm$0.10 &0.55$\pm$0.12 \\
$\AFBSM$               &0.69          &0.60          &0.59 \\
\hline%-----------------------------------------------------------------
\hline%-----------------------------------------------------------------
\end{tabular}
\caption[]{The numbers of forward ($N_{\mathrm{F}}$) and backward 
($N_{\mathrm{B}}$) events and measured asymmetry values for muon and
tau pairs. The measured asymmetry values include corrections for background 
and efficiency and are corrected to the full solid angle. The errors shown 
are the combined statistical and systematic errors; in each case the 
systematic error is less than 0.01. The values labelled 
$\AFB^{\mathrm{meas}}$ are
the measured values without the correction for interference between 
initial- and final-state radiation, those labelled $\AFB^{\mathrm{corr}}$ 
are with this correction. In some cases, the results are the same
to the quoted precision. The Standard Model predictions, $\AFBSM$, are from 
the ZFITTER~\cite{bib:zfitter} program.
}
\label{tab:mutau_afb}
\end{table}%------------------------------------------------------------

%-----------------------------------------------------------------------
\subsection{\boldmath Tau Pairs}           \label{sec:tautau}
%-----------------------------------------------------------------------
The selection of $\eetotautau$ events is based on that used in previous 
analyses~\cite{bib:OPAL-SM161,bib:OPAL-SM130}, using information from the 
central tracking detectors and electromagnetic calorimetry to identify events
with two collimated, low multiplicity jets. However, the cuts have been
optimized and unified for the different energies, giving an improved
efficiency at 130--136~GeV. An inclusive sample of events was selected
with the following cuts.
 
\begin{figure}
  \epsfxsize=0.95\textwidth
  %\epsfbox[0 0 567 709]{disk$cbhpae:[work.drw.talbot]praaa_tau.eps}
  \epsfbox[0 0 567 709]{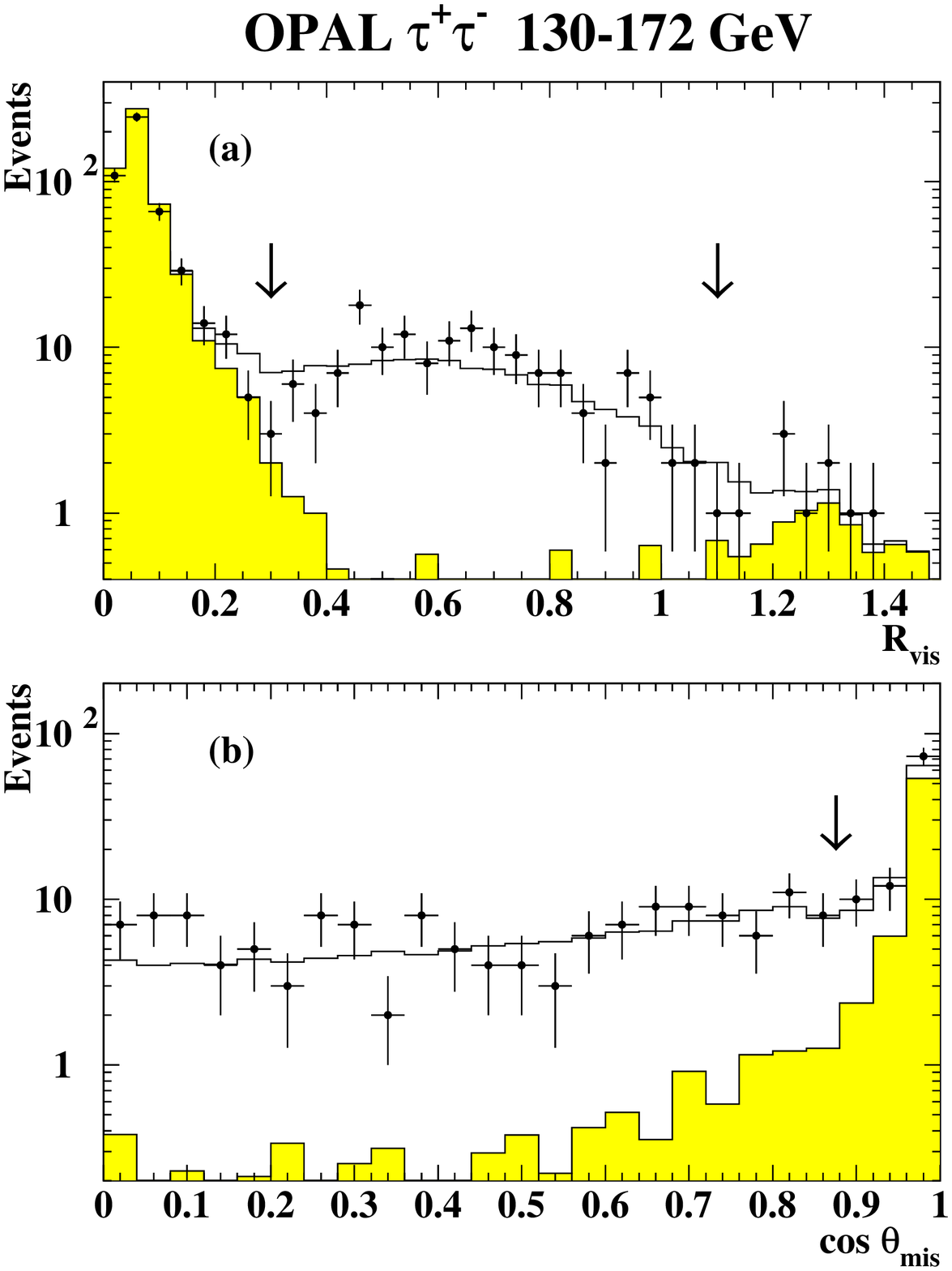}
  \caption
{
 (a) The distribution of the ratio of the visible energy to the
 centre-of-mass energy for tau pair events. Note that visible energy here
 is defined as the scalar sum of track momenta plus electromagnetic calorimeter 
 energy, with no correction for double counting.
 (b) Distribution of $\ct$ of the missing momentum vector, calculated 
 using electromagnetic clusters, for tau pair candidates. 
 Data from all energies have been included in this figure.
 In each case, the points show the data and the histogram the Monte Carlo 
 expectation, normalized to the integrated luminosity of the data, with the 
 background contribution shaded. The arrows show the positions of the cuts.
}
\label{fig:tau_rvis}
\end{figure}

\begin{itemize}
\item{}
Hadronic events were rejected by demanding low multiplicity: the
number of tracks reconstructed in the central tracking detectors had
to be at least two and at most six, and the sum of the number of tracks 
and the number of electromagnetic clusters not more than~15.
\item{}
The total energy of an event was restricted in order to reject
events from $\eetoee(\gamma)$ and two-photon processes:
the total visible energy in the event, derived 
from the scalar sum of all track momenta plus electromagnetic calorimeter 
energy, was required to be between 0.3$\roots$ and 1.1$\roots$.
In addition, the total electromagnetic calorimeter energy was required
to be less than 0.7$\roots$ and the scalar sum of track momenta less than 
0.8$\roots$. In the endcap region, $\absct > 0.7$, the upper limit on
the visible energy was reduced to 1.05$\roots$ because of the less good
electron energy resolution. 
The distribution of total visible energy, after all other cuts 
have been applied, is shown in figure~\ref{fig:tau_rvis}(a) for all 
centre-of-mass energies combined.
\item{} Background from
two-photon events was further reduced by cuts on the missing momentum
and its direction. The missing momentum in the transverse plane to the beam
axis, calculated 
using the electromagnetic calorimeter, was required to exceed 1.5~GeV.
The polar angle of the missing momentum calculated using tracks only
or electromagnetic clusters only was required to satisfy $\absct < 0.95$ and 
$\absct < 0.875$ respectively. Figure~\ref{fig:tau_rvis}(b) shows the
distribution of $\ct$ of the missing momentum vector calculated using
electromagnetic clusters after all other cuts have been applied, for
all centre-of-mass energies combined.
\item{}
Vertex and TOF cuts were imposed to remove cosmic ray events, as for
\mumu\ events. In addition, $\eetomumu$ events were removed; these
were identified by the criteria described in section \ref{sec:mumu}, except 
that the total visible energy was required to exceed 60\% of the 
centre-of-mass energy.
\item{} 
Tau pair events are characterized by a pair of narrow `jets'.
Tracks and electromagnetic clusters, each treated as separate particles,
were combined in the following way.
First the highest energy particle in the event was selected and
a cone with a half angle of 35$\degree$ was defined around it.
The particle with the next highest energy inside the cone was
combined with the first.
The momenta of the combined particles were added and the direction
of the sum was used to define a new cone,
inside which the next highest
energy particle was again looked for. This procedure was repeated
until no more particles were found inside the cone.
Similarly, starting with the highest energy particle among the
remainder, a new cone was initiated and treated in the same way.
This process continued until finally all
the particles in the event had been assigned to a cone.
\item{}
At least one charged particle was required for each cone,
and the sum of the energy in the electromagnetic calorimeter and the track 
momenta in a cone had to be more than 1\% of the beam energy.
Events which had exactly two such cones were selected as 
$\eetotautau$ candidates. 
The direction of each $\tau$ was approximated by that of the total
momentum vector of its cone of particles.
%The vectorial difference between the momenta of the two $\tau$ jets
%was used to define an event axis. 
Events were accepted if the average
value of $\absct$ for the two $\tau$ jets, $|\ct_{\mathrm{av}}|$,
satisfied $|\ct_{\mathrm{av}}| < 0.85$.
\item{}
Most of the remaining background from two-photon processes was rejected
by a cut on the acollinearity and acoplanarity angles of the two $\tau$
cones: the acollinearity angle, in degrees, was required to satisfy
\[
  \thacol < (180\degree - 2\tan^{-1}(2\mPZ\sqrt{s} / (s - \mPZ^{2}))) 
    + 10\degree
\]
and the acoplanarity angle was required to be less than 30$^{\circ}$.
The value of the cut on acollinearity was chosen such as to include the
peak from radiative return events at each energy; it is $50\degree$ at
130~GeV rising to $78\degree$ at 172~GeV.
\item{}
Remaining background from $\eetoee (\gamma)$ events was removed by
rejecting events if the ratio of the electromagnetic energy to the track 
momentum in both of the $\tau$ cones was between 0.9 and 1.1,
as expected for an electron.
\item{}
Finally, at 161 and 172~GeV, events classified as W-pair candidates
according to the criteria in reference~\cite{bib:OPAL-WW161} were rejected.
\end{itemize}

The effective centre-of-mass energy of the $\epem$ collision was 
estimated from the directions of the two $\tau$ jets, as described for
\epem\ events in section~\ref{sec:ee}. The distribution for the 
172~GeV events is shown in
figure~\ref{fig:emutau_sp}(d). A non-radiative sample of $\tautau$ events
was selected from the inclusive sample by requiring $s'/s > 0.8$.

The numbers of events selected at each energy, together with the
efficiencies and feedthrough of events from lower $s'$ into the 
$s'/s > 0.8$ samples, all determined from Monte Carlo
simulations, are shown in table~\ref{tab:tautau}. 

The remaining background, which amounts to 5--13\% in the inclusive samples
and 2--7\% in the non-radiative samples, is mainly from two-photon 
interactions; there are also contributions
from electron and muon pairs. The total background contributions are
shown in table~\ref{tab:tautau}, together with numbers of selected
events and resulting cross-sections.

The main sources of systematic uncertainty in the cross-section measurements 
arise from the efficiency and background estimation. The error
in the efficiency has been estimated using high statistics samples of
LEP1 data, that in the background by comparing data and Monte Carlo 
distributions of the selection variables after loosening some of the cuts. 

\begin{table}[htb]%-------------------------------------------------------
\centering
\begin{tabular}{|l|c|c|c|c|}
\hline%-----------------------------------------------------------------
\hline%-----------------------------------------------------------------
\multicolumn{5}{|c|}{\bf \boldmath \tautau\ ($s'/s>0.01$)} \\
\hline%-----------------------------------------------------------------
           &130.25~GeV &136.22~GeV &161.34~GeV &172.12~GeV \\
\hline%-----------------------------------------------------------------
Events &31   &25   &59   &32   \\
Efficiency (\%)  &39.4$\pm$1.1 &38.4$\pm$1.0 &33.9$\pm$0.9 &32.6$\pm$0.9  \\
Background (pb)  &0.56$\pm$0.14 &0.51$\pm$0.13 &0.29$\pm$0.08 &0.43$\pm$0.11 \\
$\sigma_{\mathrm{meas}}$ (pb) &27.7$\pm$5.0$\pm$0.9 &23.9$\pm$4.8$\pm$0.8 
                              &16.7$\pm$2.2$\pm$0.5 & 8.4$\pm$1.5$\pm$0.4 \\
$\sigma_{\mathrm{corr}}$ (pb) &27.6$\pm$5.0$\pm$0.9 &23.8$\pm$4.8$\pm$0.8 
                              &16.6$\pm$2.2$\pm$0.5 & 8.4$\pm$1.5$\pm$0.4 \\
$\sigma_{\mathrm{SM}}$ (pb) &22.0 &18.8 &11.3 & 9.6 \\
\hline%-----------------------------------------------------------------
\hline%-----------------------------------------------------------------
\multicolumn{5}{|c|}{\bf \boldmath \tautau\ ($s'/s>0.8$)}\\
\hline%-----------------------------------------------------------------
           &130.25~GeV &136.22~GeV &161.34~GeV &172.12~GeV \\
\hline%-----------------------------------------------------------------
Events &12   &12   &38   &25   \\
Efficiency (\%)  &55.2$\pm$1.5 &56.1$\pm$1.6 &56.9$\pm$1.5 &56.8$\pm$1.5  \\
Feedthrough (\%) & 8.5$\pm$0.4 & 7.2$\pm$0.4 & 4.6$\pm$0.1 & 4.2$\pm$0.1  \\
Background (pb)  &0.24$\pm$0.08 &0.22$\pm$0.08 &0.08$\pm$0.03 &0.16$\pm$0.04 \\
$\sigma_{\mathrm{meas}}$ (pb) & 6.9$\pm$2.0$\pm$0.3 & 7.3$\pm$2.1$\pm$0.3 
                              & 6.3$\pm$1.0$\pm$0.2 & 3.9$\pm$0.8$\pm$0.1 \\
$\sigma_{\mathrm{corr}}$ (pb) & 6.8$\pm$2.0$\pm$0.3 & 7.2$\pm$2.1$\pm$0.3 
                              & 6.2$\pm$1.0$\pm$0.2 & 3.9$\pm$0.8$\pm$0.1 \\
$\sigma_{\mathrm{SM}}$ (pb) & 8.0 & 6.9 & 4.4 & 3.8 \\
\hline%-----------------------------------------------------------------
\hline%-----------------------------------------------------------------
\end{tabular}
\caption[]{Numbers of selected events, efficiencies, backgrounds, 
feedthrough of events from lower $s'$ to the $s'/s > 0.8$ samples and
measured cross-sections for \tautau\ events. The errors on efficiencies
and background include Monte Carlo statistics and systematic effects. The 
cross-sections labelled $\sigma_{\mathrm{meas}}$ are the measured values 
without the correction for interference between initial- and final-state 
radiation, those labelled $\sigma_{\mathrm{corr}}$ are with this correction. 
In some cases, the results are the same to the quoted precision. The first 
error on each measured cross-section is statistical, the second systematic.
The Standard Model predictions, $\sigma_{\mathrm{SM}}$, are from the
ZFITTER~\cite{bib:zfitter} program.
}
\label{tab:tautau}
\end{table}%------------------------------------------------------------

The observed angular distribution of the $\Pgtm$ is shown in 
figure~\ref{fig:mutau_angdis}(c) for the $s'/s > 0.01$ sample and 
figure~\ref{fig:mutau_angdis}(d) for the $s'/s > 0.8$ sample, for all 
centre-of-mass energies combined. Monte Carlo events were used to correct for 
efficiency 
and background, including feedthrough of events from lower $s'/s$ into the 
non-radiative samples. The corrected angular distributions at each energy 
are given in table~\ref{tab:angdis}. The forward-backward asymmetries were 
evaluated by counting the corrected numbers of events, as for
the muons. Systematic errors were assessed by comparing different methods 
of determining the asymmetry: using tracks, electromagnetic clusters or 
both to determine the $\tau$ angles. The total systematic error, including 
the contribution from the correction for interference between initial- and 
final-state radiation, is below 0.01 in all cases, much smaller than the 
statistical errors.
The measured values are shown in table~\ref{tab:mutau_afb}.
In the same way as for the muons, we combine the 130 and 136~GeV data for 
the asymmetry measurements. From table~\ref{tab:mutau_afb} it can be
seen that for the non-radiative sample at 133~GeV there is only one event 
in the backward hemisphere which after correction for efficiency, 
background and acceptance would yield an unphysical value of the asymmetry.

Combined asymmetries from the \mumu\ and \tautau\ channels were
obtained, assuming $\mu$--$\tau$ universality, by forming a weighted average 
of the corrected numbers of forward and backward events observed in
the two channels at each energy. The combined values are shown in
table~\ref{tab:mutau_afb}. 
 
%-----------------------------------------------------------------------
\subsection{\boldmath The Fraction $R_{\rm b}$ of \bbbar\ Events} \label{sec:rb}
%-----------------------------------------------------------------------
To measure $\Rb$, the ratio of the cross-section for \bbbar\ production
to the hadronic cross-section, we have performed b-flavour tagging
for the hadronic events with $s'/s > 0.8$, selected as described in 
section~\ref{sec:mh}. In addition we rquire at least seven tracks that
pass standard track quality requirements, and the polar
angle of the thrust direction to fulfil $\absct < 0.9$ for the 161 and 
172~GeV data, $\absct < 0.8$ for the 130--136~GeV data. This acceptance cut
ensures that a large proportion of tracks are within the acceptance of
the silicon microvertex detector, which had a different geometry for
the two sets of data.

The b-tagging technique is based on the relatively long lifetime 
($\sim$ 1.5 ps)
of bottom hadrons, which allows the detection of secondary vertices 
significantly separated from the primary vertex. The primary vertex
for each event was reconstructed using a $\chi^2$ minimization method
incorporating as a constraint the average beam spot position, determined 
from tracks and the LEP beam orbit measurement system. Although the beam
spot is less precisely determined than at \LEPone, the resulting error on the
primary vertex position is still small compared to the errors on the
reconstructed secondary vertex positions.
The secondary vertex reconstruction was the same as adopted 
in \cite{bib:rb_new}, but the minimum number of 
tracks forming a vertex was reduced from four to three. Vertices were
reconstructed in the $x$--$y$ plane.
Tracks used for secondary vertex reconstruction were required to have a 
momentum greater than 500 MeV. In addition, the impact parameter in the
$x$--$y$ plane relative
to the reconstructed primary vertex was required to satisfy
$|d_0|\!<\!0.3$\,cm,
and its error
$\sigma_{d_0}\!<\!0.1$\,cm.
This mainly removes badly measured tracks
and, for example, tracks from $\rm K^0$ or $\Lambda$ decays.

\begin{figure}
\begin{center}
\epsfxsize=0.95\textwidth
%\epsfbox[0 0 567 709]{rbfig_paper.eps}
\epsfbox[0 0 567 709]{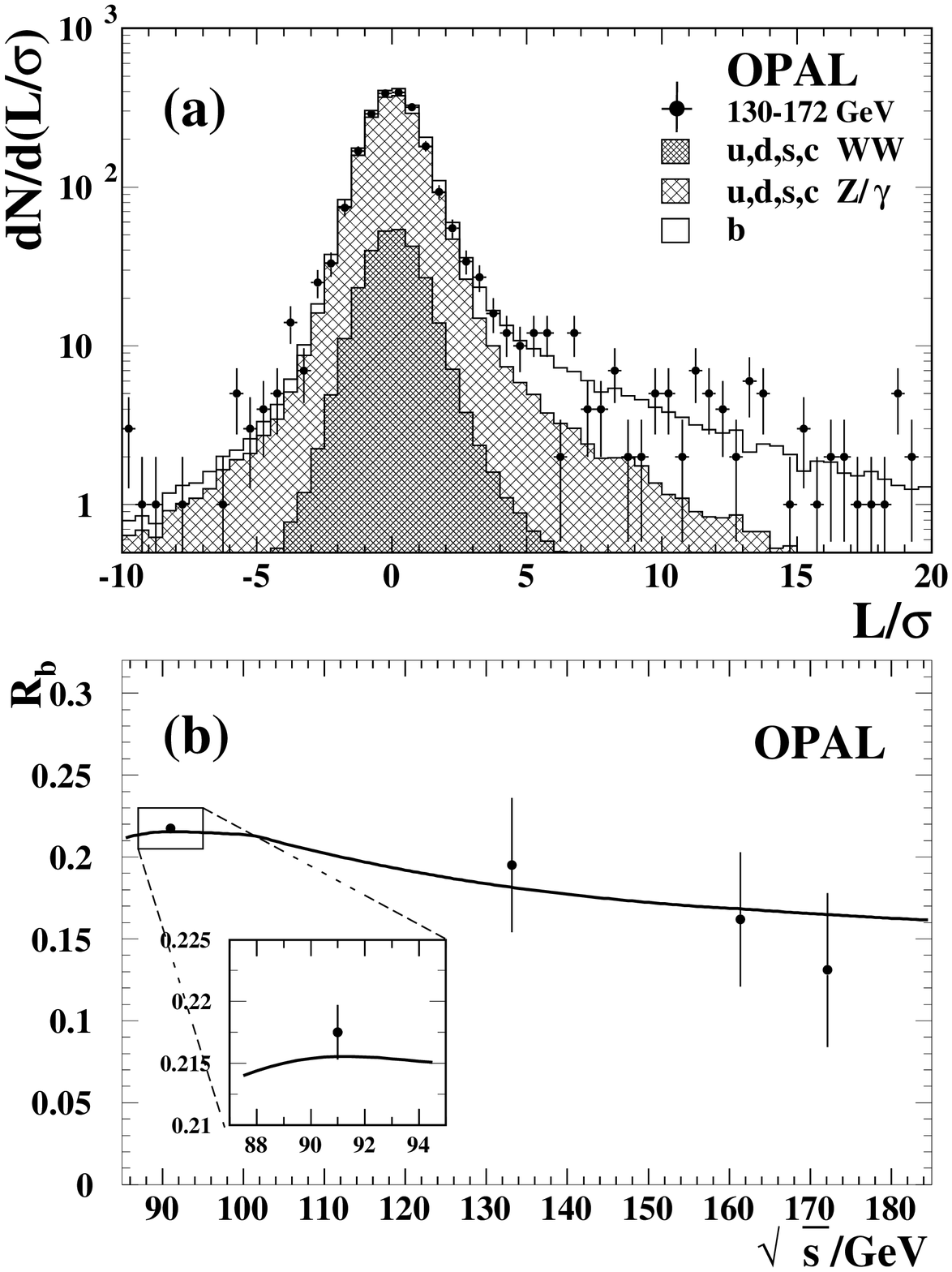}
\end{center}
\caption
{(a) Decay length significance distribution for all centre-of-mass
energies combined.
The points show the data, the histogram the Monte Carlo prediction. 
(b) $R_{\rm b}$\ 
as a function of the centre-of-mass energy.
The points show the measurements presented here, and the 
value~\cite{bib:rb_new} 
obtained on the \PZ\ peak. The errors are statistical and systematic, 
summed in quadrature. The solid line is the ZFITTER prediction for 
$s'/s > 0.8$.
}
\label{fig:rb}
\end{figure}

%Secondary vertex finding was carried out separately for each reconstructed jet
%in an event. In a first iteration, all the tracks in a given jet were 
%fitted to a common vertex point in the $x$-$y$ plane.
%If one or more tracks contributed $\Delta\chi^2\!>\!4$ to the overall 
%$\chi^2$ for the secondary vertex fit, then the track with the largest 
%$\Delta\chi^2$ was removed and the fit repeated. This process was continued 
%until all tracks contributed $\Delta\chi^2\!<\!4$ or until fewer than 
%three tracks remained, in which case the secondary vertex reconstruction 
%fails for this particular jet.
For each reconstructed secondary vertex, the decay length
$L$ was defined as the distance
of the secondary vertex from the primary vertex
in the plane transverse to the beam direction,
constrained by the direction of the total momentum vector of the
tracks assigned to the secondary vertex.
The decay length was taken to be positive if the secondary vertex was 
displaced from the primary vertex in the same hemisphere as the momentum 
sum of the charged particles at the vertex, and negative otherwise.
The distribution of decay length significance,
defined as $L$ divided by its error $\sigma_L$,
combining data from all centre-of-mass energies,
is shown in figure~\ref{fig:rb}(a), superimposed on the Monte Carlo
simulation. 
  
A `folded tag'~\cite{bib:rb_new} was used in this analysis in order to 
reduce the light flavour component and the sensitivity to detector resolution
uncertainties. 
%In this method, which is fully described in 
%reference~\cite{bib:rb_new}.
Each hadronic event is divided into two 
hemispheres by the plane perpendicular to the thrust axis, and the 
hemispheres are examined separately.
Each hemisphere is assigned a `forward tag' if it contains a secondary
vertex with a decay length significance $\Lsig > 3$, or a `backward tag'
if it contains a vertex with a decay length significance $\Lsig < -3$.
Neglecting background in the hadronic sample, the difference between
the number of forward and backward tagged 
hemispheres $N_t - \overline{N}_t$ in a sample of $N_{\mathrm{had}}$ 
hadronic events can be expressed as: 
\begin{equation} \nonumber
{N_t-\overline{N}_t = 
      2 N_{\mathrm{had}}[\epsb\Rb + \epsc\Rc +\epsuds(1-\Rb - \Rc)] }
\label{eq:rb}
\end{equation}
where $\epsb$, $\epsc$ and $\epsuds$ are the differences between the
forward and backward tagging efficiencies. The difference for \bbbar\
events $\epsb$ is about a factor of five bigger than that for \ccbar\
events $\epsc$, and a factor of fifty bigger than that for light quark
events $\epsuds$. $\Rc$ is the ratio of the cross-section for $\ccbar$ 
production to the hadronic cross-section and was computed using ZFITTER.
Due to the limited statistics compared with the LEP1 data, a double tag
technique cannot be applied in this analysis and one has to rely on a
single tag method. Therefore
the hemisphere tagging efficiency differences were determined from Monte 
Carlo, and are shown in table~\ref{tab:Rb}; the efficiencies vary only
slightly with energy. The errors on these are predominantly systematic. 
The largest contributions to the systematic errors come from the
uncertainties in Monte Carlo modelling of b and c fragmentation
and decay, and from track parameter resolution. The b and c fragmentation
and decay parameters were estimated by following the prescriptions of
reference~\cite{bib:rblep1}. The effect of track parameter resolution
was evaluated by varying the resolution in the transverse plane by
20\%, in analogy to the procedure described in reference~\cite{bib:rb_new}.
At the three centre-of-mass energy points the expected contribution from
four-fermion background was subtracted, as described above for hadronic
events. Within this background, only W-pair events are expected to
contribute to the tagged sample. 
The probability for a W-pair event to be tagged was estimated from 
Monte Carlo to be (7.8$\pm$0.4)\% at 161~GeV and (8.3$\pm$0.4)\% at
172~GeV. The errors reflect the uncertainty of charm fragmentation in
the W hadronic decay. After four-fermion background subtraction, b
purities of the tagged sample of the order of 70\% are obtained.

The numbers of selected events, tagged hemispheres and resulting
values of \Rb\ are shown\footnote{The results
presented here supersede those in reference~\cite{bib:OPAL-SM161}. In 
particular the statistical errors on \Rb\ in reference~\cite{bib:OPAL-SM161} 
were calculated incorrectly.} in table~\ref{tab:Rb}.
The systematic error on \Rb\ is dominated by the uncertainty on the
tagging efficiencies.
The other important systematic contributions result from Monte Carlo 
statistics and 
detector resolution. To check the understanding of the systematic errors, the 
analysis was repeated on data collected at the \PZ\ peak during 1996.
The resulting measurement of \Rb\ is in excellent agreement with the OPAL 
published value~\cite{bib:rb_new}, differing by (0.7$\pm$1.7)\%, where the 
error is purely statistical. 

\begin{table}[hbt]
\begin{center}
\begin{tabular}{|l|c|c|c|}
\hline%-----------------------------------------------------------------
\hline%-----------------------------------------------------------------
\multicolumn{4}{|c|}{\bf \boldmath \bbbar} \\
\hline%-----------------------------------------------------------------
           &133.17~GeV &161.34~GeV &172.12~GeV \\
\hline%-----------------------------------------------------------------
Events           &255   &328   &296   \\
Forward tags          &61    &76    &66    \\
Backward tags         &10    &20    &25    \\
$\epsb$             &0.414$\pm$0.023 &0.402$\pm$0.021 &0.395$\pm$0.022 \\
$\epsc$             &0.075$\pm$0.006 &0.079$\pm$0.006 &0.080$\pm$0.006 \\
$\epsuds$           &0.0059$\pm$0.0015 &0.0074$\pm$0.0019 &0.0085$\pm$0.0021 \\
$R_{\mathrm{b}}^{\mathrm{meas}}$ &0.199$\pm$0.040$\pm$0.013 
     &0.168$\pm$0.040$\pm$0.011  &0.136$\pm$0.048$\pm$0.010 \\ 
$R_{\mathrm{b}}^{\mathrm{corr}}$ &0.195$\pm$0.039$\pm$0.013 
     &0.162$\pm$0.039$\pm$0.011  &0.131$\pm$0.046$\pm$0.010 \\
$R_{\mathrm{b}}^{\mathrm{SM}}$   &0.182  &0.169  &0.165  \\
\hline%-----------------------------------------------------------------
$\sigma_{\mathrm{b\overline{b}}}$ (pb) &12.5$\pm$2.6$\pm$0.9 
&5.8$\pm$1.4$\pm$0.4  &3.5$\pm$1.4$\pm$0.3 \\
$\sigma_{\mathrm{b\overline{b}}}^{\mathrm{SM}}$ (pb) &12.7 &5.7 &4.6  \\
\hline%-----------------------------------------------------------------
\hline%-----------------------------------------------------------------
\end{tabular}
\caption {
Numbers of selected events, forward and backward tags, tagging efficiency
differences and measured values of \Rb. The values labelled
$R_{\mathrm{b}}^{\mathrm{meas}}$ have not been corrected for interference 
between initial- and final-state radiation, those labelled 
$R_{\mathrm{b}}^{\mathrm{corr}}$ include this correction. The value
of the $\bbbar$ cross-section, after correction for interference, is
also given. The errors on the tagging efficiency differences include 
Monte Carlo statistics and systematic effects, the latter being dominant. 
The first error on \Rb\ or $\protect\sigma_{\mathrm{b\overline{b}}}$ is 
statistical, the second systematic.  The Standard Model predictions, 
$R_{\mathrm{b}}^{\mathrm{SM}}$, 
$\protect\sigma_{\mathrm{b\overline{b}}}^{\mathrm{SM}}$,
are from the ZFITTER~\cite{bib:zfitter} program.
}
\label{tab:Rb}
\end{center}
\end{table}

The measured values of \Rb\ at each energy are compared to the Standard Model 
prediction in figure \ref{fig:rb}(b). Values for the $\bbbar$ cross-section,
derived from the measurements of the hadronic cross-section and \Rb, are
given in table~\ref{tab:Rb}.

%-----------------------------------------------------------------------
\section{Comparison with Standard Model Predictions}    \label{sec:sm}
\begin{figure}
  \epsfxsize=\textwidth
  %\epsfbox[0 0 567 680]{xs.eps}
  \epsfbox[0 0 567 680]{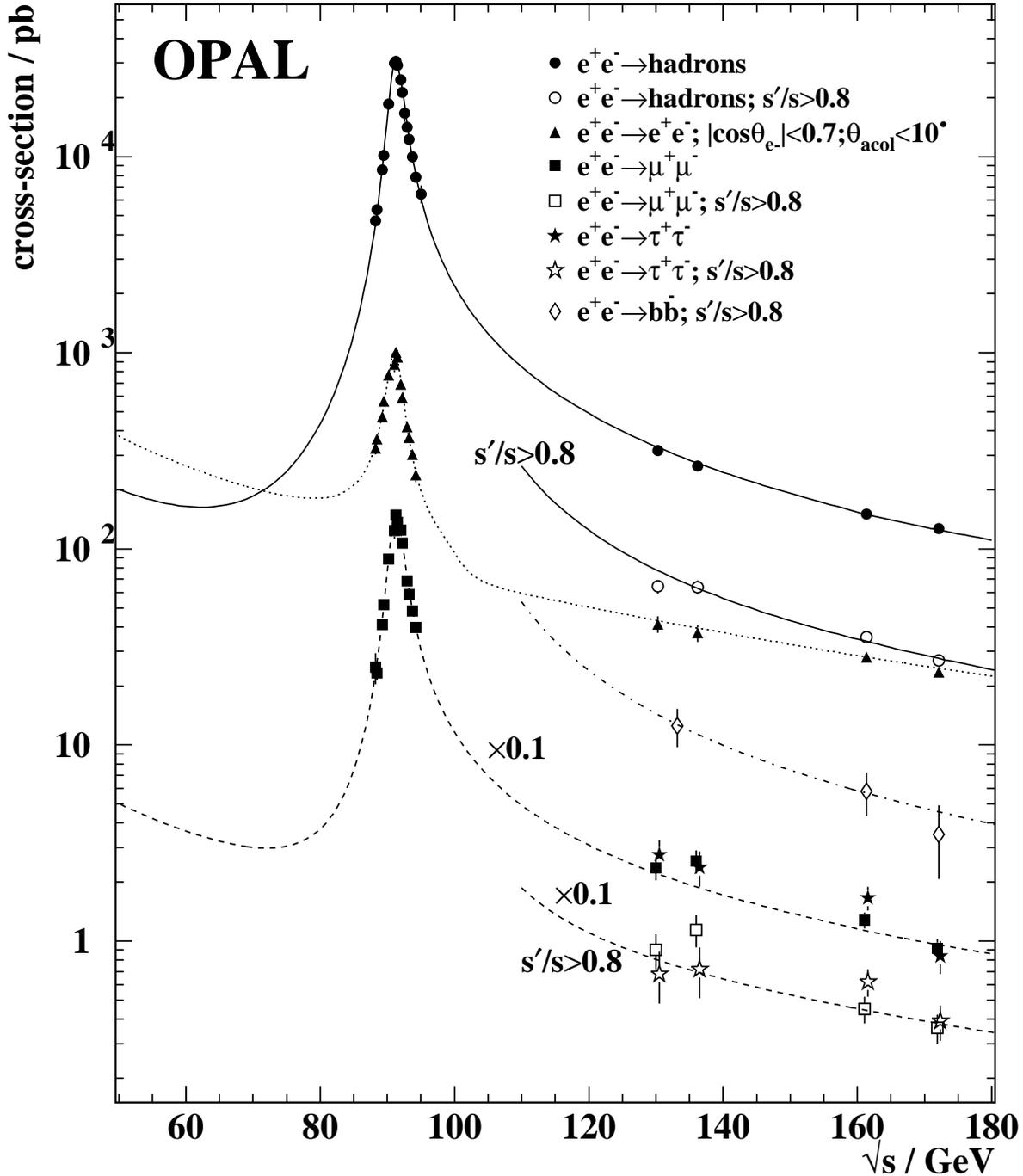}
  \caption
{
    Measured total cross-sections ($s'/s>0.01$) for different final
    states at lower energies~\cite{bib:OPAL-LS90,bib:OPAL-LS91,bib:OPAL-LS92},
    and this analysis. Cross-section measurements for hadrons, \bbbar, 
    muon and tau pairs for $s'/s>0.8$ from this analysis are also shown.
    The cross-sections for \Pgmp\Pgmm\ and \Pgtp\Pgtm\ 
    production have been reduced by a factor of ten for clarity. The 
    curves show the predictions of ZFITTER for hadronic (solid), 
    \bbbar\ (dot-dashed),
    \Pgmp\Pgmm\ and \Pgtp\Pgtm\ (dashed) final states and that of ALIBABA 
    for the \Pep\Pem\ final state (dotted). 
}
\label{fig:xsec}
\end{figure}

\begin{figure}
\epsfxsize=\textwidth
%\epsfbox[0 0 567 567]{afb.eps}
\epsfbox[0 0 567 567]{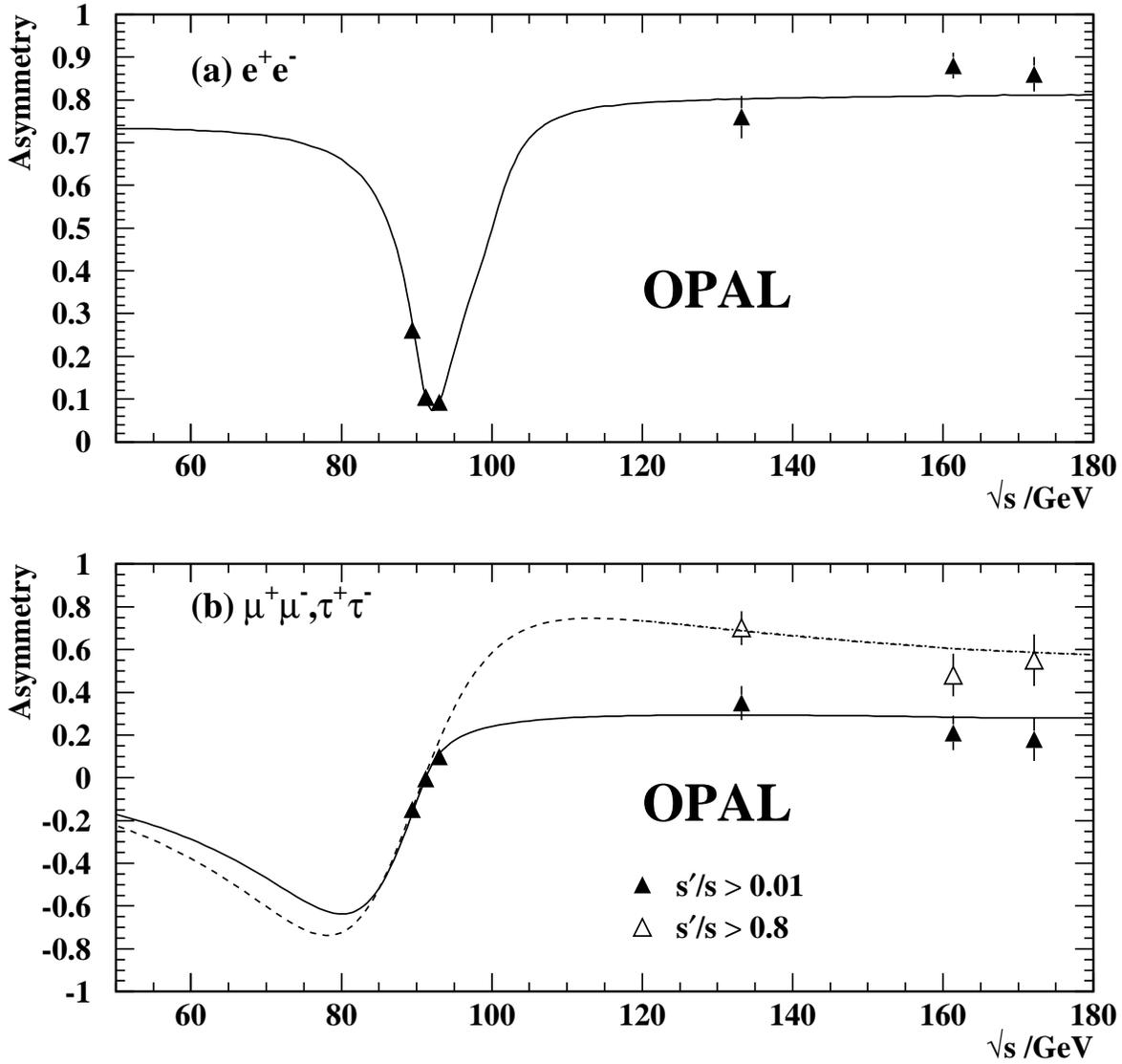}
\caption
{ (a)
 Measured forward-backward asymmetry for electron pairs with 
 $|\cos\theta_{\Pem}|<0.7$ and $\thacol<10\degree$, as a function
 of $\protect\roots$. The curve shows the prediction of ALIBABA.
 (b) Measured asymmetries for all ($s'/s>0.01$) and non-radiative
     ($s'/s>0.8$) samples as functions of $\protect\roots$
     for \Pgmp\Pgmm\ and \Pgtp\Pgtm\ events (combined).
     The curves show ZFITTER predictions for $s'/s>0.01$ (solid) and
     $s'/s>0.8$ (dotted), as well as the Born-level expectation
     without QED radiative effects (dashed). 
     The expectation for $s'/s>0.8$ lies extremely close to the Born curve,
     such that it appears indistinguishable on this plot.
}
\label{fig:afb}
\end{figure}

\begin{figure}
\epsfxsize=\textwidth
%\epsfbox[0 0 567 567]{mh_angdis.eps}
\epsfbox[0 0 567 567]{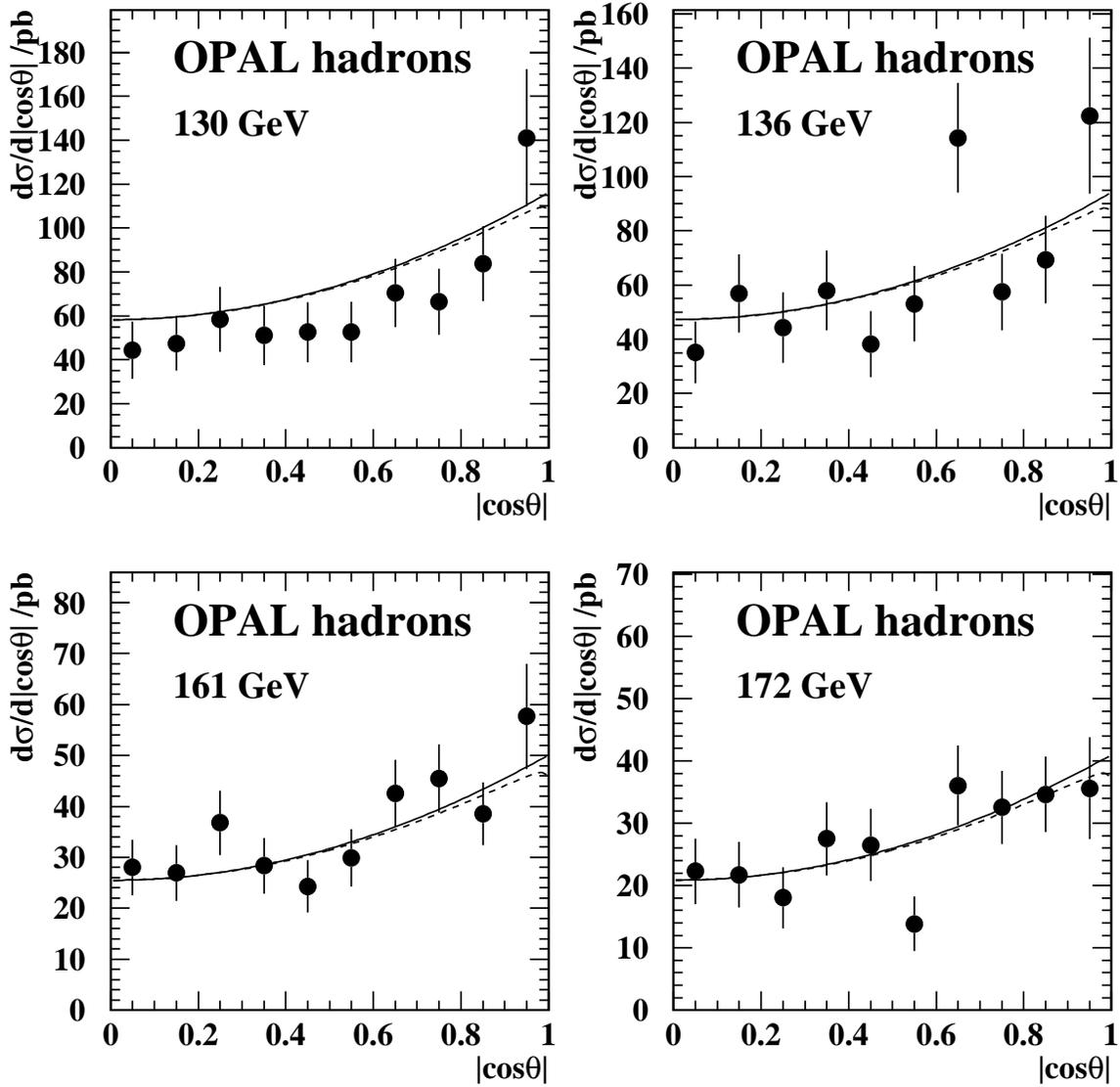}
\caption
{ Angular distributions for hadronic events with $s'/s > 0.8$. The
  data points show the measurements corrected to the direction of the
  primary quark/antiquark, and corrected to no interference between
  initial- and final-state radiation. The curves show the predictions
  of the ZFITTER program with no interference between initial-
  and final-state radiation (solid) and with interference (dashed).
}
\label{fig:mh_angdis}
\end{figure}

\begin{figure}
\epsfxsize=\textwidth
%\epsfbox[0 0 567 567]{ee_angdis.eps}
\epsfbox[0 0 567 567]{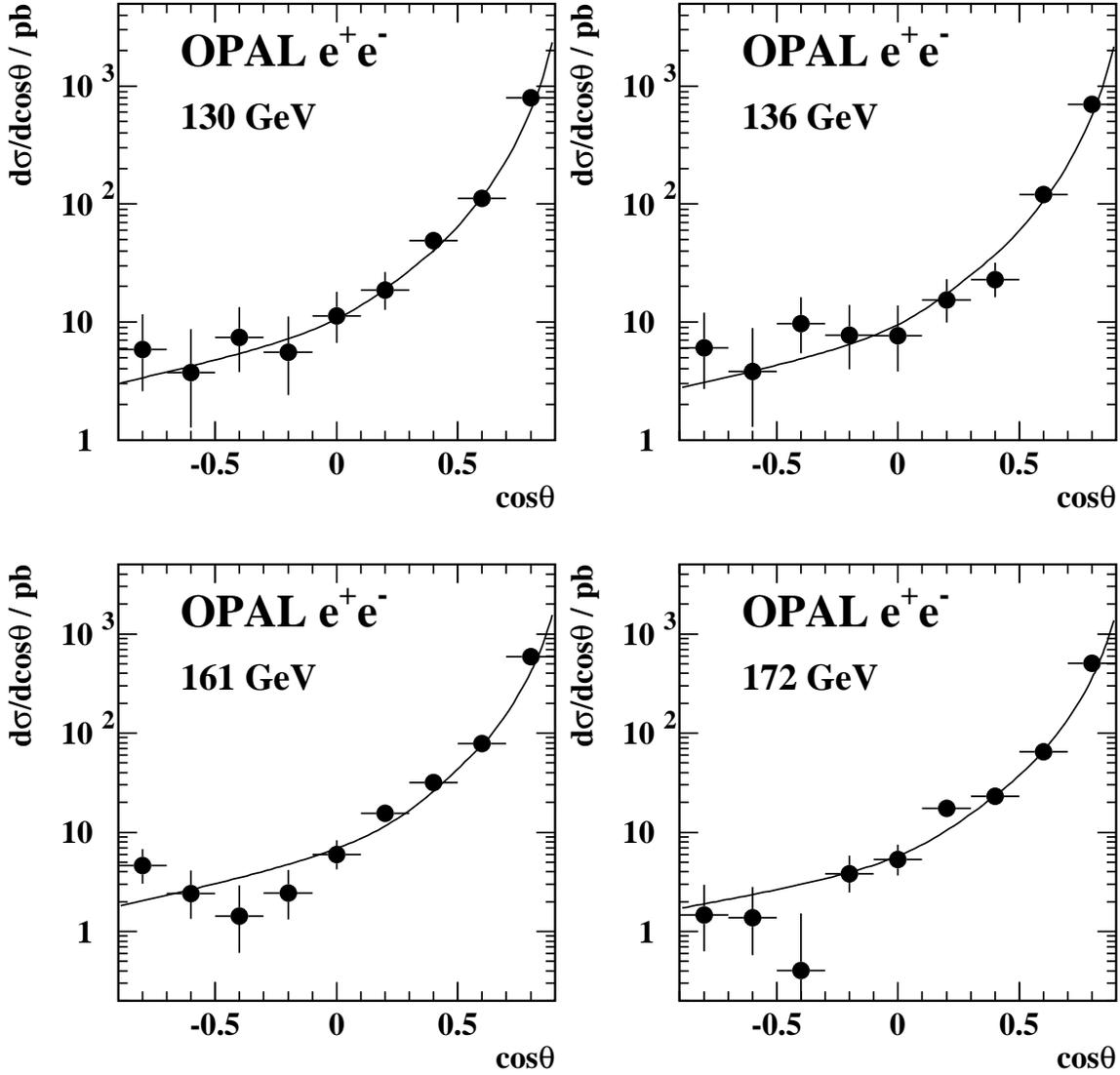}
\caption
{ Angular distributions for $\epem$ events with $\thacol < 10\degree$. 
  The points show the data, the curves the predictions of the ALIBABA 
  program.
}
\label{fig:ee_angdis}
\end{figure}

\begin{figure}
\epsfxsize=\textwidth
%\epsfbox[0 0 567 567]{mutau_cor_angdis_new.eps}
\epsfbox[0 0 567 567]{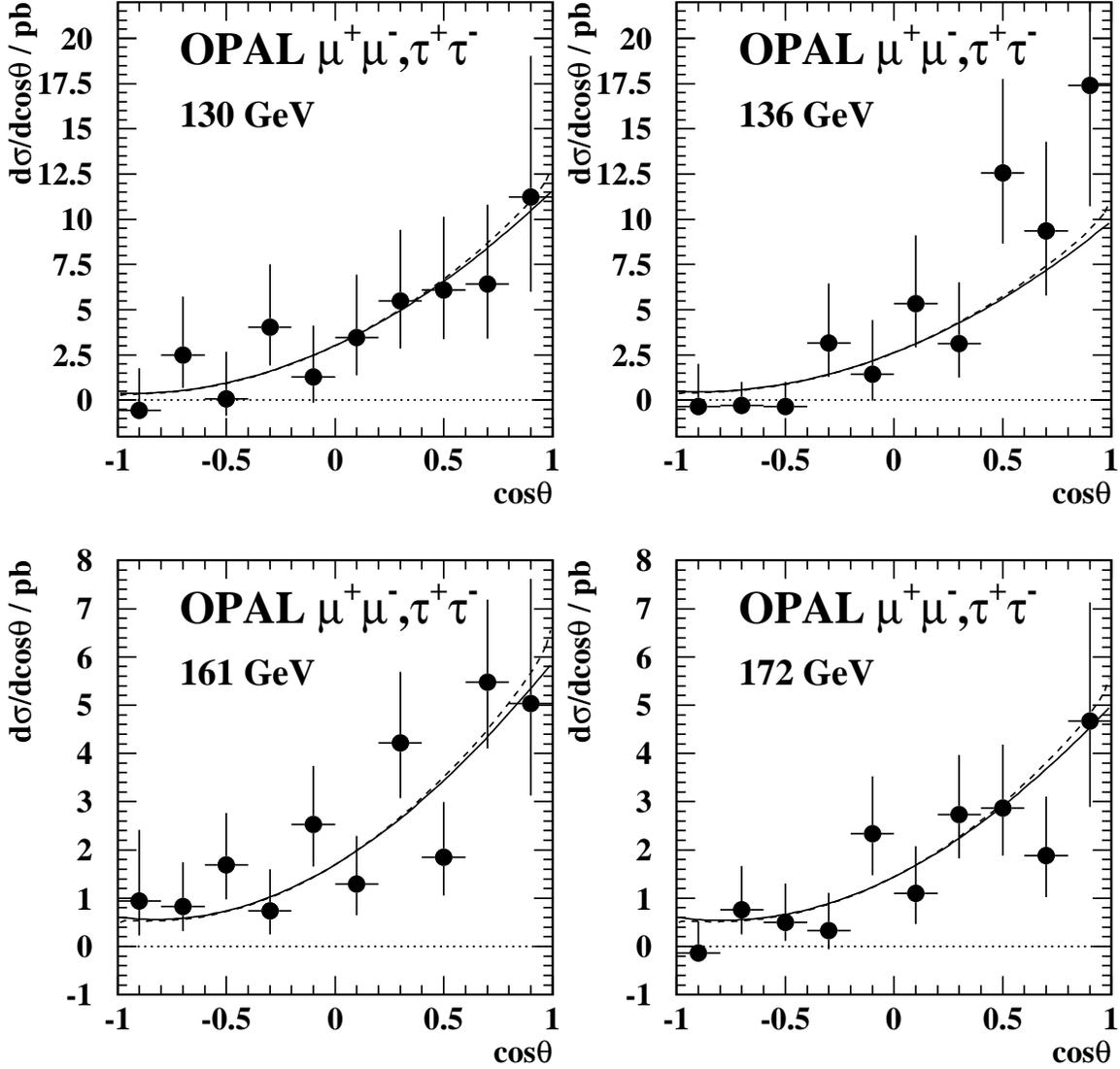}
\caption
{ Angular distributions for $\mumu$ and $\tautau$ events with 
  $s'/s > 0.8$. The data points show the measurements, which have 
  been formed as a weighted average of the $\mumu$ and $\tautau$ 
  distributions corrected to no interference between initial- and 
  final-state radiation. The curves show the predictions of
  the ZFITTER program with no interference between initial-
  and final-state radiation (solid) and with interference (dashed).
}
\label{fig:mutau_cor_angdis}
\end{figure}

%-----------------------------------------------------------------------
We compare our measurements with Standard Model predictions taken from the 
ALIBABA program for electron pairs, and the ZFITTER
program for all other final states, with input parameters $\mPZ$=91.1863~GeV,
$\mtop$=175~GeV, $\mHiggs$=300~GeV, $\alphaem(\mPZ)$=1/128.896 and
$\alphas(\mPZ)$=0.118. We use ZFITTER version 5.0, with a small modification 
to the code to ensure that the $s'$ cut is applied to fermion pair emission 
in the same way as it is to photon emission\footnote{We use default flag
settings, except {\tt BOXD}=1, {\tt CONV}=1, {\tt INTF}=0 and {\tt FINR}=0.
The effect of the {\tt BOXD} flag is to include the contribution of
box diagrams, which are significant at LEP2 energies; the setting of the
other flags was discussed in sections~\ref{sec:ifsr} and~\ref{sec:4f}.}. 
Measured values of cross-sections, presented in
tables~\ref{tab:mh},~\ref{tab:ee},~\ref{tab:mumu},~\ref{tab:tautau} 
and~\ref{tab:Rb}, are shown in
figure~\ref{fig:xsec}. The measurements are consistent with the Standard 
Model expectations. The asymmetry measurements, presented in
tables~\ref{tab:ee_afb} and~\ref{tab:mutau_afb}, are shown in 
figure~\ref{fig:afb},
while the corrected angular distributions for hadrons are shown in 
figure~\ref{fig:mh_angdis} and for electron pairs in figure~\ref{fig:ee_angdis}.
We have combined the differential cross-sections for muon and tau pairs,
and show the average in figure~\ref{fig:mutau_cor_angdis}. The measured
angular distributions and asymmetry values are in satisfactory
agreement with the Standard Model expectations.

\begin{figure}
\epsfxsize=\textwidth
%\epsfbox[0 0 567 567]{rplotpro.eps}
\epsfbox[0 0 567 567]{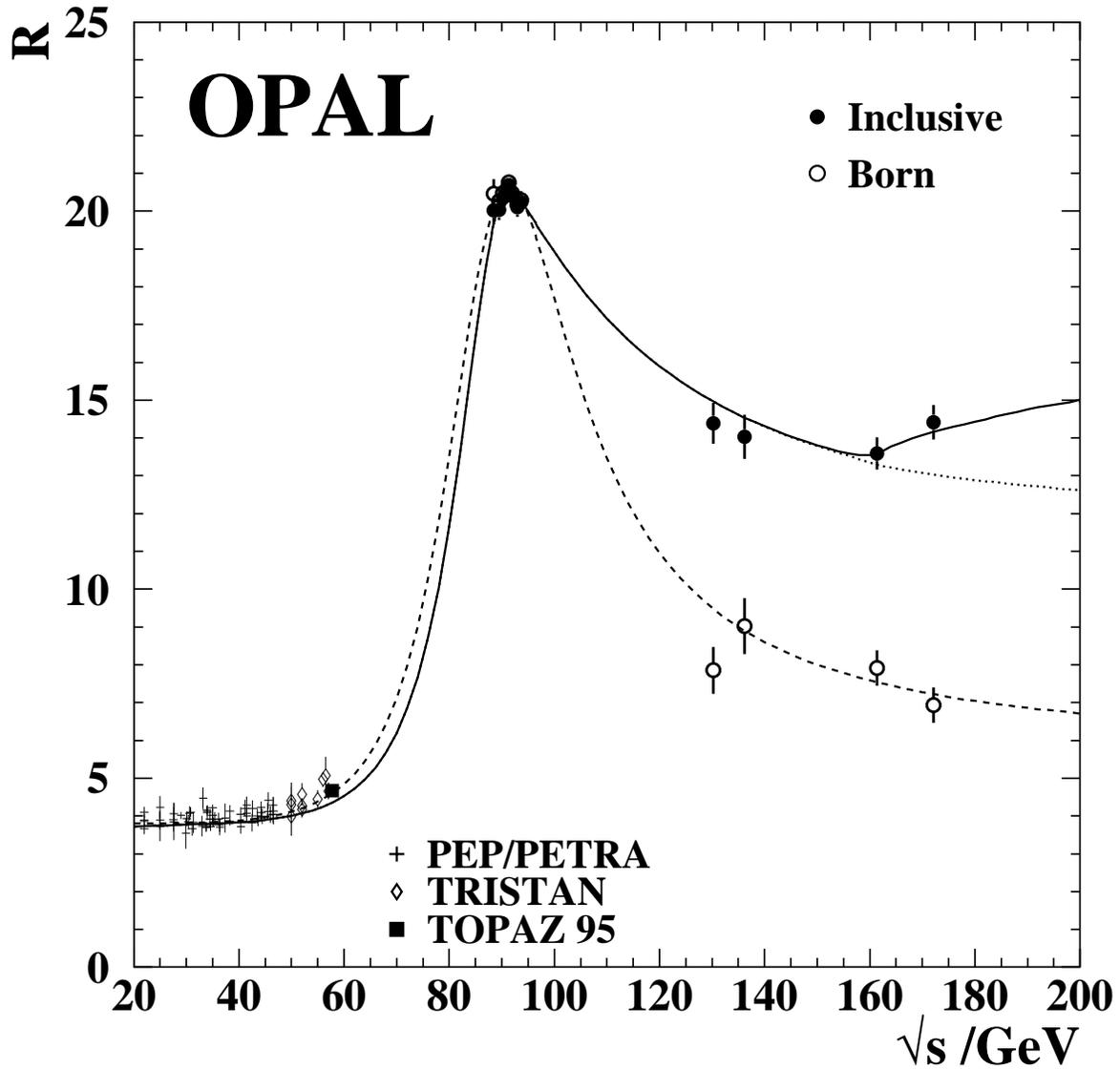}
\caption
{Ratio of measured hadronic cross-sections to theoretical muon
 pair cross-sections as a function of centre-of-mass energy. 
 Values are shown for the inclusive cross-section, $\sigma(\qqbar X)$
 and for the Born level cross-section, as described in the text. The 
 dotted and dashed curves show the predictions of ZFITTER for these 
 cross-sections, 
 while the solid curve also includes the contributions from W-pairs 
 calculated using GENTLE and from Z-pairs calculated
 using FERMISV. Measurements at lower energies are from 
 references~\cite{bib:OPAL-LS90,bib:OPAL-LS91,bib:OPAL-LS92,bib:rdata}.
}
\label{fig:rplot}
\end{figure}

In figure~\ref{fig:rplot} we show $R_{\mathrm{inc}}$, defined as the ratio of 
measured hadronic cross-section to the theoretical muon pair cross-section, as 
a function of centre-of-mass energy. The muon pair cross-sections are
calculated using ZFITTER, as described above. The hadronic cross-sections
used here are somewhat different from the measurements presented in 
section~\ref{sec:mh}. We use an inclusive cross-section, 
$\sigma(\qqbar\mathrm{X})$, which is measured in a
similar manner to the inclusive hadronic cross-section described
above, but without subtraction of the W-pair contribution. The observed
cross-section is corrected using an efficiency which includes the effect
of W-pair events to give a total cross-section which is the sum of
the two-fermion cross-section plus the cross-section for W-pair production
with at least one of the W bosons decaying hadronically. This cross-section
is thus an inclusive measurement of hadron production in $\epem$
annihilation, in which production thresholds (e.g. for WW, ZZ or new
particles) can be seen. The measured values of this cross-section 
and the ratio $R_{\mathrm{inc}}$ are
given in table~\ref{tab:rplot}. In figure~\ref{fig:rplot}
the measured values of $R_{\mathrm{inc}}$
are compared with the prediction of ZFITTER, which does not include
W-pair production, and also with a theoretical prediction
$R_{\mathrm{inc}}^{\rm SM}$ including the
expected contributions from WW and ZZ events, calculated using 
GENTLE~\cite{bib:gentle} and FERMISV~\cite{bib:fermisv} respectively.
The effect of W-pair production is clear.

In figure~\ref{fig:rplot} and table~\ref{tab:rplot} we also show 
$R_{\mathrm{Born}}$, the ratio of
the measured hadronic cross-section for $s'/s > 0.8$, corrected to the Born 
level, to the theoretical muon pair cross-section at the Born level.
The correction of the measured cross-section is performed using ZFITTER,
and for both the numerator and denominator `Born level' means the
improved Born approximation of ZFITTER. Far below the Z resonance,
this ratio becomes the usual $R_{\gamma}$ that has been measured by
many experiments at lower energy. Some of these low energy 
measurements~\cite{bib:rdata} are also shown in figure~\ref{fig:rplot}.
The measurements close to the Z peak have been corrected to our
definition of $R_{\mathrm{Born}}$ for this figure.

\begin{table}[hbt]
\begin{center}
\begin{tabular}{|l|c|c|c|c|}
\hline%-----------------------------------------------------------------
\hline%-----------------------------------------------------------------
\multicolumn{5}{|c|}{\bf \boldmath $\qqbar\mathrm{X}$} \\
\hline%-----------------------------------------------------------------
           &130.25~GeV &136.22~GeV &161.34~GeV &172.12~GeV \\
\hline%-----------------------------------------------------------------
$\sigma_{\mathrm{corr}}$ (pb) &317$\pm$11$\pm$5  &264$\pm$10$\pm$4
                              &153$\pm$4$\pm$2   &138$\pm$4$\pm$2  \\
$R_{\mathrm{inc}}$            &14.4$\pm$0.5      &14.0$\pm$0.6
                              &13.6$\pm$0.4      &14.4$\pm$0.5 \\
$R_{\mathrm{inc}}^{\rm SM}$   &15.0   & 14.5 &  13.6 & 14.2 \\
\hline%-----------------------------------------------------------------
 $R_{\mathrm{Born}}$          & 7.9$\pm$0.6      & 9.0$\pm$0.7
                              & 7.9$\pm$0.5      & 6.9$\pm$0.5 \\
 $R_{\mathrm{Born}}^{\rm SM}$ & 9.5   &  8.9 &   7.5 &  7.2   \\
\hline%-----------------------------------------------------------------
\hline%-----------------------------------------------------------------
\end{tabular}
\caption {Measured cross-sections for $\qqbar$X, after correcting
 for interference between initial- and final-state radiation. The
 first error shown is statistical, the second systematic.
 Also listed are the ratios $R_{\mathrm{inc}}$ and $R_{\mathrm{Born}}$,
 as defined in the text, where statistical and systematic errors have
 been combined, and their respective Standard Model predictions. 
}
\label{tab:rplot}
\end{center}
\end{table}

%-----------------------------------------------------------------------
\subsection{Influence on Electroweak Precision Measurements} \label{sec:blob}
%-----------------------------------------------------------------------

In references~\cite{bib:OPAL-SM130,bib:OPAL-SM161} non-radiative 
data above the \PZ\ resonance were used to constrain the size of the 
interference terms between photon exchange and \PZ\ exchange processes, which 
have amplitudes of similar magnitude. Using the ZFITTER~\cite{bib:zfitter} and 
SMATASY~\cite{bib:smatasy} programs, we have repeated the model-independent 
fits to OPAL data described in 
references~\cite{bib:OPAL-SM130,bib:OPAL-SM161}, 
including all the measurements of the non-radiative hadronic cross-section 
and combined \Pgmp\Pgmm\ and \Pgtp\Pgtm\ asymmetry presented here. In
addition, we have included the measurements of the muon and tau pair
non-radiative cross-sections; these were not used in the fits in
references~\cite{bib:OPAL-SM130,bib:OPAL-SM161}. In the fit, 
the parameters $\jtoth$ and $\jfbl$, determining the sizes of the hadronic 
and leptonic $\Pgg\PZ$-interference respectively, have been left free
(see reference~\cite{bib:OPAL-SM130} for more discussion of these parameters
and details of the fit). In the Standard Model, $\jtoth$ and $\jfbl$
have the values $0.216\pm 0.011$ and $0.799\pm 0.001$ respectively, 
for a top quark mass of 175~GeV and Higgs boson mass of 300~GeV, where
the uncertainties in this prediction come from varying $\mHiggs$ in the 
range 70--1000~GeV, $\alphas(\mPZ)$ in the range 0.112--0.124,
and  $\mPZ$, $\mtop$, and $\alphaem(\mPZ)$ in ranges taken 
from reference~\cite{bib:LEPEWWG}.

\begin{figure}
\epsfxsize=\textwidth
%\epsfbox[0 0 567 567]{blob.eps}
\epsfbox[0 0 567 567]{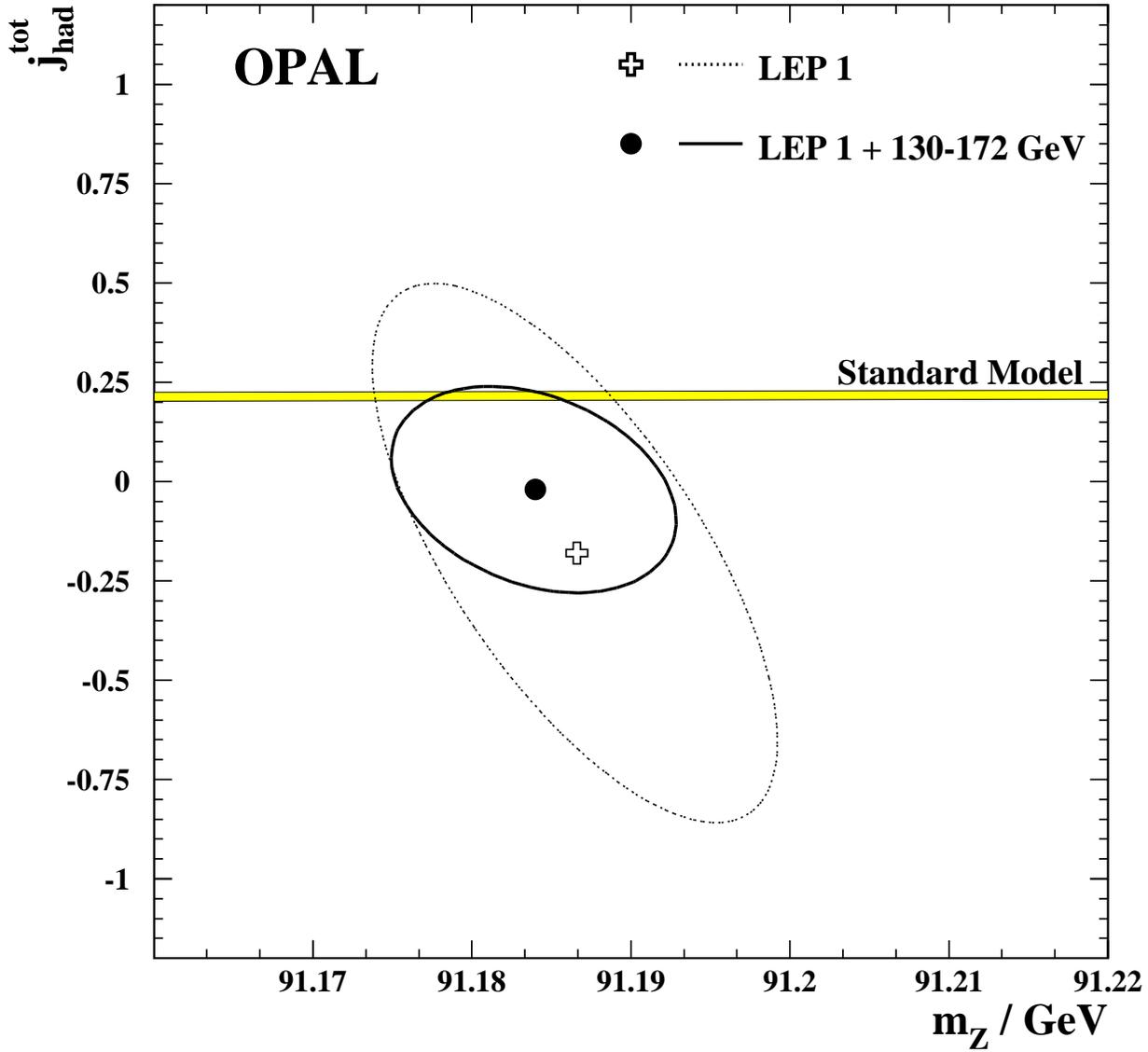}
\caption
{Central values and one standard deviation contours (39\% probability
 content) in the $\protect\jtoth$ vs. $\protect\mPZ$ plane resulting from
 model-independent fits to the OPAL data samples described in
 section~\protect{\ref{sec:blob}}. The horizontal band shows the
 Standard Model expectation $\jtoth = 0.216 \pm 0.011$ for a top quark
 mass of 175$\pm$5~GeV and a Higgs mass range of 70-1000~GeV. 
}
\label{fig:blob}
\end{figure}

The results of the fit, which has a $\chi^2$ of 81.5 for 112
degrees of freedom, are given in table~\ref{tab:implic}.
For comparison, the table also shows the
results of the fit presented in~\cite{bib:OPAL-SM130} to OPAL data
collected at \LEPone\ alone, and the Standard Model predictions for $\jtoth$
and $\jfbl$.
%and to \LEPone\ data plus non-radiative hadronic 
%cross-sections and combined \Pgmp\Pgmm\ and \Pgtp\Pgtm\ asymmetries at 
%130, 136 and 161~GeV. 
Since the $\Pgg\PZ$-interference vanishes on the $\PZ$ peak,
the inclusion of data far away from the $\PZ$ resonance
considerably reduces the uncertainty of $\jtoth$~\cite{bib:OPAL-SM130}. 
The inclusion of the 130-161~GeV data reduced the uncertainty on
$\jtoth$ by 55\%; a further improvement of about 16\% is observed by
including the 172~GeV data presented here. As shown in
reference~\cite{bib:OPAL-SM130},
this improvement is much larger than that which would be obtained by
the inclusion of the full \LEPone\ off-peak data. The high energy data
also reduce the correlation between fitted values of $\jtoth$ and the 
\PZ\ mass, as can be seen in table~\ref{tab:implic} and 
figure~\ref{fig:blob}.

\begin{table}[htb]%-------------------------------------------------------
\newcommand{\whole}{\phantom{.0}}
\centering
\small
\begin{tabular}{|l|c|c|c|c|}
\hline%-----------------------------------------------------------------
\hline%-----------------------------------------------------------------
OPAL data sample & $\jtoth$ & $\mPZ$ & $\jtoth$, $\mPZ$ % & $\mPZ$ (GeV) 
& $\jfbl$ \\
            &         & (GeV) & correlation      % & ($\jtoth=0.216$, fixed)
            & \\
\hline%-----------------------------------------------------------------
\LEPone\ (1989-92)~\cite{bib:OPAL-LS92,bib:OPAL-LS91,bib:OPAL-LS90} 
&  $-0.18 \pm 0.68$  & $91.187 \pm 0.013$ & $-0.70$ 
&  $0.684 \pm 0.053$ \\
\LEPone\ + these data
&  $-0.02 \pm 0.26$  & $91.184 \pm 0.009$ & $-0.31$ 
&  $0.715 \pm 0.043$ \\
\hline%-----------------------------------------------------------------
Standard Model  & 0.22$\pm$0.01 & --- & --- & 0.799$\pm$0.001 \\
\hline%-----------------------------------------------------------------
\hline%-----------------------------------------------------------------
\end{tabular}
\caption[]{
  Fitted values of the hadronic $\Pgg\PZ$-interference parameter, 
  $\jtoth$, the $\PZ$ mass, $\mPZ$, and the leptonic $\Pgg\PZ$-interference 
  parameter, $\jfbl$, using different OPAL data samples.
  The $\mPZ$ values are quoted for the $s$-dependent \PZ-width.
}
\label{tab:implic}
\end{table}%-------------------------
 
%-----------------------------------------------------------------------
\subsection{\boldmath Energy Dependence of $\alphaem$} \label{sec:alphaem}
%-----------------------------------------------------------------------
The cross-section and asymmetry measurements for $s'/s > 0.8$ have been
used to investigate the energy dependence of the electromagnetic 
coupling $\alphaem$ within the framework of the
Standard Model. Here the main information does not come from
the $\gamma$-Z interference, studied in the fit described above, 
but from the pure $s$-channel photon exchange in the leptonic cross-sections.

For final state muon or tau pairs with $s'\!\sim\!s$,
the photon exchange process, which is proportional to $\alphaemsq(Q^2)$,
dominates the \PZ\ exchange by factors of 2--4 at centre-of-mass
energies of 130--172~GeV. The leptonic forward-backward asymmetries,
reflecting the $\gamma$-Z interference, 
are somewhat less sensitive since they depend only linearly on $\alphaem(Q^2)$.
For hadronic final states at  $s'\sim s$ the ratio between \PZ\ and
photon exchange is about a factor of five larger than that for leptons, so that
they are dominated by \PZ\ exchange, even at the highest energies presented
here.
In contrast to the charged leptons, quarks have a sizeable
vector coupling to the Z, which depends on $\sin^2\theta_{\mathrm{W}}(Q^2)$, 
which in turn 
is closely related to $\alphaem(Q^2)$ via the Fermi constant $G_{\mathrm{F}}$.
 For hadrons, changes to $\alphaem$ affect
the photon amplitude and \PZ\ amplitude (via the vector coupling) in
opposite directions. At 133~GeV, the hadronic cross-section is still
sensitive to $\alphaem$ via the \PZ\ amplitude, while at larger
centre-of-mass energies its sensitivity is considerably reduced due
to the relatively increased photon amplitude.

To fit for $\alphaem(\sqrt{s})$ at each centre-of mass energy, we form the 
$\chi^2$
between our measurements and the Standard Model predictions calculated
as a function of $\alphaem(\sqrt{s})$ using ZFITTER, keeping all other
ZFITTER input parameters fixed to the values given above.
The dependence of the vector coupling on $\alphaem$ is then effectively
included in ZFITTER. 
In addition to performing a fit for each centre-of-mass
energy we also perform a combined fit to all energies, in which
$\alphaem$ runs with energy with a slope which corresponds to its
fitted value. In this case, we quote as the result the value of
$\alphaem$ at a centre-of-mass energy $\sqrt{s}$= 157.42~GeV, 
corresponding to the luminosity weighted average of $1/s$.

As inputs to the fits, we use the measured values of hadronic, muon
pair and tau pair cross-sections, \Rb, and the combined muon and tau
asymmetry values, for $s'/s > 0.8$, giving five measurements at
each energy. The measurements at 130 and 136~GeV have been combined for
this analysis. 
%The mean centre-of-mass energy, weighted according to
%the expected centre-of-mass energy dependence of the measurements,
%is 133.02~GeV.
The correlation coefficients between all the measurements
are given in table~\ref{tab:correl}. The most significant correlations 
are between measurements of the same quantity at different energies arising
from common efficiency and background systematics. The error on luminosity 
measurements gives rise to correlations between cross-section measurements 
at the same energy, and also, through the systematic error, to small
correlations between cross-sections at different energies. The errors on the 
correction for interference between initial- and final-state radiation are 
assumed to be fully correlated.

\begin{table}[htbp]
\begin{sideways}
\begin{minipage}[b]{\textheight}{\small
\begin{center}\begin{tabular}{|l|r|r|r|r|r|r|r|r|r|r|r|r|r|r|r|}
\hline%-----------------------------------------------------------------
\hline%-----------------------------------------------------------------
&\multicolumn{5}{c|}{133.17~GeV} &\multicolumn{5}{c|}{161.34~GeV}
&\multicolumn{5}{c|}{172.12~GeV} \\
\hline%-----------------------------------------------------------------
&$\sigma(\qqbar)$ &\Rb &$\sigma(\mu\mu)$ &$\sigma(\tau\tau)$ &$\AFB(\ell\ell)$
&$\sigma(\qqbar)$ &\Rb &$\sigma(\mu\mu)$ &$\sigma(\tau\tau)$ &$\AFB(\ell\ell)$
&$\sigma(\qqbar)$ &\Rb &$\sigma(\mu\mu)$ &$\sigma(\tau\tau)$ &$\AFB(\ell\ell)$ 
\\
\hline%-----------------------------------------------------------------
$\sigma(\qqbar)$ &1.000   &--0.004 &0.005   &0.003   &--0.003 &0.135   &--0.006 
                 &--0.001 &0.000   &--0.003 &0.118   &--0.005 &--0.001 &--0.000
                 &--0.003 \\
\Rb              &        &1.000   &0.001   &0.001   &0.001   &--0.004 &0.085
                 &0.001   &0.001   &0.001   &--0.003 &0.067   &0.001   &0.001
                 &0.002 \\
$\sigma(\mu\mu)$ &        &        &1.000   &0.003   &0.001   &--0.000 &0.001
                 &0.012   &0.001   &0.001   &--0.000 &0.001   &0.014   &0.001
                 &0.002 \\ 
$\sigma(\tau\tau)$&       &        &        &1.000   &0.001   &--0.000 &0.001
                 &0.001   &0.022   &0.001   &--0.000 &0.001   &0.001   &0.020
                 &0.001 \\
$\AFB(\ell\ell)$ &        &        &        &        &1.000   &--0.002 &0.002
                 &0.001   &0.001   &0.004   &--0.002 &0.001   &0.001   &0.001
                 &0.004 \\
\hline%-----------------------------------------------------------------
$\sigma(\qqbar)$ &        &        &        &        &        &1.000   &--0.006
                 &0.002   &0.002   &--0.003 &0.122   &--0.005 &--0.001 &0.000
                 &--0.003 \\
\Rb              &        &        &        &        &        &        &1.000
                 &0.001   &0.001   &0.002   &--0.005 &0.057   &0.001   &0.001 
                 &0.002 \\
$\sigma(\mu\mu)$ &        &        &        &        &        &        &     
                 &1.000   &0.003   &0.001   &--0.000 &0.001   &0.014   &0.001
                 &0.002 \\
$\sigma(\tau\tau)$&       &        &        &        &        &        &       
                 &        &1.000   &0.001   &0.000   &0.001   &0.001   &0.025 
                 &0.001 \\
$\AFB(\ell\ell)$ &        &        &        &        &        &        &       
                 &        &        &1.000   &--0.002 &0.001   &0.001   &0.001
                 &0.004 \\
\hline%-----------------------------------------------------------------
$\sigma(\qqbar)$ &      &       &      &        &       &       &       &
                 &      &       &1.000 &--0.004 &0.001  &0.001  &--0.003 \\
\Rb              &      &       &      &        &       &       &       &
                 &      &       &      &1.000   &0.001  &0.001  &0.002 \\
$\sigma(\mu\mu)$ &      &       &      &        &       &       &       &
                 &      &       &      &        &1.000  &0.002  &0.002 \\
$\sigma(\tau\tau)$&     &       &      &        &       &       &       &
                 &      &       &      &        &       &1.000  &0.001 \\
$\AFB(\ell\ell)$ &      &       &      &        &       &       &       &
                 &      &       &      &        &       &       &1.000 \\
\hline%-----------------------------------------------------------------
\hline%-----------------------------------------------------------------
\end{tabular}\end{center}}
\caption[]{Correlation coefficients between the various measured
quantities ($s'/s > 0.8$) at each energy.
\label{tab:correl}}
\end{minipage}
\end{sideways}
\end{table}

\begin{figure}
\epsfxsize=\textwidth
%\epsfbox[0 0 567 567]{alph_topsig.eps}
\epsfbox[0 0 567 567]{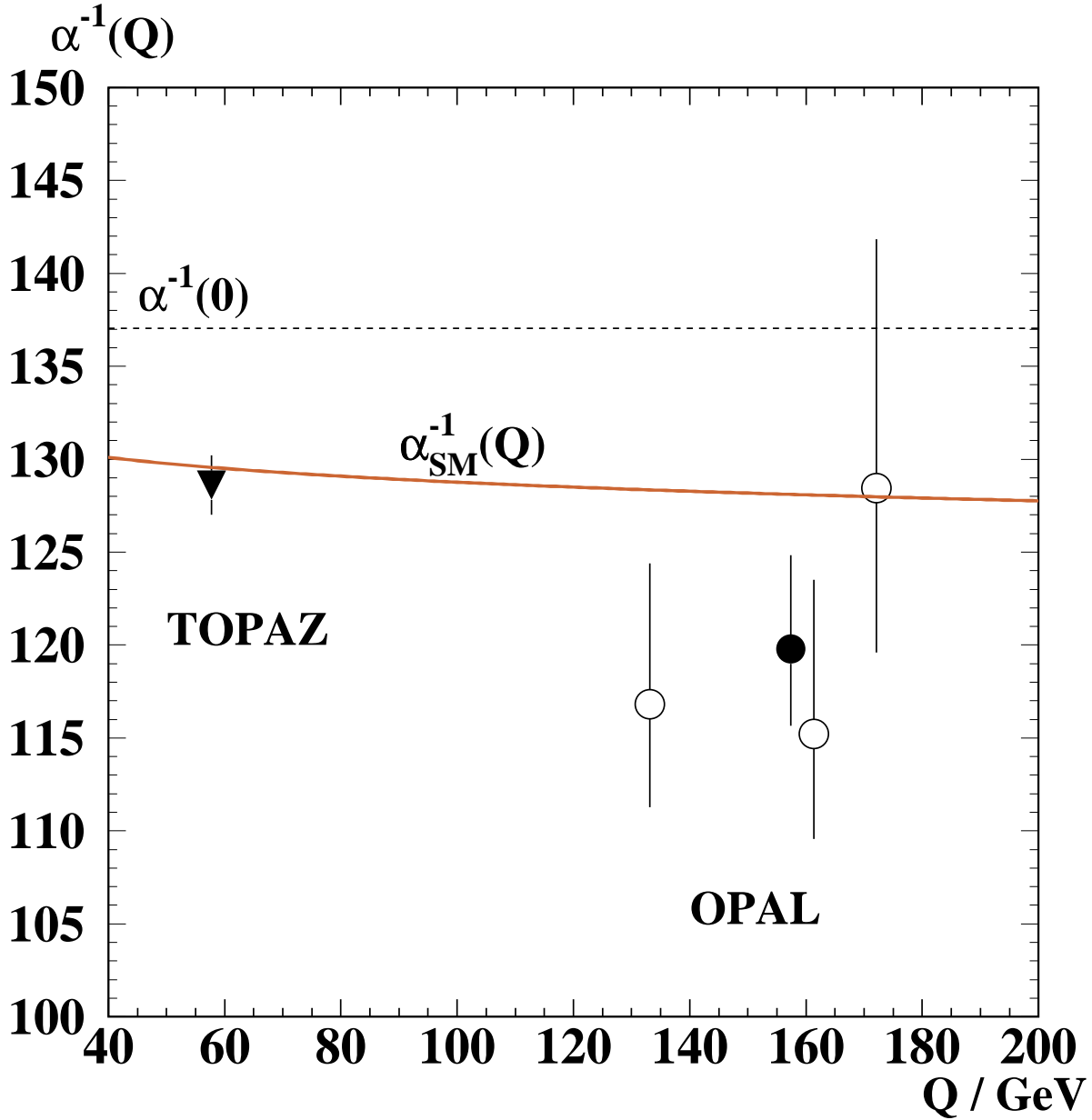}
\caption
{Fitted values of $1/\alphaem$ as a function of Q, which is
 $\protect \sqrt{s}$ for the OPAL fits. The open circles show the
 results of the fits at each centre-of-mass energy, the closed
 circle the result of the combined fit. The value obtained by
 the TOPAZ experiment~\cite{bib:alrun} is also shown for comparison.
 The solid line shows the Standard Model expectation, with the thickness
 representing the uncertainty, while the value of 1/$\alphaem(0)$ is shown 
 by the dashed line.
}
\label{fig:alphaem}
\end{figure}

The results of the fits, for the separate centre-of-mass energies and
the combined fit, are shown in table~\ref{tab:alphaem1}, with the resulting 
values of $\alphaem(\sqrt{s})$ plotted in figure~\ref{fig:alphaem}. The
values are consistent with expectation. The asymmetry in the
errors arises because the dependence of the asymmetries on $\alphaem$
is linear whereas the cross-sections and \Rb\ have a quadratic dependence.
In table~\ref{tab:alphaem2} we show, for each measurement, the normalized 
residual before and after the combined fit and an estimate of the sensitivity 
to the value of $\alphaem$. The normalized residual is the difference between 
measured and predicted value divided by the measurement error, while the 
sensitivity estimate used here is 
%the difference between the predicted value 
%before and after the fit divided by the error on the measurement. 
the number of experimental standard deviations by which the prediction
for the quantity changes if $1/\alphaem(157.42~\GeV)$ is varied by
$\pm1.0$ from its Standard Model value. 
We see that the measurements of lepton cross-sections are the most sensitive 
to the value of $\alphaem$, while the influence of the hadronic
variables is considerably smaller, as explained above.
From the combined fit, we obtain a value of 
$$1 / \alphaem(157.42~\GeV) = 119.9^{+5.1}_{-4.1},$$
where the error arises from 
the errors on the measurements. The error due to uncertainties in the input 
parameters to ZFITTER is negligible in comparison, being at most 0.04 for 
a variation of $\mHiggs$ in the range 70--1000~GeV. 

The fit described above uses measurements of cross-sections which 
depend on the measurement of luminosity, which itself assumes the
Standard Model running of $\alphaem$ from $(Q^2 = 0)$ to typically
$Q^2 = (3.5~\GeV)^2$, where $1/\alphaem\simeq 134$. Therefore it
measures running only from $Q\simeq 3.5~\GeV$ onwards. To become
independent of this assumption, we have repeated the combined fit
replacing the cross-sections for hadrons, muon and tau pairs with
the ratios $\sigma(\mu\mu)/\sigma(\qqbar)$ and 
$\sigma(\tau\tau)/\sigma(\qqbar)$.
This is possible since, above the Z peak, leptons and hadrons have a
very different sensitivity to $\alphaem$, as explained above. The
increased correlations between the input quantities have been taken
into account. 
This fit yields a result of 
$1 / \alphaem(157.42~\GeV)$ = $119.9^{+6.6}_{-5.4}\pm 0.1$, with a 
$\chi^2/$d.o.f.\ of 7.5/11, very close to the result obtained above but
with somewhat larger errors. The second error comes from varying
$\alphas(\mPZ)$ in the range 0.112 to 0.124. This measured value
of $1/\alphaem$ is 2.6 standard deviations below the low energy
limit of 137.0359979$\pm$0.0000032~\cite{bib:CAGE}, indicating
the running of $\alphaem$ from $Q^2 = 0$ independent of the luminosity
measurement. There has been only one previous measurement~\cite{bib:alrun} 
where the running of $\alphaem$ at large $Q^2$ has been measured dominantly 
from the photon propagator and not, as on the Z peak~\cite{bib:banerjee}, 
from the Standard Model relation between the weak and electromagnetic 
coupling constants.
Using a slope
corresponding to the fitted value, the combined fit result corresponds 
to a measurement of $1 / \alphaem(\mPZ)$ = $121.4^{+6.0}_{-4.9}\pm0.1$.
This measurement is completely statistics limited,  essentially
independent of  $\mHiggs$, and does not rely on assumptions about the
running of $\alphaem$ at low $Q^2$, which is the dominant uncertainty
in the current value of $1/\alphaem(\mPZ)$ of 128.90$\pm$0.09~\cite{bib:EJBP}.
If the error can be reduced by more than one
order of magnitude with future statistics, a measurement such as this
will lead to an improved knowledge of $1/\alphaem(\mPZ)$.

%%--------new table 13 ----------------------------------
\begin{table}[htbp]%----------------------------------------------------
\centering
\begin{tabular}{|c||c|c||c|c|}
\hline%-----------------------------------------------------------------
\hline%-----------------------------------------------------------------
      &\multicolumn{2}{c||}{Fit} &\multicolumn{2}{c|}{Standard Model} \\
\hline%-----------------------------------------------------------------
$\sqrt{s}$ (GeV) &$1/\alphaem$ &$\chi^2$/d.o.f. &$1/\alphaem$ 
&$\chi^2$/d.o.f. \\
\hline%-----------------------------------------------------------------
133.17   &116.9$^{+7.6}_{-5.6}$  &4.2/4  &128.3  &6.2/5 \\
161.34   &115.5$^{+8.2}_{-5.6}$  &3.4/4  &128.1  &5.3/5 \\
172.12   &128.7$^{+13.5}_{-8.9}$ &0.9/4  &128.0  &0.9/5 \\
\hline%-----------------------------------------------------------------
%%%$\mPZ$   &121.4$^{+4.6}_{-3.8}$  &10.6/14 &128.9  &12.9/15 \\
157.42   &119.9$^{+5.1}_{-4.1}$  &10.6/14 &128.1  &12.9/15 \\
\hline%-----------------------------------------------------------------
\hline%-----------------------------------------------------------------
\end{tabular}
\caption[]{Results of fits for $\alphaem$. The first three rows show
the fits to data at each centre-of-mass energy, the last row the combined
fit to all energies. The Standard Model values of $1/\alphaem$, and
the $\chi^2$ between the measurements and the Standard Model predictions
are also given for comparison.
}
\label{tab:alphaem1}
\end{table}%------------------------------------------------------------
%%--------end of new table 13--------------------------------------

\begin{table}[htbp]%----------------------------------------------------
\centering
\begin{tabular}{|c|c|c|c|c|}
\hline%-----------------------------------------------------------------
\hline%-----------------------------------------------------------------
$\sqrt{s}$ (GeV)  &Measurement &$\chi$(SM) &$\chi$(fit) &Sensitivity \\
\hline%-----------------------------------------------------------------
133.17  &$\sigma(\qqbar)$     &--1.60     &--1.29     &0.047 \\
        &\Rb                  &0.31       &0.42       &0.011 \\
        &$\sigma(\mu\mu)$     &1.85       &1.37       &0.045 \\
        &$\sigma(\tau\tau)$   &--0.35     &--0.79     &0.042 \\
        &$\AFB(\ell\ell)$     &0.13       &0.33       &0.021 \\
\hline%-----------------------------------------------------------------
161.34  &$\sigma(\qqbar)$     &0.84       &0.80       &0.006 \\
        &\Rb                  &--0.16     &--0.02     &0.016 \\
        &$\sigma(\mu\mu)$     &0.09       &--0.59     &0.068 \\
        &$\sigma(\tau\tau)$   &1.74       &1.29       &0.046 \\
        &$\AFB(\ell\ell)$     &--1.24     &--1.03     &0.023 \\
\hline%-----------------------------------------------------------------
172.12  &$\sigma(\qqbar)$     &--0.33     &--0.44     &0.003 \\
        &\Rb                  &--0.73     &--0.60     &0.014 \\
        &$\sigma(\mu\mu)$     &--0.39     &--1.08     &0.070 \\
        &$\sigma(\tau\tau)$   &0.07       &--0.46     &0.054 \\
        &$\AFB(\ell\ell)$     &--0.34     &--0.14     &0.022 \\
\hline%-----------------------------------------------------------------
\hline%-----------------------------------------------------------------
\end{tabular}
\caption[]{Normalized residuals $\chi$ and sensitivities for the $\alphaem$ 
fit to the combined data. For each measured quantity we give the residual 
before and after the fit, and the sensitivity, as defined in the text.
}
\label{tab:alphaem2}
\end{table}%------------------------------------------------------------

%-----------------------------------------------------------------------
\section{Constraints on New Physics}     \label{sec:new_phys}
%-----------------------------------------------------------------------
%-----------------------------------------------------------------------
\subsection{Limits on Four-fermion Contact Interactions}     \label{sec:ci}
%-----------------------------------------------------------------------
In this section we use our measurements of non-radiative cross-sections
and angular distributions at 130--172~GeV to 
place limits on possible four-fermion contact interactions. This analysis
is similar to those presented previously~\cite{bib:OPAL-CI,bib:OPAL-SM161}.
Because the Standard Model cross-sections decrease as $1/s$, the
sensitivity of the measurements to the contact interaction increases with 
centre-of-mass energy and so the inclusion of data at 172~GeV is expected
to give improved limits. In addition, the analysis of the hadronic 
cross-section in 
references~\cite{bib:OPAL-CI,bib:OPAL-SM161} assumed that the
contact interaction is flavour blind. Here the study is extended to the
case where the new physics couples exclusively to one up-type quark or
one down-type quark.

\subsubsection{Theoretical Expectation}
\begin{figure}
\begin{center}
  \epsfig{file=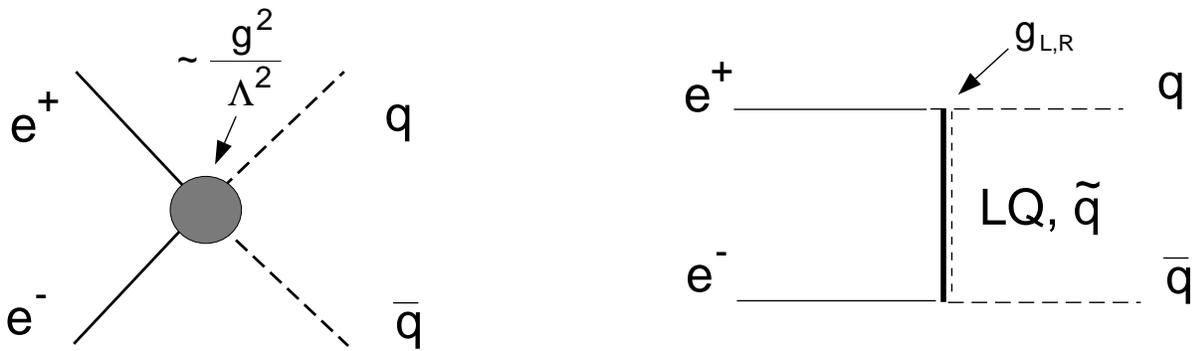,height=5cm}
\caption{Feynman diagram representing a generic contact interaction (left) 
and the exchange of a leptoquark or a squark in a t-channel diagram 
in an $\epem$ collision. 
}
\label{fig:feynman} 
\end{center}
\end{figure}

The Standard Model could be part of a more general 
theory characterized by an energy scale $\Lambda$. The consequences of the 
theory would be observed at energies well below $\Lambda$ as a deviation 
from the Standard Model which could be described by an effective contact 
interaction, as depicted in figure~\ref{fig:feynman}. In the context of 
composite models of leptons and quarks,
the contact interaction is regarded as a remnant of the binding force 
between the substructure of fermions. If electrons were composite, 
such an effect would appear in Bhabha scattering ($\eetoee$).
If the other leptons and quarks shared the same type of substructure, 
the contact interaction would exist also in the processes 
$\eetomumu,\eetotautau$ and $\eetoqq$. Alternatively, a four-fermion
contact interaction could be a good description of deviations from the
Standard Model due to the exchange of a new very heavy boson of mass 
$\mX$ if $\mX\gg\sqrt{s}$.
More generally, the contact interaction is considered to be
a convenient parametrization to describe possible deviations from 
the Standard Model which may be caused by some new physics.
The concept of contact interactions with a universal energy scale
$\Lambda$ is also used in ep and p$\overline{\mathrm{p}}$ collisions to 
search for substructure of quarks or new heavy particles coupling to quarks 
or gluons.
 
In the framework of a contact interaction~\cite{bib:Eichten} the Standard Model 
Lagrangian for $\epem\to\ff$ is extended by a term describing a new 
effective interaction with an unknown coupling constant $\mathrm{g}$ and an 
energy scale $\Lambda$:
\begin{eqnarray}\label{eq-contact}
{\cal L}^{\mathrm{contact}} & = 
         & \frac{\mathrm{g}^2}{(1 + \delta)\Lambda^2}
         \sum_{i,j=\mathrm{L,R}}\eta_{ij}[\bar{\Pe}_i\gamma^{\mu}{\Pe}_i]
                                       [\bar{\rm f}_j\gamma_{\mu}{\rm f}_j]
\end{eqnarray} 
where $\delta = 1$ for $\epem\to\epem$ and $\delta = 0$ otherwise.
Here $\mathrm{e}_L (f_L)$ and $\mathrm{e}_R (f_R)$ 
are chirality projections of electron (fermion) 
spinors, and $\eta_{ij}$ describes the chiral structure of 
the interaction. The parameters $\eta_{ij}$ are free in these models, but
typical values are between $-1$ and $+1$, depending on the type of theory 
assumed~\cite{bib:contacttable}. Here we consider the same set of 
models as in reference~\cite{bib:OPAL-CI}. In addition we consider three
other models: LL + RR ($\eta_{\mathrm{LL}}=\eta_{\mathrm{RR}}=\pm
1,\,\eta_{\mathrm{LR}}=\eta_{\mathrm{RL}}=0$) and
LR + RL ($\eta_{\mathrm{LL}}=\eta_{\mathrm{RR}}=0,\,
\eta_{\mathrm{LR}}=\eta_{\mathrm{RL}}=\pm1$)~\cite{bib:lqalt,bib:CInew1} which
are parity-conserving combinations consistent with the limits from
atomic parity violation experiments, and $\overline{\cal{O}}_{\mathrm{DB}}$
($\eta_{\mathrm{LL}}=\pm 1,\,\eta_{\mathrm{RR}}=\pm 4,
\,\eta_{\mathrm{LR}}=\eta_{\mathrm{RL}}=\pm 2 $)~\cite{bib:CInew2}.

Equation~\eqref{eq-contact} can be rewritten in terms of the parameter 
$\varepsilon=(\mathrm{g}^2/4\pi) / \Lambda^2$. In this case the differential 
cross-section can be expressed as 
\begin{equation}\label{eq-sig_contact}
{{\rm d} \sigma \over {\rm d} \cos \theta } = 
   \sigma_{\mathrm SM}(s,t) + C_2^0(s,t) {\varepsilon} 
                            + C_4^0(s,t) {\varepsilon^2}\ .
\end{equation}
Here $t=-s(1-\cos\theta)/2$ and $\theta$ is the polar angle of the 
outgoing fermion 
with respect to the $\Pe^-$ beam direction.
The $C_2^0$ term describes the interference between the 
Standard Model and the contact interaction, the $C_4^0$ 
term is the pure contact interaction contribution. 
Their exact form 
depends on the type of fermion in the final state and the particular 
model chosen, and is given, for example, in 
reference~\cite{bib:OPAL-CI}\footnote{
Equation~(2) in reference~\cite{bib:OPAL-CI} has a typographical error:
the factor $4s$ on the left-hand side should be replaced by $2s$.}.
If the underlying process is the exchange of a new heavy scalar 
particle, X, then equation~\eqref{eq-contact} and 
equation~\eqref{eq-sig_contact} 
are good approximations of the process so long as $\mX\gg\sqrt{s}$. 
Limits on the energy scale of the new interaction are extracted 
assuming $\mathrm{g}^2/4 \pi = 1$.

In order to compare the models with the data the lowest order 
expression given in equation~\eqref{eq-sig_contact} was modified to include
electroweak and QCD effects, and experimental cuts and acceptances 
were taken into account.
The Standard Model cross-section $\sigma_{\mathrm{SM}}$ has been 
calculated using ALIBABA for the \epem\ final state and ZFITTER
for all other final states. Standard Model parameters 
were fixed to the values given in section~\ref{sec:sm}.
The errors on these quantities are negligible compared to the 
statistical precision of the data.

The dominant part of the electroweak corrections is due 
to initial-state radiation, which was taken into account by
convolving the theoretical cross-section with the 
effects of photon radiation according to~\cite{bib:init}. 
Initial-state radiation was calculated up to order $\alpha^2$ 
in the leading log approximation with soft photon exponentiation, 
and the order $\alpha$ leading log final state QED correction was
applied. Other electroweak corrections were taken into account 
by evaluating the cross-sections with the appropriate value of
$\sin^2\theta_{\mathrm{W}}^{\mathrm eff}$. For the hadronic
cross-section, QCD effects were taken into account by multiplying
the electroweak-corrected cross-section by 
$\delta_{\mathrm{QCD}}=1 + \alphas/\pi + 1.409(\alphas/\pi)^2$.

%As an illustration of the expected sensitivity, in figure~\ref{fig:theo}(a) 
%the ratio of total hadronic cross-section including a contact interaction 
%to the Standard Model cross-section is shown for one of the models, the
%LL case. {\em Is this supposed to be for 172 GeV or all energies
%combined? Do we want to keep this plot} 

\subsubsection{Analysis and Results}
We have fitted the data presented here on the angular distributions for 
the non-radiative \mbox{$\eetoee$,} $\eetomumu$, $\eetotautau$ processes, the 
non-radiative cross-sections for $\eetoqq$, and the measurements of $\Rb$.
In the case of the leptonic angular distributions, a maximum likelihood 
fit was used, where the total sample of candidate events was fitted with 
the theoretical prediction plus the background. 
The number of events predicted in each bin of $\ct$ is given by  
\begin{equation}\label{eq-npred}
N_{k}^{\rm pred}(\varepsilon, r,\ct) = (1+r)\left[
                      \sigma_{k}(\varepsilon) E_{k}(\ct) 
           + \sigma_{\mathrm bgd}^{{\mathrm{pred}},k}(\ct)\right] L_{k}~,
%N_{k}^{\rm pred}(\varepsilon, r,\ct) = (1+r)\left[
%                      \sigma_{k}(\varepsilon) E_{k}(\ct) L_{k} 
%                     + N_{\mathrm bgd}^{{\mathrm{pred}},k}\right]~,
\end{equation}
where $\sigma_{k}(\varepsilon)$ is the cross-section at the centre-of-mass 
energy point $k$, which is a function of the free parameter $\varepsilon$,
$E_{k}(\ct)$ is the correction factor for experimental efficiency, 
$\sigma_{\mathrm bgd}^{{\mathrm{pred}},k}(\ct)$ is the predicted 
background cross-section, 
and $L_k$ is the integrated luminosity. The background cross-section
is evaluated including efficiency and detector acceptances for the 
different background processes.
The factor $r$ is a scaling factor which takes uncertainties
from systematic errors into account. 
It allows the overall normalisation 
to vary within bounds set by the appropriate systematic 
errors~\cite{bib:OPAL-CI}.
We derive $95\%$ confidence limits from the values of $\varepsilon$
corresponding to a change in the likelihood of $1.92$ with respect to
the minimum. 
In the case of the hadronic and $\bbbar$ cross-sections, we used a
$\chi^2$ fit to the measured values, incorporating the correlations
between the measurements as for the $\alphaem$ fit. In this case the
95\% confidence limits correspond to a change in $\chi^2$ of 3.84.
In the case of the hadronic cross-section, we fit both for the case 
where the new physics couples equally to all five flavours and under 
the assumption that the new interaction couples only to one flavour.

\begin{figure}
\begin{center}
\epsfxsize=0.86\textwidth
%\epsfbox[50 90 545 792]{ccfitbar.ps}
\epsfbox[50 90 545 792]{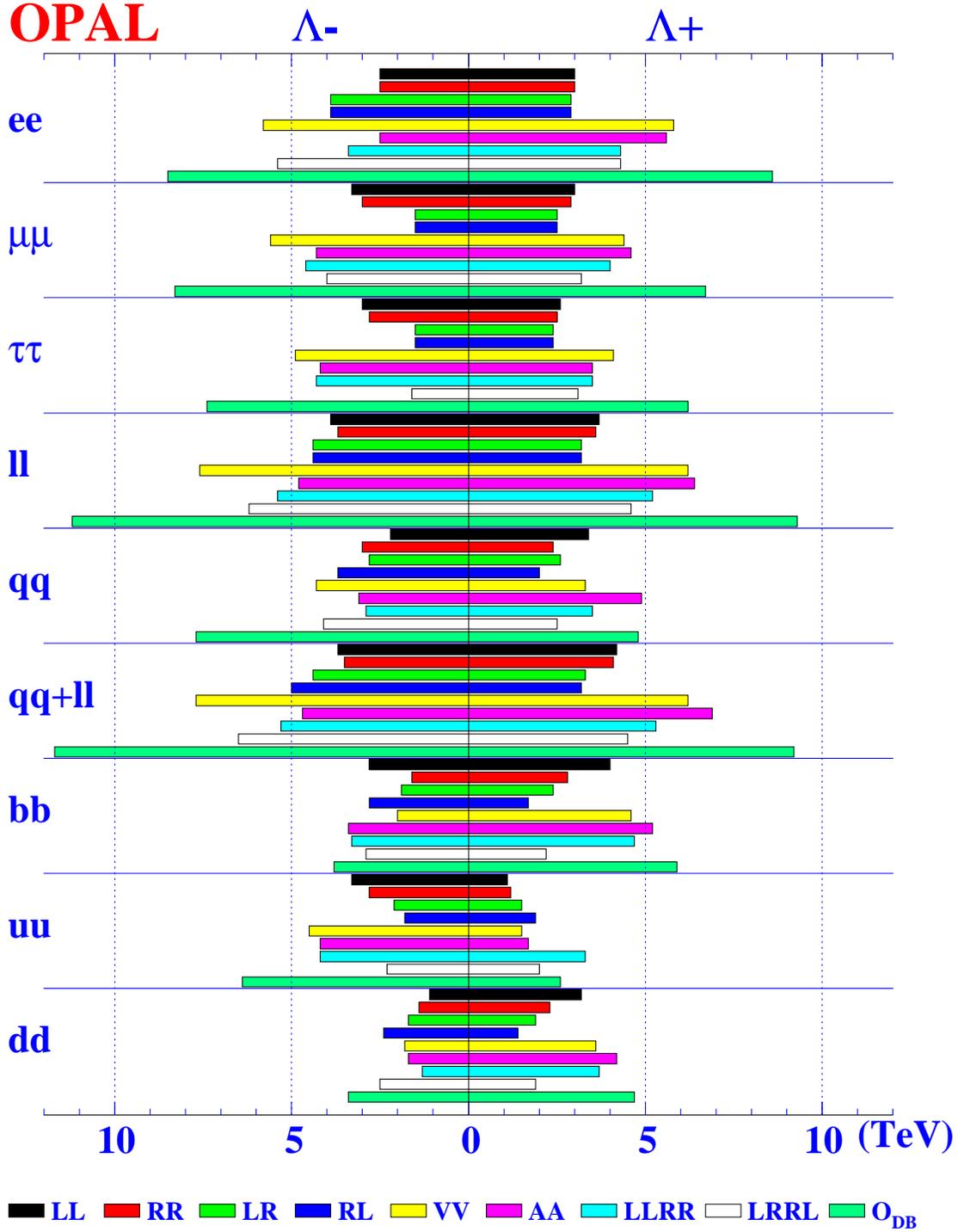}
\caption{95\% confidence level limits on the energy scale $\Lambda$
resulting from the contact interaction fits. For each channel, the
bars from top to bottom indicate the results for models LL to 
$\overline{\cal{O}}_{\mathrm{DB}}$ in the order given in the key.
}
\label{fig:ccres} 
\end{center}
\end{figure}

\begin{figure}
\begin{center}
\epsfxsize=0.86\textwidth
%\epsfbox[0 0 567 567]{xvseps.eps}
\epsfbox[0 0 567 567]{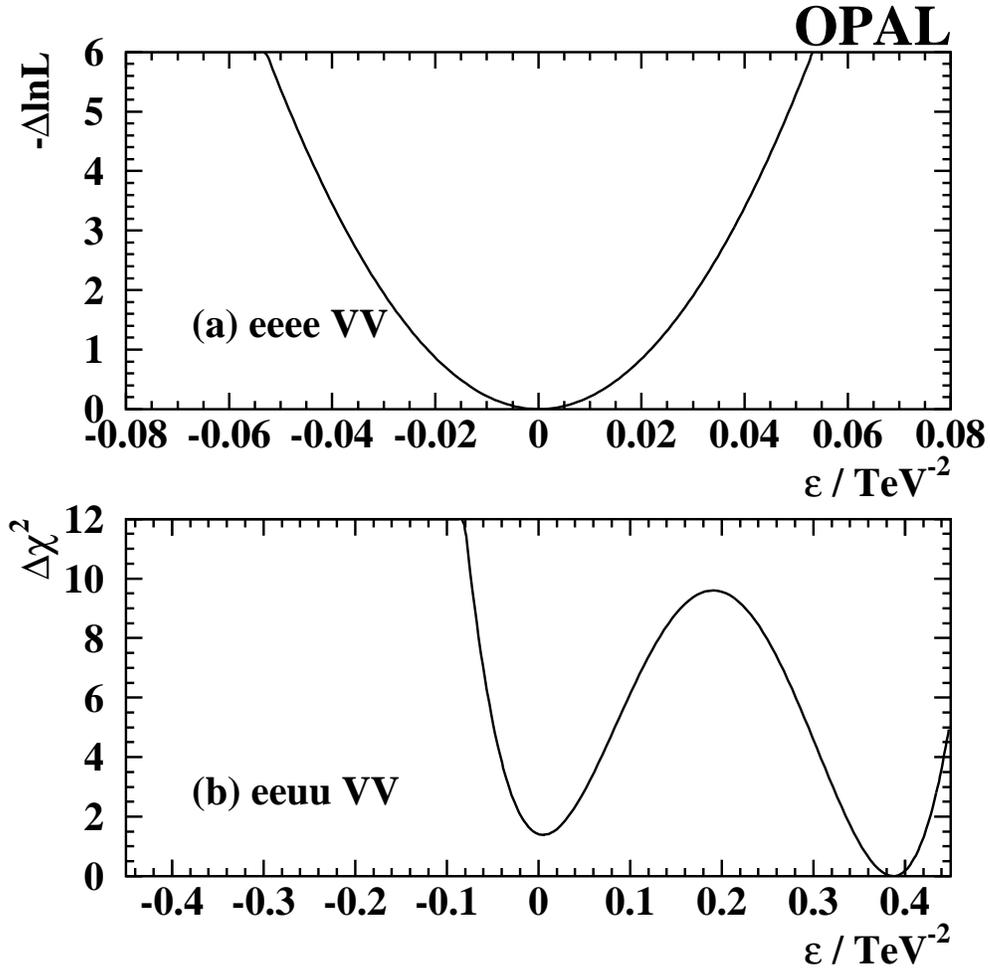}
\caption{(a) Negative log likelihood curve for the contact interaction
 fit to the \epem\ angular distribution for the VV model.
 (b) $\chi^2$ curve for the contact interaction fit to
 the hadronic cross-section, assuming coupling to only one up-type
 quark, for the same model.
}
\label{fig:xvseps} 
\end{center}
\end{figure}

The results are shown in table~\ref{tab:ccres} and are illustrated 
graphically in figure~\ref{fig:ccres}; the notation for
the different models is identical to reference~\cite{bib:OPAL-CI}.
The values for $\bbbar$ are obtained by fitting the cross-sections
for $\bbbar$ production, and there is no requirement on whether or
not the new interaction couples to other flavours. By contrast,
those for up-type quarks and down-type quarks are obtained by fitting 
the hadronic cross-sections assuming the new interaction couples only to one
flavour.
%We see that 
%the inclusion of the 172~GeV data has increased the sensitivity factor 
%$\lambda$\footnote{The sensitivity estimate $\lambda$~\cite{bib:OPAL-CI} is 
%defined in terms of the one sided 95\% confidence level upper limit on 
%$\epsilon$: $\lambda = 1/\sqrt{1.64 \sigma_{\epsilon}}$
%where $\sigma_{\epsilon}$ is the one standard deviation parabolic error
%on $\epsilon$.}, which indicates the limit on $\Lambda$ which would be
%obtained if the data were in perfect agreement with the Standard Model
%prediction, by typically 1~TeV, and the limits on $\Lambda$ are generally 
%close to the sensitivity estimate. 
Most of the fitted values of $\varepsilon$ show no significant 
deviation from zero, indicating agreement with the Standard Model.
However, the quadratic form of the cross-section 
(equation~\eqref{eq-sig_contact}) can lead to two local minima, and
in a few cases the local minimum near zero gives the higher $\chi^2$
or $-\log$(likelihood).
As an example, in figure~\ref{fig:xvseps} we show the negative log
likelihood curve for the fit to the \epem\ angular distributions, and
the $\chi^2$ curve for the fit to the hadronic cross-sections assuming
couplings to one up-type quark only, in both cases for the VV model.
In the former case the positive interference between the Standard Model and 
contact interaction amplitudes results in a curve with only a single
minimum, whereas in the latter case the negative interference results
in two local minima. As can be seen in figure~\ref{fig:xvseps}(b),
one of these is near zero, but the second one gives the lower $\chi^2$.
In this case, we quote 0.39 as the central value of $\varepsilon$,
while the limits on $\Lambda$ are derived from the points
$\varepsilon = 0.44$ and $\varepsilon = -0.05$ where $\Delta\chi^2$ is
3.84 above the right-hand and left-hand minimum respectively.

The two sets of values $\Lambda_+$ and $\Lambda_-$ shown in 
table~\ref{tab:ccres} correspond to 
positive and negative values of $\varepsilon$ respectively,
reflecting the two possible signs of $\eta_{ij}$ in equation~\eqref{eq-contact}.
%reflecting the two possible signs of interference between the contact 
%interaction and the Standard Model.
As before, the data are particularly sensitive to the 
VV and AA models; the combined data give limits on $\Lambda$ in the
range 4.7--7.7~TeV for these models, roughly 1~TeV higher than
those for 130--161~GeV data alone. For the other models the limits
generally lie in the range 2--5~TeV, approximately 0.5~TeV above those from 
previous data. Those for the $\overline{\cal{O}}_{\mathrm{DB}}$ model
are larger (3--12~TeV) because the values of the $\eta$ parameters are
larger for these models.
The limits obtained here are complementary to those obtained in
p$\overline{\mathrm{p}}$ collisions~\cite{bib:CDF_CI}. They are
superior to published contact interaction bounds from ep collisions at
HERA~\cite{bib:HERA_CI}, and severely constrain interpretations of the 
recently reported excess of events at high momentum transfer at 
HERA~\cite{bib:HERA}.

\begin{table}[htbp]
\begin{sideways}
\begin{minipage}[b]{\textheight}{\small 
\begin{center}\begin{tabular}{|cc|c|c|c|c|c|c|c|c|c|}
\hline
Channel &      &  LL  &  RR  &  LR  &  RL  &  VV  &  AA  &  
               LL+RR  &LR+RL & $\overline{\cal O}_{\mathrm{DB}}$ \\
        &      &  \scriptsize{ $[\pm1,0,0,0]$}  & 
                  \scriptsize{ $[0,\pm1,0,0]$}  & 
                  \scriptsize{ $[0,0,\pm1,0]$}  &
                  \scriptsize{ $[0,0,0,\pm1]$}  & 
                  \scriptsize{ $[\pm1,\pm1,\pm1,\pm1]$} &
                  \scriptsize{ $[\pm1,\pm1,\mp1,\mp1]$} &
                  \scriptsize{ $[\pm1,\pm1,0,0]$} & 
                  \scriptsize{ $[0,0,\pm1,\pm1]$} &
                  \scriptsize{ $[\pm1,\pm4,\pm2,\pm2]$} \\
\hline
$\ee$    &$\epsz$& $-0.041_{-0.062}^{+0.068}$ & $-0.039_{-0.063}^{+0.069}$ &
                   $ 0.017_{-0.043}^{+0.048}$ & $ 0.017_{-0.043}^{+0.048}$ &
                   $ 0.000_{-0.015}^{+0.015}$ & $-0.033_{-0.046}^{+0.033}$ &
                   $-0.022_{-0.035}^{+0.036}$ & $ 0.009_{-0.022}^{+0.023}$ &
                   $ 0.000_{-0.007}^{+0.007}$ \\ 
         &$\lamp$& 3.0 & 3.0 & 2.9 & 2.9 & 5.8 & 5.6 & 4.3 & 4.3 & 8.6 \\
         &$\lamm$& 2.5 & 2.5 & 3.9 & 3.9 & 5.8 & 2.5 & 3.4 & 5.4 & 8.5 \\
\hline
$\mumu$  &$\epsz$& $ 0.013_{-0.053}^{+0.051}$ & $ 0.014_{-0.059}^{+0.056}$ &
                   $ 0.043_{-0.079}^{+0.064}$ & $ 0.043_{-0.079}^{+0.064}$ &
                   $ 0.010_{-0.021}^{+0.021}$ & $-0.004_{-0.026}^{+0.026}$ &
                   $ 0.007_{-0.028}^{+0.028}$ & $ 0.023_{-0.040}^{+0.038}$ &
                   $ 0.004_{-0.009}^{+0.009}$ \\
         &$\lamp$& 3.0 & 2.9 & 2.5 & 2.5 & 4.4 & 4.6 & 4.0 & 3.2 & 6.7 \\
         &$\lamm$& 3.3 & 3.0 & 1.5 & 1.5 & 5.6 & 4.3 & 4.6 & 4.0 & 8.3 \\
\hline
$\tautau$&$\epsz$& $ 0.027_{-0.065}^{+0.063}$ & $ 0.030_{-0.073}^{+0.068}$ &
                   $ 0.018_{-0.404}^{+0.088}$ & $ 0.018_{-0.404}^{+0.088}$ & 
                   $ 0.010_{-0.025}^{+0.026}$ & $ 0.011_{-0.034}^{+0.035}$ &
                   $ 0.014_{-0.034}^{+0.035}$ & $ 0.009_{-0.057}^{+0.051}$ &
                   $ 0.005_{-0.011}^{+0.011}$ \\
         &$\lamp$& 2.6 & 2.5 & 2.4 & 2.4 & 4.1 & 3.5 & 3.5 & 3.1 & 6.2 \\
         &$\lamm$& 3.0 & 2.8 & 1.5 & 1.5 & 4.9 & 4.2 & 4.3 & 1.6 & 7.4 \\
\hline
$\lept$  &$\epsz$& $ 0.002_{-0.035}^{+0.035}$ & $ 0.002_{-0.038}^{+0.038}$ &
                   $ 0.023_{-0.036}^{+0.036}$ & $ 0.023_{-0.036}^{+0.036}$ &
                   $ 0.005_{-0.011}^{+0.011}$ & $-0.009_{-0.017}^{+0.017}$ &
                   $ 0.001_{-0.018}^{+0.019}$ & $ 0.012_{-0.018}^{+0.018}$ &
                   $ 0.002_{-0.005}^{+0.005}$ \\
         &$\lamp$& 3.7 & 3.6 & 3.2 & 3.2 & 6.2 & 6.4 & 5.2 & 4.6 & 9.3 \\
         &$\lamm$& 3.9 & 3.7 & 4.4 & 4.4 & 7.6 & 4.8 & 5.4 & 6.2 &11.2 \\
\hline
$\qqbar$ &$\epsz$& $-0.131_{-0.045}^{+0.097}$ & $ 0.080_{-0.145}^{+0.057}$ &
                   $-0.019_{-0.074}^{+0.131}$ & $ 0.176_{-0.070}^{+0.041}$ &
                   $ 0.042_{-0.075}^{+0.029}$ & $-0.069_{-0.022}^{+0.050}$ &
                   $-0.054_{-0.038}^{+0.094}$ & $ 0.103_{-0.080}^{+0.032}$ &
                   $ 0.029_{-0.027}^{+0.009}$ \\
         &$\lamp$& 3.4 & 2.4 & 2.6 & 2.0 & 3.3 & 4.9 & 3.5 & 2.5 & 4.8 \\
         &$\lamm$& 2.2 & 3.0 & 2.8 & 3.7 & 4.3 & 3.1 & 2.9 & 4.1 & 7.7 \\
\hline
combined &$\epsz$& $-0.004_{-0.035}^{+0.032}$ & $-0.006_{-0.038}^{+0.034}$ &
                   $ 0.023_{-0.036}^{+0.036}$ & $ 0.026_{-0.033}^{+0.037}$ &
                   $ 0.005_{-0.011}^{+0.011}$ & $-0.011_{-0.017}^{+0.016}$ &
                   $-0.001_{-0.018}^{+0.018}$ & $ 0.013_{-0.018}^{+0.018}$ &
                   $ 0.002_{-0.005}^{+0.005}$ \\
         &$\lamp$& 4.2 & 4.1 & 3.3 & 3.2 & 6.2 & 6.9 & 5.3 & 4.5 & 9.2 \\
         &$\lamm$& 3.7 & 3.5 & 4.4 & 5.0 & 7.7 & 4.7 & 5.3 & 6.5 &11.7 \\
\hline
$\bbbar$ &$\epsz$& $-0.021_{-0.049}^{+0.043}$ & $-0.065_{-0.240}^{+0.116}$ &
                   $-0.045_{-0.146}^{+0.143}$ & $ 0.052_{-0.105}^{+0.214}$ &
                   $-0.021_{-0.176}^{+0.039}$ & $-0.013_{-0.032}^{+0.026}$ &
                   $-0.015_{-0.036}^{+0.031}$ & $ 0.031_{-0.093}^{+0.109}$ &
                   $-0.018_{-0.035}^{+0.030}$ \\
         &$\lamp$& 4.0 & 2.8 & 2.4 & 1.7 & 4.6 & 5.2 & 4.7 & 2.2 & 5.9 \\
         &$\lamm$& 2.8 & 1.6 & 1.9 & 2.8 & 2.0 & 3.4 & 3.3 & 2.9 & 3.8 \\
\hline
$\uubar$ &$\epsz$& $ 0.018_{-0.059}^{+0.069}$ & $ 0.555_{-0.102}^{+0.075}$ &
                   $-0.022_{-0.132}^{+0.356}$ & $-0.132_{-0.114}^{+0.266}$ &
                   $ 0.387_{-0.035}^{+0.029}$ & $ 0.287_{-0.051}^{+0.037}$ &
                   $ 0.010_{-0.035}^{+0.039}$ & $-0.051_{-0.081}^{+0.215}$ &
                   $ 0.126_{-0.018}^{+0.014}$ \\
         &$\lamp$& 1.1 & 1.2 & 1.5 & 1.9 & 1.5 & 1.7 & 3.3 & 2.0 & 2.6 \\
         &$\lamm$& 3.3 & 2.8 & 2.1 & 1.8 & 4.5 & 4.2 & 4.2 & 2.3 & 6.4 \\
\hline
$\ddbar$ &$\epsz$& $-0.701_{-0.065}^{+0.081}$ & $-0.311_{-0.103}^{+0.340}$ &
                   $ 0.033_{-0.310}^{+0.154}$ & $ 0.317_{-0.178}^{+0.097}$ &
                   $-0.216_{-0.044}^{+0.074}$ & $-0.287_{-0.037}^{+0.051}$ &
                   $-0.487_{-0.046}^{+0.058}$ & $ 0.151_{-0.165}^{+0.076}$ &
                   $-0.045_{-0.026}^{+0.073}$ \\
         &$\lamp$& 3.2 & 2.3 & 1.9 & 1.4 & 3.6 & 4.2 & 3.7 & 1.9 & 4.7 \\
         &$\lamm$& 1.1 & 1.4 & 1.7 & 2.4 & 1.8 & 1.7 & 1.3 & 2.5 & 3.4 \\
\hline

\end{tabular}\end{center}}
\caption[foo]{\label{tab:ccres}
Results of the contact interaction fits to the angular distributions for
non-radiative $\eetoee$, $\eetomumu$, $\eetotautau$, the cross-sections
for $\eetoqq$ and the measurements of $\Rb$ presented here. The combined 
results include all leptonic angular distributions and the hadronic 
cross-sections. The numbers in square brackets are the values of
[$\eta_{\mathrm{LL}}$,$\eta_{\mathrm{RR}}$,$\eta_{\mathrm{LR}}$,
$\!\eta_{\mathrm{RL}}$] which define the models.
$\epsz$ is the fitted value of $\varepsilon = 1/\Lambda^{2}$, 
$\Lambda_{\pm}$ are the 95\% confidence level limits.
The units of $\Lambda$ are TeV, those of $\epsz$ are $\mathrm{TeV}^{-2}$.
}
\end{minipage}
\end{sideways}
\end{table}

%-----------------------------------------------------------------------
\subsection{Limits on Heavy Particles}           \label{sec:lq}
%-----------------------------------------------------------------------
The  contact   interaction analysis  is an   appropriate framework for
searching for effects arising from the exchange of a new particle with
mass   $\mX\gg\sqrt{s}$.     For    lower  mass    ranges,  $\sqrt{s}$
\raisebox{2pt}{\mbox{$<$}}\makebox[-8pt]{\raisebox{-3pt}{$\sim$}\,} $\
\, \mX < \Lambda$, we search  for signs of  new physics not within the
framework  of   the  contact  interaction,   but under  the   explicit
assumption that  the new phenomena are due  to a heavy particle, which
couples to leptons and quarks. 
Such a particle could be a leptoquark~\cite{bib:lqbas} or
a   squark   in      supersymmetric   theories    with      $R$-parity
violation~\cite{bib:rpvbas}.  Beyond the
kinematic limit for direct production, such a new particle might be
seen through a change of the total cross-section in the process
$\epem\to\qqbar$ via a $t$-channel exchange diagram as depicted in
figure~\ref{fig:feynman}.  
The    allowed   leptoquark
states  can be  classified according  to  spin and weak isospin $I$. 
We  denote scalar particles  $\mathrm{S}_{\it I}$ and
vector particles  $\mathrm{V}_{\it  I}$.  Isomultiplets with different
hypercharges    are    distinguished   by        a   tilde,    as   in
reference~\cite{bib:lqsig}.  The  coupling strength  of the leptoquark
is given by a coupling constant $\mathrm{g_{L}}$ or $\mathrm{g_{R}}$,
where L and  R refer to the chirality  of the lepton.   The two scalar
states $\mathrm{S}_0$ and $\mathrm{S}_{1/2}$ and the two vector states
$\mathrm{V}_{0}$ and $\mathrm{V}_{1/2}$  can in principle couple to
both left and right chiralities. The  product
$\mathrm{g_{L}} \cdot  {g_{R}}$  is constrained very  strongly by
low energy processes~\cite{bib:lqsig}.  Therefore in this analysis only
one  coupling at a time is assumed to be  non-zero.  

In a $t$-channel reaction the
exchange of $\mathrm{S}_{0}$ and $\mathrm{\tilde{S}_{1/2}}$ are
equivalent to the  exchange of an  $R$-parity violating down-type squark
and up-type antisquark respectively. The 
coupling to electrons is given by the term $\lambda'_{1jk}L_1Q_jD_k^c$   
of the superpotential~\cite{bib:rpvsup}, where  the  indices
denote the family of the particles involved. 
$L_1$ and $Q_j$ are the SU(2) doublet
lepton and quark superfields and  $D_k^c$ denotes a down-type
antiquark singlet superfield. The coupling constant
$\lambda'_{1jk}$ is equivalent to $\mathrm{g_{L}}$ in the
leptoquark case since the lepton involved is left-handed.
In the limit of very
large masses, $\mX\gg\sqrt{s}$, the leptoquark coupling constant
$\mathrm{g_{LQ}}$ is related to the contact interaction coupling
constant $\mathrm{g_{CI}}$ by ${\mathrm{g_{LQ}}}/\mX =
\mathrm{g_{CI}}/\Lambda$. 
% Beyond the
% kinematic limit for direct production, such a new particle might be
% seen through a change of the total cross-section in the process
% $\epem\to\qqbar$ via a $t$-channel exchange diagram as depicted in
% figure~\ref{fig:feynman}.  

The cross-section for
$\epem\rightarrow\qqbar$ including $t$-channel exchange of a
leptoquark has been calculated by several authors, for example
references~\cite{bib:lqsig,bib:lqalt,bib:dreiner}. It can be written as 
\begin{equation}\label{eq-lqsig}
  \frac{\mathrm{d}\sigma}{\mathrm{d}\ct} = \frac{N_{\mathrm{c}}}{128\pi s}
\sum_{i,k=\mathrm{L,R}} \rho_{ik}|f_{ik}|^{2} 
\end{equation}
where the colour factor $N_{\mathrm{c}} = 3$, $\rho_{ik}$ are the spin density
matrix elements and $f_{ik}$ are the helicity amplitudes for the
process
$\mathrm{e}_{\it i}^-\mathrm{e^+}\rightarrow
\mathrm{q}_{\it k}\overline{\mathrm{q}}$. 
The helicity amplitudes used here are taken from
reference~\cite{bib:lqsig}. 

%The $t$-channel exchange of a heavy up-type 
%squark in $R$-parity violating supersymmetric theories gives a form for the
%cross-section identical with that for a $\tilde{\mathrm{S}}_{1/2}$ 
%leptoquark with coupling $\mathrm{g_{L}}$, whereas an $R$-parity violating
%down-type squark is equivalent to the $\mathrm{S}_0$ leptoquark with
%coupling $\mathrm{g_{L}}$~\cite{bib:lqsig,bib:dreiner}. Hence the limits
%presented here for these leptoquark states can also be interpreted in
%terms of squark exchange. 
 
In order to compare the model with data the lowest order expression
given above was evaluated taking electroweak corrections into account as 
described in section~\ref{sec:ci} for the contact interaction analysis.  
No QCD corrections to the heavy particle
propagator or vertex corrections were included. Standard Model
cross-sections were calculated using ZFITTER. 

\subsubsection{Analysis and Results}
In order to  derive a  limit on the  mass  and  coupling of a   heavy
particle which couples to quarks and leptons, we calculate the $\chi^2$ 
between the measured non-radiative hadronic cross-sections and the model
predictions. The predicted cross-section depends on the mass of the heavy 
particle and its coupling constant.
Varying the mass in steps of 25~GeV, for each value we find  the coupling 
constant which
minimizes $\chi^2$,  and determine the  95\% confidence  limit on the
coupling corresponding to a change in $\chi^2$  of 3.84.  We assume the
presence of only one leptoquark multiplet at  a time. As mentioned above,
only one coupling, either $\mathrm{g_{L}}$  or $\mathrm{g_{R}}$, is assumed
to be non-zero. In the case  of  $\mathrm{S_1}$,
$\mathrm{V_1}$  with      $\mathrm{g_{L}}$, and $\mathrm{S_{1/2}}$,
$\mathrm{V_{1/2}}$ with $\mathrm{g_{R}}$,   two states of  the  isospin
multiplet contribute to the hadronic  cross-section.  Their masses are
assumed to be degenerate. 
%We assume one new particle which can couple to only one quark. 
%For some of the leptoquark states this is an up-type quark, for some
%a down-type quark, while other states can couple to either, but with
%different chiral couplings. 

We perform the analysis for two cases. In the first case we consider
all non-radiative hadronic events, assuming a non-vanishing coupling to
one quark family. In the second case we use the cross-sections for $\bbbar$
production as described    in  section~\ref{sec:rb} considering    all
possible leptoquark couplings to the b quark. 

\begin{figure}
\begin{center}
\epsfxsize=0.83\textwidth
%\epsfbox[50 90 545 792]{limits_scalar_bw.ps}
\epsfbox[50 90 545 792]{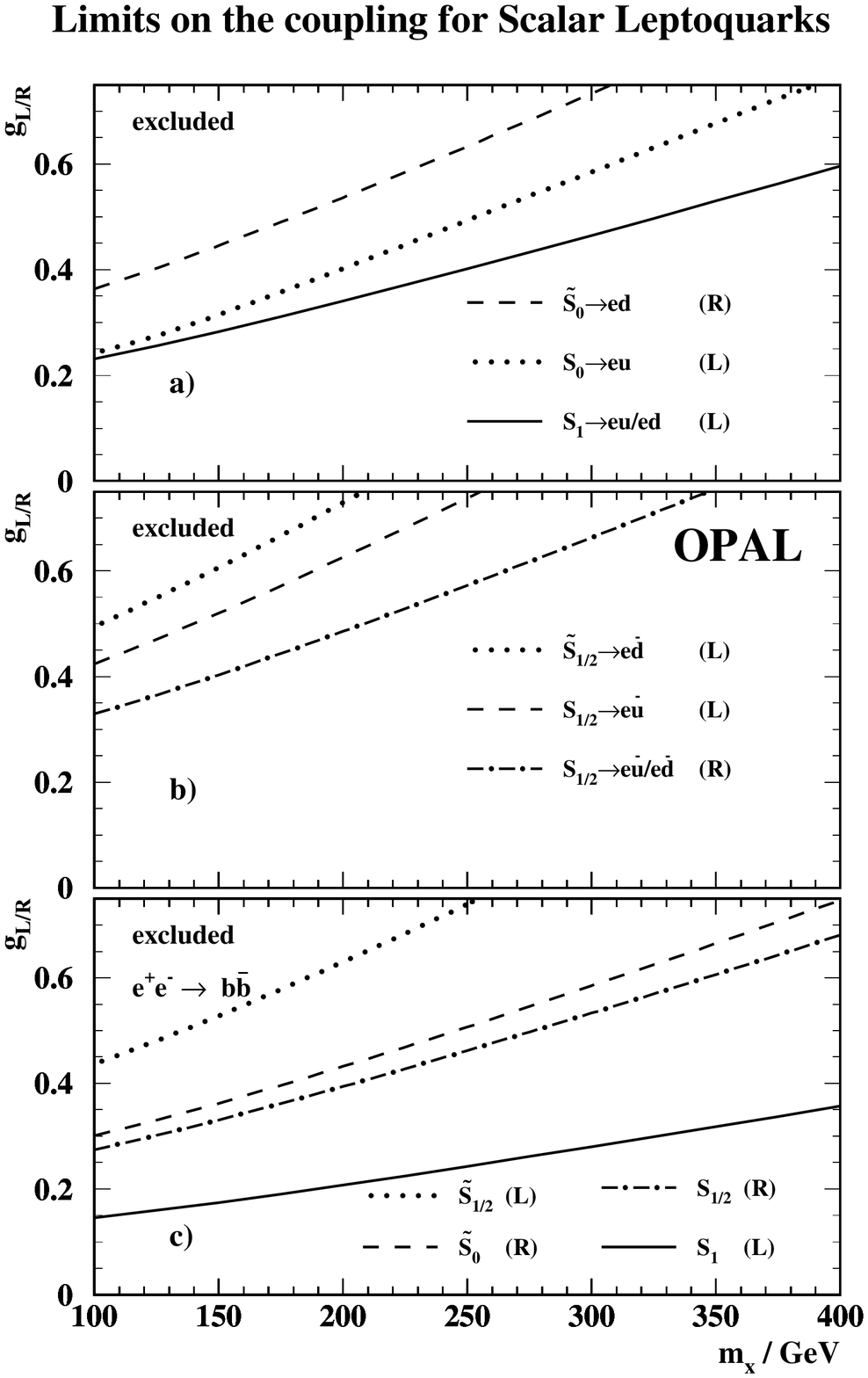}
\caption{95\% confidence exclusion limits on $\mathrm{g_{L}}$ or 
$\mathrm{g_{R}}$ as a function of $\protect\mX$, for various possible 
scalar leptoquarks. (a) and (b) show limits on leptoquarks coupling 
to a single quark family, derived from the hadronic cross-sections.
(c) shows limits on leptoquarks coupling to b quarks only, derived from
the \Pb\Pab\ cross-sections. The excluded regions are above the curves
in all cases.
The letter in parenthesis after the different leptoquark types indicates
the chirality of the lepton involved in the interaction. The limits on
the $\mathrm{S}_{0}$ and $\tilde{\mathrm{S}}_{1/2}$ leptoquarks can 
be interpreted as those on $R$-parity violating 
$\tilde{\mathrm{d}}_{\mathrm{R}}$ and $\tilde{\mathrm{u}}_{\mathrm{L}}$ 
squarks respectively.
}
\label{fig:limits_scalar} 
\end{center}
\end{figure}

\begin{figure}
\begin{center}
\epsfxsize=0.83\textwidth
%\epsfbox[50 90 545 792]{limits_vector_bw.ps}
\epsfbox[50 90 545 792]{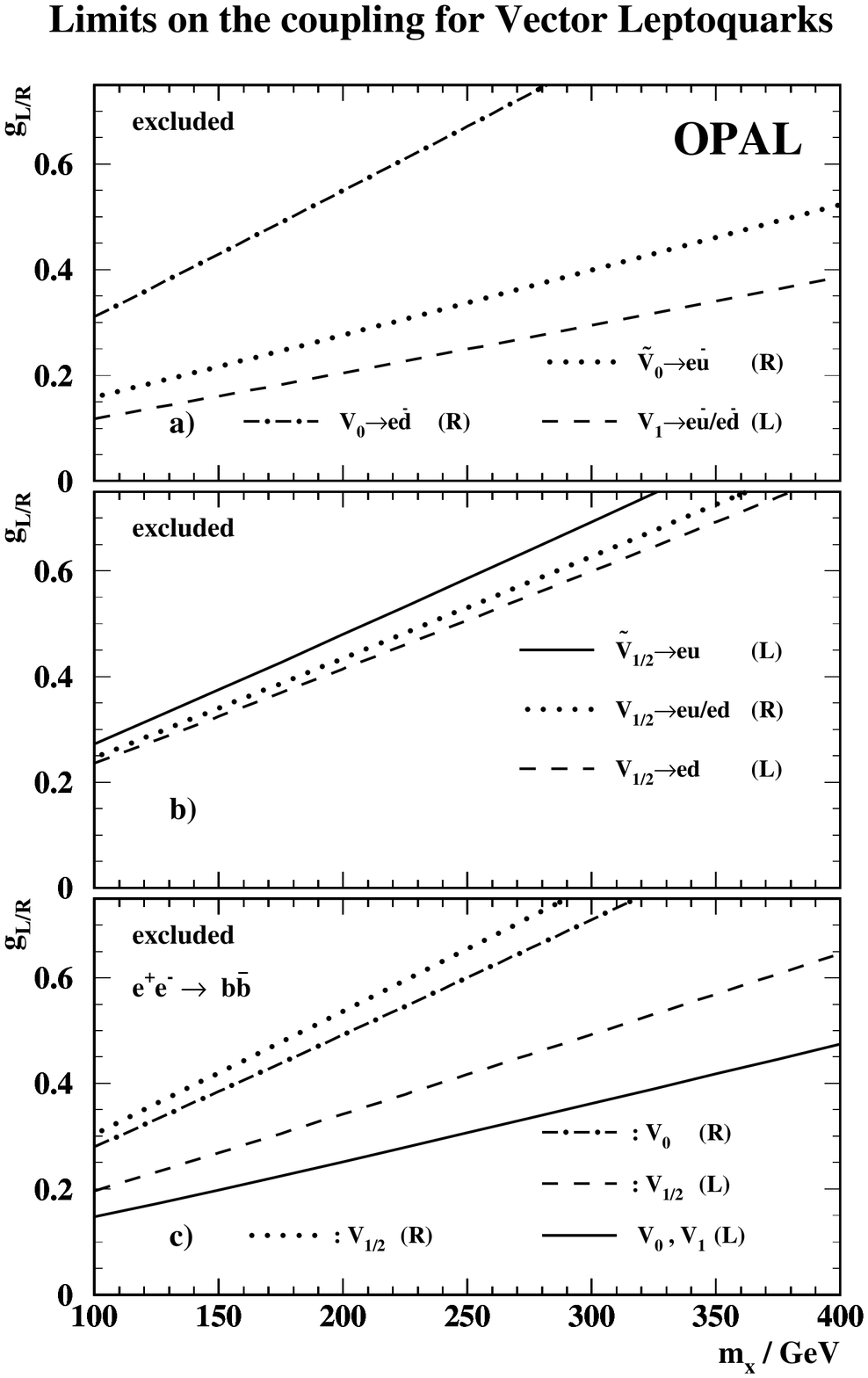}
\caption{95\% confidence exclusion limits on $\mathrm{g_{L}}$ or 
$\mathrm{g_{R}}$ as a function of $\protect\mX$, for various possible 
vector leptoquarks. (a) and (b) show limits on leptoquarks coupling
to a single quark family, derived from the hadronic cross-sections.
(c) shows limits on leptoquarks coupling to b quarks only, derived from 
the \Pb\Pab\ cross-sections. The excluded regions are above the curves
in all cases.
The letter in parenthesis after the different leptoquark types indicates
the chirality of the lepton involved in the interaction.
}
\label{fig:limits_vector} 
\end{center}
\end{figure}

The results from the analysis of the hadronic cross-sections are
presented in figure~\ref{fig:limits_scalar}(a) and (b) for the possible 
scalar leptoquark states. The $95\%$ confidence limits are shown as a
function of the mass $\mX$ and the coupling constant $\mathrm{g_L}$
or $\mathrm{g_R}$ of the new particle. 
%The regions above the curves are 
%exluded. 
The limits on $\mathrm{g_L}$
for $\mathrm{S}_{0}$ are equivalent to limits on the Yukawa couplings
$\lambda'_{1jk}$ of supersymmetric models with $R$-parity violation with
$j=1,2$ and $k=1,2,3$, where $k$ is the family index of the exchanged squark.
The limits on $\mathrm{\tilde{S}}_{1/2}$ derived from the hadronic
event sample are also limits on $\lambda'_{1jk}$ with
$j=1,2,3$ and $k=1,2$, where $j$ is the family index of the exchanged squark.
Results from the analysis of the  hadronic cross-sections for the
vector leptoquark states are shown in figure~\ref{fig:limits_vector}(a)
and (b). 
We do not show limits on
the $\rm S_0$ ($\rm V_0$) leptoquark with coupling $\rm g_R$ ($\rm g_L$)
because the effect of these particles on the
hadronic cross-section at these energies is too small. 
%In the case
%of the $\rm V_0$ leptoquark with coupling $\rm g_L$ the limit is very
%close to that of the ${\rm \tilde{V}_0}$ leptoquark with coupling
%$\rm g_R$, except for a small band at high coupling constants which
%cannot be excluded.

Figures~\ref{fig:limits_scalar}(c) and~\ref{fig:limits_vector}(c) give
the limits on scalar and vector leptoquarks respectively,
derived in the analysis of the $\bbbar$ cross-sections.
%Because of the assumption of 
%coupling to only one quark, 
Most of these limits are considerably more 
stringent than those from the total cross-section analysis. The limits
are weaker for the $\mathrm{V}_{1/2}$ leptoquark with coupling 
$\mathrm{g_R}$ and the $\mathrm{V}_{1}$ leptoquark with coupling
$\mathrm{g_L}$ because in these cases two leptoquarks of the isospin
multiplet contribute to the hadronic cross-section, but only one of
these couples to b quarks. The limit on $\mathrm{\tilde{S}}_{1/2}$ is
equivalent to a limit on $\lambda'_{1j3}$ with $j=1,2,3$.

As can be seen in the figures our analysis is sensitive to
leptoquark masses much higher than the beam energy. 
Direct searches at the Tevatron can exclude
scalar  and vector leptoquarks with Yukawa couplings down to 
${\cal O}(10^{-7})$ up to masses of $\approx 225$~GeV \cite{bib:tevlim}.
%\footnote{Preliminary exclusion limits  
%for larger masses have been presented in \cite{bib:lqprel}.}
Our limits extend this excluded region for large couplings. 

As mentioned above, in the limit of very large mass, the leptoquark 
and contact interaction couplings are related by ${\mathrm{g_{LQ}}}/\mX =
\mathrm{g_{CI}}/\Lambda$. Where the cross-section formula of a leptoquark
exchange corresponds to one of the investigated contact interaction models we
have checked the consistency of the two analyses by calculating 
limits on the leptoquark coupling at masses of several TeV. We found
excellent agreement. 

\subsection{Search for Charginos and Neutralinos Decaying into Light Gluinos}
%=======================================================================
 In general, a comparison between the observed number of non-radiative 
 hadronic events and the Standard Model prediction may be used to constrain 
 possible new particle production whose signature is similar to that of the 
 non-radiative hadronic events. The upper limit on the excess in the observed
 number of
 events over the Standard Model prediction was determined to be 49.9 events 
 at 95\% confidence level, summing over all centre-of-mass energies and 
 taking into account the correlated systematic errors between different 
 centre-of-mass energies. 
 The numbers of observed non-radiative 
 hadronic events are listed in table~\ref{tab:mh}, and
 the Standard Model prediction was calculated using
 ZFITTER with input parameters given in section~\ref{sec:sm}.
 An example with a particular model of new particle production
 is presented here.
 
 In supersymmetric extensions of the Standard Model, the supersymmetric 
 partners of the gluons are Majorana fermions called gluinos. Experiments 
 at the TEVATRON have excluded a gluino of mass up to about 150~GeV, 
 independent of squark mass~\cite{bib:CDFsgsq,bib:D0sgsq}, 
 if missing energy
 is a relevant signature. However, a relatively long-lived very light gluino 
 with mass less than 1~GeV might not be completely excluded by any
 experiment~\cite{bib:Farrar}.     
 
 The supersymmetric partners of the weak gauge and Higgs bosons mix to form
 charginos and neutralinos. Chargino and neutralino pairs 
 ($\chp_i \chm_j$, $i,j=1,2$ and $\nt_i\nt_j$, $i,j=1,2,3,4$) 
 can be produced in $\epem$ annihilation via $s$-channel virtual $\gamma$ or  
 Z, or in $t$-channel slepton exchange. The cross-section for chargino
 pair production is typically a few picobarns at $\sqrt{s} =172~\GeV$, 
 although, if the sneutrino mass is less than 100~GeV, the cross-section can 
 be significantly reduced by interference between the $s$-channel and 
 $t$-channel diagrams. In the usual SUSY scenario, $\chpm_i$ decays into a 
 virtual W boson and a neutralino. The cross-section for neutralino pair
 production is typically a fraction of picobarn at $\sqrt{s} = 172$~GeV 
 for each process.
 Normally $\nt_i$ ($i=2,3,4$) are assumed to decay into a virtual 
 Z boson and a lighter neutralino. The lightest neutralino is usually 
 expected to be the lightest supersymmetric particle and is stable, 
 however this is unimportant for the present analysis.
 
 Searches for charginos and neutralinos at various experiments have been
 carried out with an implicit assumption that light gluinos do not 
 exist, so that the decay of charginos or neutralinos leads to
 signatures of large missing 
 energy~\cite{LEP15-opal,LEP15-chargino,LEP16-opal,CDFD0-chargino}. 
 If light gluinos exist and squark masses are comparable to the W
 boson mass, charginos and neutralinos decay dominantly into 
 $\qqbar' \gluino$ (where the \Pq\ and $\Paq'$ are the same flavour
 in the neutralino case) through a virtual squark by the strong 
 interaction~\cite{bib:Farrar}. For larger squark masses,
 $m_{\tilde{\mathrm{q}}}\simeq 1.5\mPW$, the squark propagator reduces
 the decay width into $\qqbar'\gluino$, and this becomes comparable to
 the $\mathrm{W}^{*}\nt$ or $\PZ^{*}\nt$ mode\footnote{The two body decays 
 $\chpm_i\rightarrow\snu\ell^{\pm}$ or $\nt_i\rightarrow\snu\nu$ are in 
 principle possible if $\snu$ is lighter than $\chpm_i$ or $\nt_i$
 respectively. However, assuming $\squark$ and $\snu$ masses are almost 
 degenerate in the parameter space considered in this analysis, as is the
 case in the Minimal Supersymmetric Standard Model, 
 $\chpm_i\rightarrow\squark\Paq'$ or $\nt_i\rightarrow\squark\Paq$ followed 
 by $\squark\rightarrow\Pq\gluino$ will be the dominant decay.}.
 Since the gluino hadronizes into
 many hadrons, the signature of such chargino or neutralino events
 is very similar to ordinary $\qqbar$ events or WW hadronic decay events. 
 Therefore, as proposed in reference~\cite{bib:Farrar}, we have used the 
 hadronic cross-section measurements for $s'/s > 0.8$ presented in
 section~\ref{sec:mh} to place limits on any possible contribution from
 chargino or neutralino pair production in which gauginos decay into  
 $\qqbar'\gluino$. This method has the advantage of being insensitive to
 details of gluino fragmentation which would cause a significant
 uncertainty in a search for an excess of multijet events. 
 
 The SUSY partners of SU(3), SU(2) and U(1) 
 gauge bosons of the Standard Model (gauginos) are assumed to have the same 
 mass at the Grand Unification mass scale. Masses of these gauginos at the 
 weak scale are determined by the renormalization group equations. As a 
 result, the ratios of the SU(3), SU(2) and U(1) gaugino masses at the weak 
 scale ($M_3 : M_2: M_1$) are proportional to the strength of the gauge 
 couplings at the weak scale ($\alpha_3 : \alpha_2 : \alpha_1$).   
 Therefore, if the gluino, which is the SU(3) gaugino, is very light 
 ($\leq 1$~GeV) at the weak scale, all the other gauginos are also very light.
 We assume $M_3 = M_2 = M_1 = 0$.  Then the chargino and neutralino pair
 cross-sections depend on three SUSY parameters: the Higgsino mixing
 parameter $\mu$, the ratio of the two vacuum expectation values $\tan\beta$  
 and the sneutrino mass $m_{\snu}$. We have calculated the cross-sections
 using the formulae in reference~\cite{bib:Bartl}. We assume that the 
 sneutrino mass is larger than the current limit of 43~GeV~\cite{bib:msnu}.
 
 To  determine the limits, the total number of $\chp_i\chm_j$ ($i,j=1,2$) 
 and $\nt_i\nt_j$ ($i,j=2,3,4$) events expected at each centre-of-mass energy 
 was calculated from the cross-sections, the integrated luminosity and
 the detection efficiency. We assume that the branching fractions of    
 $\chpm_{1,2}\rightarrow\qqbar'\gluino$ and 
 $\nt_{2,3,4}\rightarrow\qqbar\gluino$
 are 100\% and that the detection efficiency for chargino events with 
 $s'/s >0.8$ is the same as that for $\qqbar$ events at each centre-of-mass 
 energy\footnote{Note that no cuts have been applied to remove hadronic 
 W-pair events. If such cuts had been made, this might
 have removed chargino pairs. Therefore the assumption that the 
 efficiency is the same is justified. 
% The missing energy caused by possible 
% invisible decay products of light gluino bound states~\cite{bib:Farrar}
% is not expected to be large enough to increase the inefficiency of the
% $s'$ cut significantly.}. 
%mjk Andrea did some checks with WW-->K0l + K0l + X 
%mjk dont know if we should be more specific than "not expected to be" 
%mjk what about adding:
 A small decrease of ${\cal O}(5-10\%)$ on the $s'$ cut 
 efficiency is e.g.\ predicted in
 Monte Carlo simulations of WW$\to\PKL\PKL$+X, for high-mass
 \PKL\PKL\ pairs, when the energy deposits in the hadron calorimeter
 are ignored. Assuming similar fragmentation behaviour for gluinos 
 and strange quarks~\cite{bib:Farrar}, the missing energy caused by possible 
 invisible decay products of light gluino bound states
 is therefore not expected to be large enough to reduce the 
 efficiency of the  $s'$ cut significantly.}. 
%mjk  
  We make a similar assumption about the detection
  efficiency for neutralino pairs, except when $m_{\nt_2} < 10$~GeV, where
  the decay multiplicity could be low. We conservatively assume the
  efficiency is zero for $\nt_2\nt_2$ events if $m_{\nt_2} < 10$~GeV.
  If $m_{\nt_2} < 3$~GeV, the decay $\nt_2\rightarrow\gamma\nt_1$ may
  be significant, which could affect our detection efficiency; therefore
  we have conservatively assumed the efficiency to be zero for
  $\nt_2 \nt_j$ events ($j=3,4$) if $m_{\nt_2} < 3$~GeV.
 
\begin{figure}
\begin{center}
\epsfxsize=\textwidth
%\epsfbox[100 250 567 650]{sglim_bw.ps}
\epsfbox[100 250 567 650]{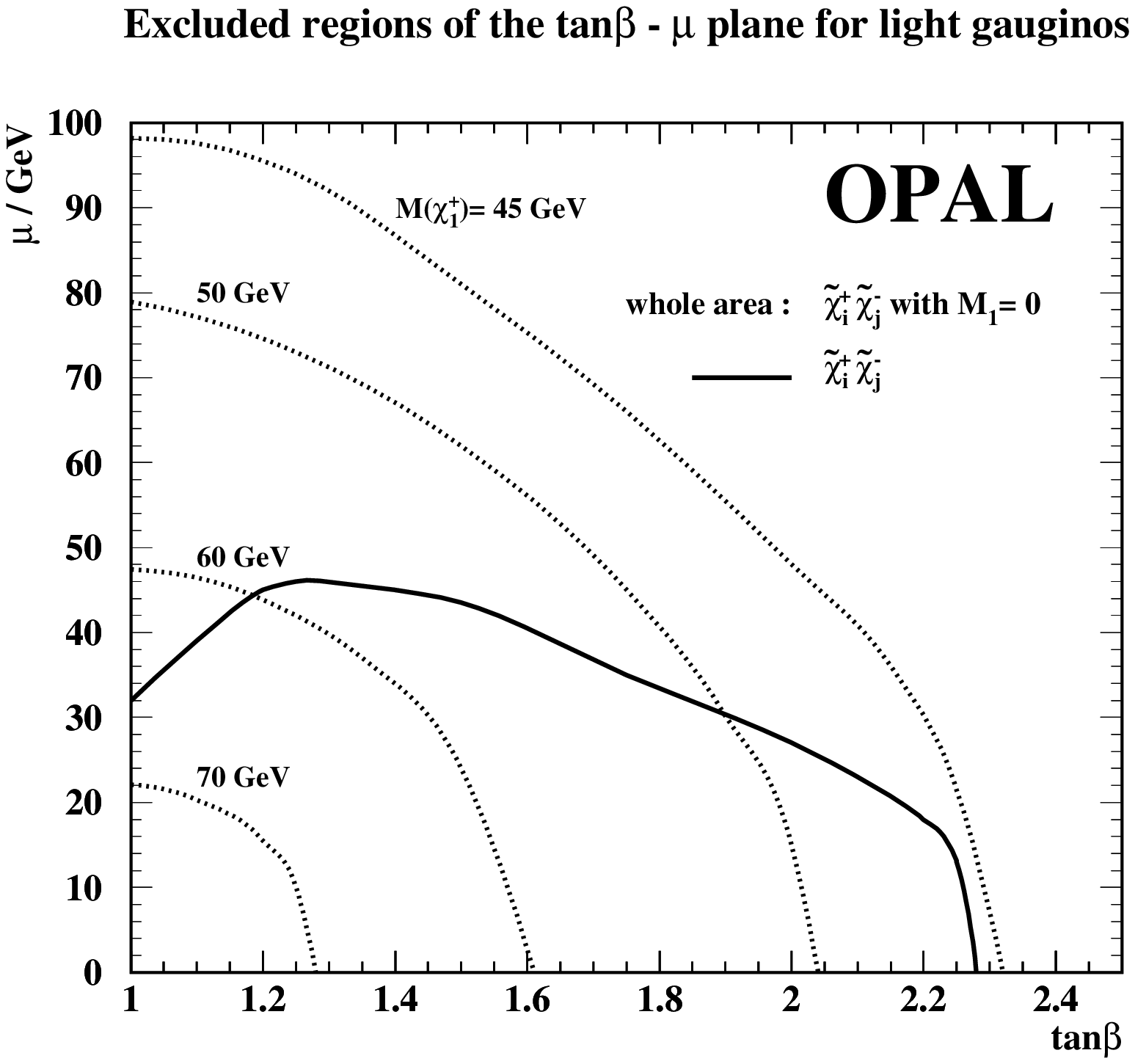}
\caption{The 95\% confidence level lower limit of $\mu$ as a function of 
$\tan \beta$ calculated assuming $M_2 = M_3 = 0$, with no restriction
on $M_1$. The region above the solid curve is excluded.
The sneutrino mass is assumed to be larger than the current limit 
of 43~GeV, and the decay branching fraction of 
$\chpm\rightarrow\qqbar'\gluino$ is assumed to be 100\%.
The contours for chargino masses $m_{\chpm} =$ 45~GeV, 50~GeV, 60~GeV and 
70~GeV are also shown with dotted lines. 
With the additional restriction $M_1 = 0$, the existence of light
gluinos is excluded everywhere in the $\mu$--$\tan\beta$ plane.
}
\label{fig:sglim} 
\end{center}
\end{figure}

 We first consider the case where the difference between the observed
 and expected numbers of hadronic events is due to the production of charginos
 ($\chp_i\chm_j, i,j=1,2$) only. This corresponds to a SUSY model
 with the looser condition $M_3 = M_2 = 0$ 
 with no restriction on $M_1$~\cite{bib:Farrar}.  
 The 95\% confidence level lower limit obtained on the $\mu$ parameter
 as a function of $\tan \beta$ for this case is shown in 
 figure~\ref{fig:sglim}.
 
 We then consider $M_3 = M_2 = M_1 = 0$~\cite{bib:Farrar} and calculate the 
 total number of chargino and neutralino events with the assumptions above.
 We derive limits on the SUSY parameter
 $\mu$ as a function of $\tan\beta$.
 The entire $\mu$--$\tan \beta$ region is excluded.  Therefore in this
 particular model these excluded parameter 
 regions imply the exclusion of a light gluino
 under the assumptions that the branching fractions of    
 $\chpm_{1,2}\rightarrow\qqbar'\gluino$ and 
 $\nt_{2,3,4}\rightarrow\qqbar\gluino$
 are 100\% and that the detection efficiency for chargino events 
 and neutralino events (with $m_{\nt_2} > 10$~GeV) 
 is the same as that for $\qqbar$ events with $s'/s >0.8$ at each 
 centre-of-mass 
 energy.  This limit would continue to hold even if smaller branching ratios
% decrease the total number of expected events by 77\%, i.e., for
% $m_{\tilde{\mathrm{q}}}\simeq 70$~GeV.
%mjk  you mean by 23%, not by 77%??????
  or a slightly reduced efficiency of the $s'$ cut 
  were to decrease the total number of expected events by 23\%.
%mjk i would omit the "i.e. ..."
%mjk it puzzles me a bit, because we say in the beginning
%mjk that we asuume m_squark \sim m_W , 70 GeV *is* \sim m_W...
%mjk so what is teh diference to the initial assumption?

%%-----------------------------------------------------------------------
\section{Conclusions}    \label{sec:sum}
%%-----------------------------------------------------------------------
 
We have presented new measurements of the production of events with
two-fermion hadronic and leptonic final states in \epem\ collisions at
a centre-of-mass energy of 172~GeV, and updated similar measurements
at 130--161~GeV. Special attention has been paid to the treatment of
the interference between initial- and final-state radiation, and the
contribution from four-fermion production. The measured rates and 
distributions are all consistent with the Standard Model expectations.
In a model-independent fit to the \PZ\ lineshape, the inclusion of data at
172~GeV provides an improved constraint on the size of the
interference between \PZ\ and photon amplitudes. Within the framework
of the Standard Model, the data have been used to measure the
electromagnetic coupling constant, giving $1/\alphaem(157.42~\GeV) =
119.9^{+5.1}_{-4.1}$, compatible with expectation. 

We have used these data to place limits on possible deviations from
the Standard Model represented by effective four-fermion contact
interactions. Limits are obtained on the energy scale $\Lambda$ generally 
in the range 2--7~TeV, assuming $\mathrm{g}^2/4\pi = 1$.  We have
searched for the effect of a new heavy particle which might be exchanged 
in the $t$-channel. Limits are obtained on the coupling constants 
$\mathrm{g_{L}}$, $\mathrm{g_{R}}$ between typically 0.2 and 0.6 for masses 
below $\sim200~\GeV$. These limits can be interpreted both 
as limits on leptoquarks, or in some cases as limits on squarks in 
supersymmetric theories  with $R$-parity violation. Compared to previous 
searches we are able to improve existing limits in particular in the high  
mass region $\mX>250\,\GeV$.  Limits have also been placed on chargino
and neutralino decays to light gluinos in supersymmetric extensions of 
the Standard Model.

\clearpage
%%-----------------------------------------------------------------------
\section*{Acknowledgements}
%%-----------------------------------------------------------------------
We would like to thank F.~Caravaglios, H.~Dreiner, G.F.~Giudice and
R.~R\"uckl for helpful discussions concerning the use of these measurements 
in searching for effects of the $t$-channel exchange of new particles, 
and G.~Farrar for help in setting limits on light gluino production.
We are grateful to G.~Altarelli for comments and suggestions on the
contact interactions.

We particularly wish to thank the SL Division for the efficient operation
of the LEP accelerator at all energies
 and for
their continuing close cooperation with
our experimental group.  We thank our colleagues from CEA, DAPNIA/SPP,
CE-Saclay for their efforts over the years on the time-of-flight and trigger
systems which we continue to use.  In addition to the support staff at our own
institutions we are pleased to acknowledge the  \\
Department of Energy, USA, \\
National Science Foundation, USA, \\
Particle Physics and Astronomy Research Council, UK, \\
Natural Sciences and Engineering Research Council, Canada, \\
Israel Science Foundation, administered by the Israel
Academy of Science and Humanities, \\
Minerva Gesellschaft, \\
Benoziyo Center for High Energy Physics,\\
Japanese Ministry of Education, Science and Culture (the
Monbusho) and a grant under the Monbusho International
Science Research Program,\\
German Israeli Bi-national Science Foundation (GIF), \\
Bundesministerium f\"ur Bildung, Wissenschaft,
Forschung und Technologie, Germany, \\
National Research Council of Canada, \\
Hungarian Foundation for Scientific Research, OTKA T-016660, 
T023793 and OTKA F-023259.\\
 
\clearpage 
%---------------------------------------------------------------------
%       References
%---------------------------------------------------------------------

%-----------------------------------------------------------------------
%       Figures
%-----------------------------------------------------------------------
%
\clearpage
%%%%%%%%%%%%%%%%%%%%%%%%%%%%%%%%%%%%%%%%%%%%%%%%%%%%%%%%%%
%
%%%%%%%%%%%%%%%%%%%%%%%%%%%%%%%%%%%%%%%%%%%%%%%%%%%%%%%%%%
%
%%%%%%%%%%%%%%%%%%%%%%%%%%%%%%%%%%%%%%%%%%%%%%%%%%%%%%%%%%
%
%%%%%%%%%%%%%%%%%%%%%%%%%%%%%%%%%%%%%%%%%%%%%%%%%%%%%%%%%%
%
%%%%%%%%%%%%%%%%%%%%%%%%%%%%%%%%%%%%%%%%%%%%%%%%%%%%%%%%%%
%
%%%%%%%%%%%%%%%%%%%%%%%%%%%%%%%%%%%%%%%%%%%%%%%%%%%%%%%%%%
%
%%%%%%%%%%%%%%%%%%%%%%%%%%%%%%%%%%%%%%%%%%%%%%%%%%%%%%%%%%
%
%%%%%%%%%%%%%%%%%%%%%%%%%%%%%%%%%%%%%%%%%%%%%%%%
%
\clearpage
%%%%%%%%%%%%%%%%%%%%%%%%%%%%%%%%%%%%%%%%%%%%%%%%%%%%%%%%%%
%
%%%%%%%%%%%%%%%%%%%%%%%%%%%%%%%%%%%%%%%%%%%%%%%%%%%%%%%%%%
%
%%%%%%%%%%%%%%%%%%%%%%%%%%%%%%%%%%%%%%%%%%%%%%%%%%%%%%%%%%
%
%%%%%%%%%%%%%%%%%%%%%%%%%%%%%%%%%%%%%%%%%%%%%%%%%%%%%%%%%%
%
%%%%%%%%%%%%%%%%%%%%%%%%%%%%%%%%%%%%%%%%%%%%%%%%%%%%%%%%%%
%
%%%%%%%%%%%%%%%%%%%%%%%%%%%%%%%%%%%%%%%%%%%%%%%%%%%%%%%%%%
%
%%%%%%%%%%%%%%%%%%%%%%%%%%%%%%%%%%%%%%%%%%%%%%%%%%%%%%%%%%
%
%%%%%%%%%%%%%%%%%%%%%%%%%%%%%%%%%%%%%%%%%%%%%%%%%%%%%%%%%%
%
%%%%%%%%%%%%%%%%%%%%%%%%%%%%%%%%%%%%%%%%%%%%%%%%%%%%%%%%%%
%
%%%%%%%%%%%%%%%%%%%%%%%%%%%%%%%%%%%%%%%%%%%%%%%%%%%%%%%%%%
%
%%%%%%%%%%%%%%%%%%%%%%%%%%%%%%%%%%%%%%%%%%%%%%%%%%%%%%%%%%
%
%%%%%%%%%%%%%%%%%%%%%%%%%%%%%%%%%%%%%%%%%%%%%%%%%%%%%%%%%%
%
%%%%%%%%%%%%%%%%%%%%%%%%%%%%%%%%%%%%%%%%%%%%%%%%%%%%%%%%%%
%

%%%%%%%%%%%%%%%%%%%%%%%%%%%%%%%%%%%%%%%%%%%%%%%%%%%%%%%%%%
%

%%%%%%%%%%%%%%%%%%%%%%%%%%%%%%%%%%%%%%%%%%%%%%%%%%%%%%%%%%%%%%%%%%%%%%%%
\end{document}